\documentclass[12pt, a4paper]{article} 

\usepackage[margin=1in]{geometry} 
\usepackage{amsfonts, amscd, amssymb, mathtools, mathrsfs, dsfont, bbm, bbding} 
\usepackage[amsmath, amsthm, thmmarks]{ntheorem} 
\usepackage{graphicx, epstopdf, xypic, color, float} 
\usepackage{indentfirst}
\usepackage{lmodern}
\usepackage[T1]{fontenc} 
\usepackage{enumerate, listings, verbatim, paralist}
\usepackage{setspace, xspace}
\usepackage[colorlinks=true, citecolor=blue]{hyperref}
\usepackage{extarrows}
\usepackage{pdfpages}
\usepackage{ulem}

\usepackage{setspace}
\AtBeginDocument{%
\addtolength\abovedisplayskip{-0.15\baselineskip}%
\addtolength\belowdisplayskip{-0.15\baselineskip}%
\addtolength\abovedisplayshortskip{-0.15\baselineskip}%
\addtolength\belowdisplayshortskip{-0.15\baselineskip}%
}

\newtheorem{thm}{Theorem} [section]
\newtheorem{prop}[thm]{Proposition}

\newtheorem{lemma}[thm]{Lemma}

\newtheorem{assmp}{Assumption}[section]
\newtheorem{remark}{Remark}[section]

\numberwithin{equation}{section} 

\renewcommand{\geq}{\geqslant}
\renewcommand{\leq}{\leqslant}

\newcommand{\citethm}[1]{Theorem \ref{#1}}
\newcommand{\citeprop}[1]{Proposition \ref{#1}}

\newcommand{\citelem}[1]{Lemma \ref{#1}}

\newcommand{\citeassmp}[1]{Assumption \ref{#1}}

\newcommand{\citefig}[1]{Figure \ref{#1}}
\newcommand{\citeapp}[1]{Appendix \ref{#1}}

\newcommand{\opfont}{\mathbb}

\newcommand{\BE}[2][]{\ensuremath{\operatorname{\opfont{E}}^{#1}\!\left[#2\right]}}
\newcommand{\bp}{\ensuremath{\opfont{P}}}

\newcommand{\BF}{\ensuremath{\mathcal{F}}}

\newcommand{\R}{\ensuremath{\operatorname{\mathbb{R}}}}

\newcommand{\dd}{\ensuremath{\operatorname{d}\! }}
\newcommand{\dt}{\ensuremath{\operatorname{d}\! t}}
\newcommand{\ds}{\ensuremath{\operatorname{d}\! s}}

\newcommand{\dx}{\ensuremath{\operatorname{d}\! x}}

\newcommand{\ddp}{\ensuremath{\operatorname{d}\! p}}

\newcommand{\setq}{\mathscr{Q}}

\newcommand{\idd}[1]{\ensuremath{\operatorname{\mathds{1}}_{#1}}}

\newcommand{\barc}{L}

\newcommand{\blue}[1]{{\color{blue}#1}}

\newcommand{\nn}{\nonumber}

\newenvironment{proofof}[1]{\noindent\emph{\textbf{Proof of #1}.}\hspace{1ex}}{\hspace*{\fill}$\Box$\newline}

\newcommand{\BV}[2][]{\ensuremath{\operatorname{\mathscr{E}}^{#1}\!\left[#2\right]}}

\newcommand{\ep}{\varepsilon}

\newcommand{\bc}{\wp}
\newcommand{\seta}{\mathscr{A}}
\newcommand{\rr}{R}
\newcommand{\qlow}{\hat{c}}
\newcommand{\hatv}{\hat{v}_{\qlow}}
\newcommand{\inv}{I}
\newcommand{\setu}[1][\eta]{\mathscr{U}_{#1}}
\newcommand{\barQ}{\overline{Q}}
\newcommand{\hatphi}{\widehat{\varphi}}
\newcommand{\barho}{\overline{\rho}}

\begin{document}
\title{Relative growth rate optimization \\
under behavioral criterion} 
\author{
Jing Peng\thanks{\rm Department of Applied Mathematics, The Hong Kong Polytechnic University, Hong Kong. Email: jing.peng@connect.polyu.hk.}
\and
Pengyu Wei\thanks{\rm Division of Banking and Finance, Nanyang Business School, Nanyang Technological University, Singapore. E-mail: pengyu.wei@ntu.edu.sg.}
\and
Zuo Quan Xu\thanks{\rm Department of Applied Mathematics, The Hong Kong Polytechnic University, Hong Kong. Email: maxu@polyu.edu.hk.}
} 

\date{\today}
\maketitle

\begin{abstract} 
This paper studies a continuous-time optimal portfolio selection problem in the complete market for a behavioral investor whose preference is of the prospect type with probability distortion. The investor concerns about the terminal relative growth rate (log-return) instead of absolute capital value. This model can be regarded as an extension of the classical growth optimal problem to the behavioral framework. 
It leads to a new type of $M$-shaped utility maximization problem under nonlinear Choquet expectation. Due to the presence of probability distortion, the classical stochastic control methods are not applicable. By the martingale method, concavification and quantile optimization techniques, we derive the closed-form optimal growth rate. We find that the benchmark growth rate has a significant impact on investment behaviors. 
Compared to Zhang et al \cite{ZJZ11} where the same preference measure is applied to the terminal relative wealth, we find a new phenomenon when the investor's risk tolerance level is high and the market states are bad. In addition, our optimal wealth in every scenario is less sensitive to the pricing kernel and thus more stable than theirs. 
\end{abstract}
\newpage

\section{Introduction}

One of the tacit rules adopted in the modern portfolio optimization models is the expected utility hypothesis on the psychology of decision-making. It dates back to the pioneering work of Merton \cite{M69,M75} in which an investor chooses a portfolio built upon a simplified market and seeks to maximize the expected utility from consumption. For models that do not consider consumption behavior, investors' targets are usually to maximize the expected utility of terminal wealth or/plus either exogenous or endogenous penalty terms. Such goals seem quite natural since the majority of investors are concerned about how much wealth they will accumulate when they decide to liquidate their portfolios. However, in practice, practitioners and institutions, especially those with long-term investment plans, do not always keep their eyes on the capital values of their portfolios. Instead, they care more about their portfolios' growth or return rates (such as seasonal or annual return, internal rate of return), and often adopt relative return rates (such as excess return rate over a stock index or interest rate) to measure their portfolios' performance.

Taking growth rate rather than capital value as a performance evaluation metric has many advantages in practice. It allows for a comparison between various portfolios because it removes the impact of the initial endowment and investment horizon. For fund managers, achieving a higher return rate represents excellent investment skills and brings fame and fortune. 
Moreover, it is difficult or almost impossible to do comparisons on portfolios' capital values between worldwide practitioners and institutions who trade different currencies, because exchange rates have so significant impacts that they may change the comparison results. By contrast, taking growth rates (such as the well-known gross domestic products of different regions) as measurement objectives will avoid the setback of using capital values. 
In this light, it seems more natural for practitioners to measure their portfolios' performances in terms of growth rates rather than capital values. 

Taking growth rate as a measurement objective is not new in the financial economics literature. It at least goes back to the seminal Markowitz's \cite{M52} single-period mean-variance model. 
In Markowitz's original formulation, the mean-variance analysis was imposed on the return rate right from the beginning. Ever since most models on mean-variance analysis focus on capital values since they are equivalent in the single-period framework. 
Recently, Dai et al. \cite{DJKX20} investigated a dynamic mean-variance portfolio choice problem on growth rate. The optimal policies under specific settings are found to be consistent with several conventional investment wisdoms that are usually contradicted by models based on capital values. 

Kelly \cite{K56} introduced the growth-optimal portfolio (GOP) that is to maximize the expected growth rate. In theory and practice, the GOP has enjoyed wide applications in a large number of areas including portfolio theory, utility theory, game theory, information theory, asset pricing theory, and insurance theory (e.g. \cite{ABHSV00,T00}). However, these models ignore the relevant risk when selecting a desirable portfolio. This research intends to partly fill the gap in a continuous-time setting by investigating a GOP model with behavioral risk control.

Another motivation for our paper is to take the behavioral elements into our model. 
It is well known that the classical expected utility theory (EUT) and Markowitz's mean-variance theory (MVT) have failed to explain many phenomena and paradoxes (such as the equity premium puzzle \cite{CC99}) in practice. The Nobel prize-winning work Kahneman and Tversky's \cite{KT79, TK92} cumulative prospect theory (CPT) has been widely regarded as a competing alternative to EUT and MVT. There are three key ingredients in the CPT: a reference point, an $S$-shaped utility\footnote{It is convex below the reference point and concave above it.} and a probability weighting (or distortion) function. In our model, the risk is measured by the CPT preference, and the measurement objective is the relative growth rate to some given random or deterministic benchmark rate. We use the well-known Kahneman and Tversky's piece-wise power utility and general smooth probability weighting function. Incorporating behavioral criteria into continuous-time portfolio choice problems has received much attention in recent years (see, e.g., Berkelaar et al. \cite{BKP04}, Jin and Zhou \cite{JZ08}, He and Zhou \cite{HZ11}, Zhang et al. \cite{ZJZ11}, Xu \cite{X16}, Hou and Xu \cite{HX16}, Mi and Xu \cite{MX21}). These works focus on capital value maximization. 
To our best knowledge, less attentions are paid to growth rate oriented portfolio choice problems under behavioral criteria. A recent progress comes from Wei and Xu \cite{WX21} which focuses on the mean-risk problem. 

Our model leads to a non-concave utility maximization problem under nonlinear Choquet expectation. Due to the presence of probability distortion, the classical stochastic control methods are no more applicable. We adopt the martingale method to transform the problem into a quantile optimization problem, which turns out to be a new type of $M$-shaped\footnote{It is first concave, then convex, and finally concave from left to the right on the real line.} utility maximization problem. We then use the concavification and quantile optimization methods to tackle it, and 
derive the optimal growth rate in closed form. 
We find that the benchmark growth rate has a significant impact on investment behaviors. 

We also present a numerical example to compare our results with Zhang et al.'s \cite{ZJZ11}. 
Our terminal relative growth rate is higher than or at least equal to Zhang et al.'s in the intermediate and bad market states. By contrast, ours is lower than theirs in the good market states. From an economic perspective, our strategy is more suitable for loss-sensitive investors than Zhang et al.'s, whereas their strategy is more suitable for those betting for extremely good market states. 

The rest of this paper is organized as follows. We present our financial market and model in Section 2 and then turn it into a static random variable optimization problem by the martingale method in Section 3. The latter is transformed into a quantile optimization problem in Section 4 and solved in Section 5. Section 6 presents a numerical example to compare our results to Zhang et al.'s \cite{ZJZ11}. Some proofs and additional analyses are placed in the Appendices. 

\section{Problem formulation}\label{Problem formulation}
Let $T>0$ denote a fixed finite investment horizon throughout this paper. Consider a complete filtered probability space $(\Omega, \mathcal{F}, \bp)$, on which a standard $n$-dimensional Brownian motion $W(\cdot) \equiv (W^1(\cdot),\ldots, W^n(\cdot))'$ is defined. We assume the uncertainty of the market entirely comes from the Brownian motion and let the information filtration $\{\BF_t\}_{0\leq t\leq T}$ be generated by the Brownian motion, augmented by all the $\bp$-null sets. Let $\mathcal{F}={\BF}_T$. For any vector or matrix $M$, we use $M'$ to denote its transpose and $\Vert M\Vert=\sqrt{\text{trace}(MM')}$ to denote the Euclidean norm.
All random variables (r.v.s) considered in this paper are assumed to be $\BF_T$-measurable.

\subsection{Market and portfolio}

The financial market consists of $n+1$ assets which are traded continuously over the investment horizon $[0, T]$. The market is frictionless, i.e., there are no trading constraints, transaction costs or taxes etc. The first asset is a bond whose price $S_0(\cdot)$ evolves according to an ordinary differential equation (ODE): 
\begin{align*}
\begin{cases}
\dd S_0(t) = r(t)S_0(t)\dt, \quad t\in[0,T], \\
S_0(0) = s_0 > 0,
\end{cases}
\end{align*}
where $r(t)$ is the instantaneous interest rate of the bond at time $t$. The remaining $n$ assets are {stocks} (also called risky assets), and their prices $S_{i}(\cdot)$, $i=1,2,\ldots,n$, are modeled by stochastic differential equations (SDEs): 
\begin{align}\label{stockprice}
\begin{cases}
\dd S_i(t) = S_i(t)\big\{\mu_i(t)\dt + \sum_{j=1}^n\sigma_{ij}(t)\dd W^j(t)\big\}, \quad t\in[0,T], \\
S_i(0) = s_i > 0,
\end{cases}
\end{align}
where $\mu_i : [0,T]\times\Omega\rightarrow \mathbb{R}$ is called the \emph{appreciation rate} of stock $i$, and $\sigma_{ij}: [0,T]\times\Omega\rightarrow \mathbb{R}$ is called the \emph{volatility coefficient} of stock $i$ with respect to $W^j$ at time $t$. Define the appreciation rate vector process $\mu=(\mu_1, \mu_2,\cdots, \mu_n)':[0,T]\times\Omega\rightarrow \mathbb{R}^n$ and 
the volatility matrix process $\sigma:=(\sigma_{ij})_{n\times n}:[0,T]\times\Omega\rightarrow \mathbb{R}^{n\times n}$. 
We impose the following standard assumption which ensures the market is free of arbitrage opportunities. 
\begin{assmp}\label{ASS2.1}
The processes $r$, $\mu$, $\sigma$ are $\{\mathcal{F}_t\}$-progressively measurable and satisfy
\[\int_{0}^{T}|r(s)| d s<+\infty, \text { a.s. }\] and \[\int_{0}^{T}\left[\Vert \mu(t)\Vert+\Vert\sigma(t)\Vert^2\right] \dd t<+\infty, \text { a.s. }\]
Moreover, there exists a unique $\{\mathcal{F}_t\}$-progressively measurable, uniformly bounded risk premium process $\theta:[0,T]\times\Omega\rightarrow \mathbb{R}^{m}$ such that \[\sigma(t)\theta(t)= (\mu_{1}(t)-r(t),\ldots, \mu_n(t)-r(t))^{\prime} \; a.s.,\; a.e. \; t\in[0,T].\]
\end{assmp}
The last statement in \citeassmp{ASS2.1} implies that the matrix $\sigma(t)$ is invertible a.s. for a.e. $t\in[0,T]$.

Consider an investor (``She'') with an initial endowment $x_0>0$. She is a small investor so her actions cannot affect asset prices in the market. Let $\pi_i(t)$ denote the \emph{proportion} of her total wealth invested in stock $i$ at time $t$, $ i = 1,\ldots, n$. The proportion invested in the bond at time $t$ is $1-\sum_{i=1}^{n}\pi_i(t)$. 
We call the vector process $\pi:=(\pi_1,\ldots,\pi_n)^{\prime}:[0,T]\times\Omega\rightarrow \mathbb{R}^n$ a \emph{portfolio} (also called a \emph{strategy}) and denote $X^{\pi}$ the agent's \emph{wealth process} with the implementation of the portfolio $\pi$. Assume that the trading of assets takes place continuously in a self-financing fashion (i.e., there is no consumption or income). Then the wealth process $X^{\pi}$ evolves according to the following SDE (see Karatazas and Shreve \cite{KS98}): 
\begin{align}\label{wealthprocess}
\begin{cases}
\dd X^{\pi}(t) =X^{\pi}(t) \big[(r(t)+ \pi(t)^{\prime}\sigma(t)\theta(t))\dt + \pi(t)^{\prime}\sigma(t) \dd W(t)\big],\quad t\geqslant 0, \\
X^{\pi}(0) =x_0>0.
\end{cases}
\end{align}
A portfolio $\pi$ is called admissible if it is an $\{\mathcal{F}_t\}$-progressively measurable process such that 
\[\int_{0}^{T} \Vert\sigma(s)^{\prime}\pi(s)\Vert^2 \ds<\infty, \text { a.s., }\]
and \eqref{wealthprocess} admits a unique strong solution $X^{\pi}$. 
From now on, we focus on admissible portfolios, the set of which is denoted by $\seta$.

Applying It\^{o}'s lemma to $\log X^{\pi}(t)$, we obtain 
\begin{align*}
\qquad X^{\pi}(t)= x_0 \exp\bigg\{\int_{0}^{t}\big(r(s) + \pi(s)^{\prime}\sigma(s)\theta(s)
&-\frac{1}{2}\Vert\sigma(s)^{\prime}\pi(s)\Vert^2 \big)\ds\\
&+\int_{0}^{t}\pi(s)^{\prime}\sigma(s) \dd W(s)\bigg\}>0. \qquad 
\end{align*}
Therefore, we have a no-bankruptcy condition inherent in the wealth process.

\par
A particular feature of our model is that the performance is measured on the \emph{terminal relative growth rate} (also called \emph{terminal relative return rate}) rather than on the terminal wealth as in most financial economics literature. The investor's objective is to minimize the behavioral criterion of the terminal relative growth rate. 
To this end, we need to introduce a \emph{growth rate process} and a \emph{benchmark growth rate} (also called the \emph{target growth rate} or simply the \emph{benchmark}).

The growth rate process $\rr^{\pi}$ associated with a portfolio $\pi$ is defined as 
$$\rr^{\pi}(t)=\log (X^{\pi}(t)/x_{0}),\quad t\geq 0.$$ 
The normalized growth rate over $[0,t]$ is then given by $t^{-1}\rr^{\pi}(t)$ for $t>0$.
By It\^{o}'s Lemma and \eqref{wealthprocess}, we have 
\begin{align}\label{returnprocess}
\begin{cases}
\dd \rr^{\pi}(t)= \big(r(t)+ \pi(t)^{\prime}\sigma(t)\theta(t)-\frac{1}{2}\pi(t)^{\prime}\sigma(t) \sigma(t)^{\prime} \pi(t)\big)\dt + \pi(t)^{\prime}\sigma(t) \dd W(t),\quad t\geqslant 0,\\
\rr^{\pi}(0)=0.
\end{cases}
\end{align}
Note that $\rr^{\pi}(t)$ may take negative values, meaning that the position of the investor slides into a loss. 

\par
The benchmark growth rate, denoted by $\bc$, is a given ${\BF}_T$-measurable random variable. The corresponding terminal benchmark wealth is $x_0e^{\bc}$.
In practice, $\bc$ can be chosen as the return of a stock index or structured product, the deposit or loan interest rate, or a constant target rate, etc. 

We assume that the investor will not use any strategy that underperforms the benchmark significantly. Mathematically, we impose the following constraint: 
\begin{align}\label{lowerbound}
\rr^{\pi}(T)-\bc \geq -c,
\end{align}
where $c>0$ is a constant that represents the \emph{risk tolerance level}. 
To exclude trivial cases, we assume
\begin{align}\label{lowerbound1}
\text{ there exists at least one admissible portfolio $\pi_{0}$ such that } \rr^{\pi_{0}}(T)-\bc > -c.
\end{align}
 We will show that this requirement is equivalent to \citeassmp{feasibility} below, which can be easily verified in terms of $\bc$, $c$, and market parameters. 

Now we present our optimal portfolio selection problem with behavioral criterion on the terminal relative growth rate: 
\begin{align}\label{Optimalcontrol0}
\max_{\pi\in\seta} & \quad \BV{u(\rr^{\pi}(T)-\bc)}\\
\mathrm{s.t.} &\quad \text{$(\pi, \rr^{\pi})$ satisfies \eqref{returnprocess} and \;}\rr^{\pi}(T)-\bc \geq -c. \nn
\end{align} 
In this model, the investor adopts the CPT preference $\BV{u(\cdot)}$ to evaluate the terminal relative growth rate. Here, $\BV{\cdot}$ is a nonlinear expectation (indeed a Choquet expectation) and $u$ is an $S$-shaped utility function, which will be specified in the next section.

\blue{It is important to point out that neither the constraints nor the objective of Problem \eqref{Optimalcontrol0} depend on the initial endowment $x_{0}$, as we consider proportional investment strategies and logarithmic returns. 
Consequently, the optimal portfolio $\pi^*$ and the corresponding growth rate $\rr^{\pi^*}$ must be independent of $x_0$, meaning that they are the same for all investors regardless of initial endowments.}

\subsection{CPT preference}
\noindent

In our model, the investor uses a CPT preference $\BV{u(\cdot)}$. It depends on a nonlinear expectation $\BV{\cdot}$ and a non-concave utility function $u$.

\par
First, we introduce the nonlinear expectation $\BV{\cdot}$ induced by a probability weighting (or distortion) function. A function $w\colon [0,1]\mapsto[0,1] $ is called a probability weighting function if it is strictly increasing and continuously differentiable with
 $w(0)=0$ and $w(1)=1$ (three types of probability weighting functions are shown in Figure \ref{figure:distortion}). 
The expectation $\BV{\cdot}$ for a r.v. $\xi$ is a Choquet expectation defined by 
\begin{equation}\label{Choquet integral}
\BV{\xi}=\int_0^{\infty}w(1-F_{\xi}(x))\dx-\int_{-\infty}^0 (1-w(1-F_{\xi}(x)))\dx,
\end{equation}
provided that at least one of the integrals is finite. The Choquet expectation $\BV{\cdot}$ is in general nonlinear, unless there is no probability weighting (i.e. $w$ is the identity function) in which case $\BV{\cdot}$ becomes the standard linear mathematical expectation $\BE{\cdot}$ and the CPT preference $\BV{u(\cdot)}$ becomes the classical EUT preference $\BE{u(\cdot)}$. Choquet expectations have many applications in statistics, economics, and finance; see Schmeidler \cite{S89}, Wakker \cite{W01} and references therein. 
\begin{remark}
Tversky and Kahneman \cite{TK92} used different probability weighting functions for the gain part (i.e., $\xi> 0$) and loss part (i.e., $\xi<0$). In this paper, we use the same probability weighting function for them. It is possible to consider different probability weighting functions for the two parts using the ``divide and conquer'' machinery developed by Jin and Zhou \cite{JZ08} to study the corresponding model. We leave this to interested readers.
\end{remark}
\begin{remark}
When there is no probability weighting, our model reduces to the classical EUT model with a non-concave utility function. The latter has been studied in Bian et al. \cite{BCX19} and Guan et al. \cite{GLXY17}, among others. 
\end{remark}
\par
Second, the CPT involves an $S$-shaped utility function $u$ which is convex on $(-\infty, 0]$ and concave on $[0,\infty)$. Following the seminal work of Tversky and Kahneman \cite{TK92}, 
we adopt the piece-wise power utility function: 
\begin{align}\label{valuefunction}
u(x)=\begin{cases}
x^{\alpha}, &\quad x\geq 0;\\
-\kappa(-x)^{\beta}, &\quad x<0,
\end{cases}
\end{align}
where $0<\alpha$, $\beta<1$ are risk aversion parameters, and $\kappa>0$ represents the degree of {loss aversion}.\footnote{The specific parameters studied in \cite{TK92} are $\alpha=\beta=0.88$ and $\kappa=2.5$. 
} The utility function $u$ is continuous and strictly increasing on $(-\infty,\infty)$, $C^\infty$ smooth except at the point $x=0$, convex on $(-\infty, 0]$ and concave on $[0,\infty)$ (thus $S$-shaped).
One can generalize our model to more general utility functions at the cost of less explicit solutions. 

From the CPT preference $\BV{u(\cdot)}$ introduced above, we see that, on one hand, it is fully nonlinear; on the other hand, it is non-concave. Due to the presence of nonlinear expectation $\BV{\cdot}$ in the target, the problem \eqref{Optimalcontrol0} is not a standard stochastic control problem (namely unlike those in Yong and Zhou \cite{YZ99}). The nonlinear expectation leads to the failure of dynamic programming and maximum principle, so the standard results and tools in the classical control theory cannot be applied to tackle it immediately. Although the delicate control theory of mean-field type may be adopted to study our problem, it seems very hard to get explicit solutions by that approach. One should notice that the Choquet expectation $\BV{\cdot}$ is essentially different from the nonlinear $g$-expectation introduced by Peng \cite{P97, P99}; see Chen et al. \cite{CCD05}. So the results and tools from the $g$-expectation theory can not be applied to our problem as well. We will provide closed-form solutions by relatively simple arguments. 

\section{Martingale method} 
As standard tools such as the dynamic programming in the control theory cannot be applied to tackle Problem \eqref{Optimalcontrol0}, we use the martingale method instead. The latter allows us to reduce the dynamic stochastic control problem to a static (but still non-concave) r.v. optimization problem. We need to assume the market is complete in order to use the martingale method.\footnote{We can deal with the problem in some special incomplete markets such as when all the market parameters are deterministic. In this case, the market is essentially equivalent to a complete one; see \cite{HZ11} for a more detailed discussion.}

We introduce the \emph{state price density process} 
\[\rho(t)=\exp\left(-\int_{0}^{t}\Big(r(s)+\tfrac{1}{2}\Vert\theta(s)\Vert^{2}\Big)\ds-\int_{0}^{t}\theta(s)'\dd W(s)\right),\quad t\in[0,T].\]
Then It\^{o}'s lemma gives
\[\dd \rho(t)=-\rho(t)\big(r(t)\dt+\theta(t)'\dd W(t)\big).\]
We denote $\rho=\rho(T)$\footnote{In the financial economics literature, $\rho(T)$ is often called the pricing kernel.} from now on. 
For any admissible portfolio $\pi$, we have, by It\^{o}'s lemma and \eqref{returnprocess},
\begin{align}
\dd\; (\rho(t)e^{\rr^{\pi}(t)})=\rho(t)e^{\rr^{\pi}(t)}\big(\pi(t)^{\prime}\sigma(t)-\theta(t)^{\prime}\big)\dd W(t).
\end{align}
Hence, the process $\rho(t)e^{\rr^{\pi}(t)}$ is a local martingale. Because it is a positive process, it is a supermartingale. Thus
\begin{align}\label{budget1}
\BE{\rho e^{\rr^{\pi}(T)}}\leq 1.
\end{align}
This is called the \emph{budget constraint} for the investor.\footnote{In the financial economics literature, the budget constraint often refers to $\BE{\rho X^{\pi}(T)}\leq x_{0}$ which is equivalent to \eqref{budget1}.} For any admissible portfolio, the corresponding terminal growth rate must obey the budget constraint. 
We can add the budget constraint into Problem \eqref{Optimalcontrol0} without changing its nature:
\begin{align}\label{Optimalcontrol}
\max_{\pi\in\seta} & \quad \BV{u(\rr^{\pi}(T)-\bc)}\\
\mathrm{s.t.} &\quad \text{$(\pi, \rr^{\pi})$ satisfies \eqref{returnprocess}}, \quad \BE{\rho e^{\rr^{\pi}(T)}}\leq 1,\quad \rr^{\pi}(T)-\bc \geq -c.\nn
\end{align}

The following result plays a key role in transforming the dynamic stochastic optimal control problem \eqref{Optimalcontrol} to a static r.v. optimization problem. 

\begin{lemma}\label{martingalemethod}
For any random variable $\xi$ with $\BE{\rho e^{\xi}}= 1$, there exists an admissible portfolio $\pi$ such that $\rr^{\pi}(T) = \xi$.
\end{lemma}
All proofs are placed in \citeapp{proofs}.

With the help of this result, solving the problem \eqref{Optimalcontrol} reduces to solving the following static r.v. optimization problem: 
\begin{align}\label{Optimalcontrol1}
\max_{\xi} & \quad \BV{u(\xi-\bc)}\\
\mathrm{s.t.} &\quad \BE{\rho e^{\xi}}\leq 1,\quad\xi-\bc \geq -c.\nn
\end{align}

\begin{lemma}\label{martingalemethod2} 
An admissible portfolio $\pi$ is optimal to Problem \eqref{Optimalcontrol} if and only if $R^{\pi}(T)$
is optimal to Problem \eqref{Optimalcontrol1}.
\end{lemma}

We can now rewrite \eqref{Optimalcontrol1} in a more compact form:
\begin{align}\label{Obj3}
\max_{\zeta} &\quad \BV{v(\zeta)}\\
\mathrm{s.t.} &\quad \BE{\zeta \eta}\leq 1,\quad \zeta\geq \qlow, \nn
\end{align}
where 
\[v(x)=u\left( \log x \right),\quad \zeta=e^{\xi-\bc},\quad \eta=\rho e^{\bc},\quad \qlow=e^{-c}.\]
Here $\eta$ is termed the \emph{benchmark adjusted pricing kernel} under the benchmark $\bc$. 
Moreover, $\zeta$ represents the terminal wealth relative to the benchmark, $X^{\pi}(T)/(x_{0}e^{\bc})$. 

In \eqref{Obj3}, we can view $\BV{v(\cdot)}$ as a preference for $\zeta$, which is neither a CPT nor a rank-dependent utility theory (RDUT) preference due to the specific shape of $v$. In fact, $v$ is $M$-shaped as shown below, as opposed to the $S$-shaped utility in CPT or the concave utility in RDUT. 
Recall that $\zeta$ stands for the terminal relative wealth, we term $v$ the \emph{relative utility function} and $\BV{v(\cdot)}$ the \emph{relative CPT preference}.

Finally, we present an equivalent condition to fulfill the requirement \eqref{lowerbound1}. 
\begin{assmp}[Feasibility conditioin]\label{feasibility}
The trio of the pricing kernel $\rho$, the benchmark $\bc$ and the risk tolerance level $c>0$ satisfies
\[ \BE{\rho e^{\bc}} < e^{c}.\]
\end{assmp}
In fact, the requirement \eqref{lowerbound1} is fulfilled if and only if \citeassmp{feasibility} holds. 
First, \eqref{lowerbound1} together with the budget constraint \eqref{budget1} clearly implies \citeassmp{feasibility} holds. On the other hand, by 
\citelem{martingalemethod}, we can find $\pi_{0}$ such that 
\[\rr^{\pi_{0}}(T)=\bc-\log \left(\BE{\rho e^{\bc}}\right),\]
which together with \citeassmp{feasibility} gives \eqref{lowerbound1} immediately. Therefore, \eqref{lowerbound1} and \citeassmp{feasibility} are equivalent. 
Henceforth, we assume \citeassmp{feasibility} holds. \citeassmp{feasibility} is also equivalent to $\qlow\BE{\eta}<1$, so we see that there are infinitely many feasible solutions to Problem \eqref{Obj3}.

\begin{remark}
If the benchmark is constant and the interest rate process is deterministic, then \citeassmp{feasibility} is equivalent to $\bc<c +\int_{0}^{T}r(s)\dd s$. This implies that the investor can not set up an extremely high target of $\bc$ for a fixed level of risk tolerance $c$; otherwise the problem may be ill-formulated, which is consistent with the saying that ``a higher return comes with a higher risk''.
\end{remark}

\section{Quantile formulation} 

In this section, we focus on Problem \eqref{Obj3}. The Choquet expectation \eqref{Choquet integral} involves two elements: the probability weighting function $w(\cdot)$ and the probability distribution function $F_{\xi}$. The former imposes great complexity if we attempt to evaluate the nonlinear expectation directly. Expressing the objective in terms of the quantile of the decision variable will enable us to treat the two elements separately. This is the motivation and advantage of the so-called \emph{quantile formulation}, which has been used by Carlier and Dana \cite{CD08} to study risk-sharing problems, Jin and Zhou \cite{JZ08} and Xia and Zhou \cite{XZ16} to solve portfolio selection problems under behavioral criterion, Hou and Xu \cite{HX16} to study mean-variance portfolio optimization with intractable claims, and Wei \cite{W18,W21} to solve utility maximization with risk constraints. Notably, the third author \cite{X16} provided a simple form of this technique. Here we apply Xu's method to solve Problem \eqref{Obj3}.

For a r.v. $\xi$ with the probability distribution $F_{\xi}$, its quantile function is defined as
\[Q_{\xi}(p):=\inf\{z\in \mathbb{R}\;|\;F_{\xi}(z)> p\},\quad p \in(0,1), \]
with the convention that $Q_{\xi}(0)=\lim_{p\to 0}Q_{\xi}(p)$
and $Q_{\xi}(1)=\lim_{p\to 1}Q_{\xi}(p)$. 
It is easy to see that $Q_{\xi}$ is increasing and right-continuous on $(0,1)$. 
On the other hand, if $Q$ is increasing and right-continuous on $(0,1)$, then $Q=Q_{\xi}$, where $\xi=Q(U)$ with $U$ being uniformly distributed on $(0,1)$. 
Hence,
\[\setq=\big\{Q:(0,1)\to\R \;\big|\; \text{$Q$ is increasing and right-continuous} \big\},\]
denotes the set of quantile functions for all r.v.s. 

In terms of the quantile function, we can express the Choquet expectation $\BV{\xi}$ as 
\[\BV{\xi}=\int_{0}^{1}Q_{\xi}(p)w'(1-p)\ddp,\]
which leads to 
\begin{align} \label{Obj4}
\BV{v(\zeta)}&=\int_{0}^{1}Q_{v(\zeta)}(p)w'(1-p)\ddp =\int_{0}^{1}v(Q_{\zeta}(p))w'(1-p)\ddp.
\end{align}
By definition, we have $\zeta\geq \qlow$ if and only if 
\begin{align}\label{quantileconstraint1}
Q_{\zeta}(0)\geq \qlow.
\end{align}
Therefore, the objective functional and constraint $\zeta\geq \qlow$ in Problem \eqref{Obj3} can be expressed in terms of the quantile function of $\zeta$. We then need to express the other constraint $ \BE{\zeta \eta}\leq 1$ in terms of the quantile function of $\zeta$. This is impossible for general $\zeta$ since $\BE{\zeta \eta}$ depends not only on the quantile of $\zeta$, but also on the joint law of $\zeta$ and $\eta$. Nevertheless, we do not need such expressions for all $\zeta$ and it suffices to obtain them for the optimal solution. 
\par 
In the neoclassical models (such as EUT) and behavioral finance models for terminal wealth (such as Jin and Zhou \cite{JZ08} and He and Zhou \cite{HZ11}), the optimal solution $\zeta$ must be anti-comonotonic with the pricing kernel $\rho$, meaning that the optimal terminal wealth will be higher if the market status is getting better (i.e., $\rho$ is getting smaller). 
This may not be the case in our model since Problem \eqref{Obj3} relies on $\eta$ instead of $\rho$. In the same spirit of Xu \cite{X14}, we can show that any optimal candidate $\zeta$ for Problem \eqref{Obj3} must be anti-comonotonic with $\eta$. Hence, $\zeta$ is not anti-comonotonic with $\rho$ unless $\rho$ and $\eta$ are comonotonic. Intuitively, the benchmark should be higher if the market status is getting better, e.g., when the benchmark is the return of a stock index. Mathematically, this means $\bc$ is decreasing in $\rho$. In particular, if the benchmark is chosen as $\bc=k\log\rho^{-1}$ with $k>0$, then $\zeta$ is indeed anti-comonotonic with $\rho$ when $k<1$ and comonotonic when $k>1$. Therefore, the benchmark may significantly influence people's investment strategy. When one setup an aggressive target (namely $k>1$), the output may be opposite to the classical observation that one should have a higher relative return in a good market scenario.

The above analysis eventually leads to the following quantile formulation of Problem \eqref{Obj3}: 
\begin{align}\label{quantile1}
\max_{Q\in\setq} &\quad \int_{0}^{1}v(Q(p))w'(1-p)\ddp \\
\mathrm{s.t.} &\quad \int_{0}^{1}Q(p)Q_{\eta}(1-p)\ddp=1,\quad Q(0)\geq \qlow.\nn
\end{align}

We now establish the relationship between Problem \eqref{quantile1} and Problem \eqref{Obj3}. 
Let $\setu$ be the set of random variables that are uniformly distributed on $(0,1)$ and comonotonic with $\eta$. 
Then, $\setu$ is not empty, and $\eta=Q_{\eta}(U)$ a.s. for any $U\in\setu$. Moreover, the distribution function of $\eta$, $F_{\eta}$, is continuous if and only if $\setu=\{F_{\eta}(\eta)\}$ (see, e.g. Xu \cite{X14}).
\begin{prop}\label{quantileformulation}
A random variable $\zeta^{*}$ is an optimal solution to Problem \eqref{Obj3}, if and only if 
$\zeta^{*}=Q^{*}(1-U)$ a.s., where $U\in\setu$ and $Q^{*}$ is an optimal solution to Problem \eqref{quantile1}. 
\end{prop} 

Following Xu \cite{X16}, we introduce the following notations to simplify the expressions in Problem \eqref{quantile1}, 
\begin{align}
\nu(p)&=1-w^{-1}(1-p),\quad p\in[0, 1],\nn \\
G(p) &=Q(\nu(p)),\quad p\in(0, 1), \nn\\
\varphi(p)&=-\int_{p}^{1}Q_{\eta}(1-\nu(s))\nu'(s)\ds,\quad p\in[0,1]. \label{varphi}
\end{align}
Then, Problem \eqref{quantile1} can be rewritten as (see \citeapp{changeofvariable})
\begin{align}\label{quantile2}
\max_{G\in\setq} &\quad \int_{0}^{1}v(G(p))\ddp \\
\mathrm{s.t.} &\quad \int_{0}^{1}G(p)\varphi'(p)\ddp=1,\quad G(0)\geq \qlow. \nn
\end{align}

As will be clear shortly, $v$ is not concave and thus Problem \eqref{quantile2} is not a concave optimization problem. Therefore, the Lagrange method can not be applied directly to tackle it. This is different from the problem in Xu \cite{X16} wherein the constraint $G(0)\geq \qlow$ is absent and the $v$ is concave. Hence, Xu's \cite{X16} result is not readily applicable and a new method is called for.

\section{Optimal solution}
We intend to use the concavification method to solve the problem \eqref{quantile2}.
Before doing so, let us first study the properties of the function $v$. 
\subsection{Shape of the relative utility $v$}\label{UOR}
\noindent
We start with the shape of $v$. 
Note that
\begin{align*}
v(x)=u(\log x)=\begin{cases}
(\log x)^{\alpha}, &\quad x\geq 1;\\
-\kappa(-\log x)^{\beta}, &\quad 0< x<1.
\end{cases}
\end{align*}
Hence we have
\[v'(x)=u'(\log x)x^{-1}>0, \quad x\neq 1,\]
and
\[v''(x)=[u''(\log x)-u'(\log x)]x^{-2},\quad x>0.
\]
We see from the above relations that $v$ would be increasing and global concave if so was $u$, which is 
the case in EUT and RDUT. By contrast, the shape of $v$ in our model is essentially different from theirs. Indeed,
\begin{itemize}
\item When $x>1$, we have $u''(\log x)<0$ and $u'(\log x)>0$, so $v''(x)<0$ and $v$ is strictly concave.
\item When $0<x<1$, we have
\begin{align*}
v''(x)&=[u''(\log x)-u'(\log x)]x^{-2}\\
&=-\kappa\beta\big((\beta-1)+(- \log x)\big)(-\log x)^{\beta-2} x^{-2}\\
&\begin{cases}
>0, &\quad e^{\beta-1}<x<1;\\
<0, &\quad 0<x< e^{\beta-1}.
\end{cases}
\end{align*}
So $v$ is strictly concave on $(0, e^{\beta-1}]$ and strictly convex on $[e^{\beta-1},1)$.
\end{itemize}
To summarize, the function $v$ is continuous, strictly increasing on $(-\infty,\infty)$, strictly concave, respectively, on $(0, e^{\beta-1}]$ and $[1,\infty)$, and strictly convex on $[e^{\beta-1},1]$. We call such \emph{concave-convex-concave-shaped} utility an \emph{$M$-shaped} utility function. 
Since $\qlow$ and $\beta$ are entirely decided by the investor's preference, both $\qlow \leq e^{\beta-1}$ (\citefig{fig3.1}) and $\qlow> e^{\beta-1}$ (\citefig{fig3.2}) are possible.
\begin{figure}[H]
\begin{center}
\begin{picture}(280,160)
\linethickness{0.6pt}
\put(40,89){\vector(1,0){190}} 
\put(50,0){\vector(0,1){150}} 
\put(215,78){$x$}
\qbezier(54,14)(57,39)(96,51)
\qbezier(96,51)(136,60)(144,88)
\qbezier(144,88)(149,120)(209,138)
\put(200,122){$v(x)$}
\put(145,78){$1$}

\put(75, 92){$e^{\beta-1}$}
\qbezier[30](80, 45)(80, 66)(80, 88)
\put(62, 92){$\hat{c}$}
\qbezier[38](65, 35)(65, 62)(65, 88) 
\end{picture}
\caption{The function $v$ with $\qlow \leq e^{\beta-1}$.}\label{fig3.1}
\label{fig4}
\end{center}
\end{figure}
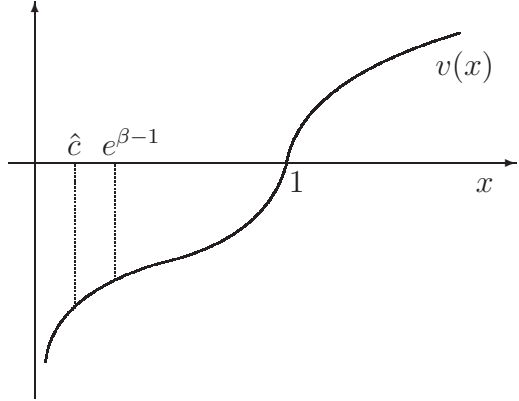

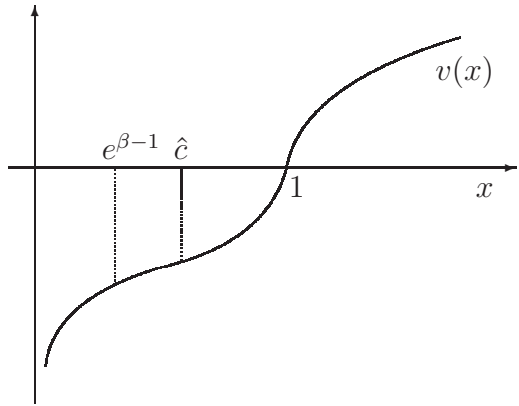
\begin{figure}[H]
\begin{center}
\begin{picture}(280,160)
\linethickness{0.6pt}
\put(40,89){\vector(1,0){190}} 
\put(50,0){\vector(0,1){150}} 
\put(215,78){$x$}
\qbezier(54,14)(57,39)(96,51)
\qbezier(96,51)(136,60)(144,88)
\qbezier(144,88)(149,120)(209,138)
\put(200,122){$v(x)$}
\put(145,78){$1$}

\put(75, 92){$e^{\beta-1}$}
\qbezier[30](80, 45)(80, 66)(80, 88)
\put(102, 92){$\hat{c}$}
\qbezier[28](105, 54)(105, 76)(105, 88) 
\end{picture}
\caption{The function $v$ with $\qlow > e^{\beta-1}$.}\label{fig3.2}
\label{fig4}
\end{center}
\end{figure}

\par
Because $v$ is $M$-shaped, Problem \eqref{quantile2} is not a concave or $S$-shaped utility optimization problem like those in Jin and Zhou \cite{JZ08} and Xu \cite{X16}. The $M$-shaped utility has been rarely studied in the financial economics literature. The earliest one we can find dates back to Friedman and Savage \cite{FS48} that initially proposed such shape of utility to explain the purchasing of both insurance and lottery tickets (the ``F-S hypothesis''). As demonstrated in \cite{FS48}, the concave segment indicates a preference for certainty (insurance) while the convex segment indicates a preference for risk (gambling). In their paper, the $M$-shaped utility was endowed with a reasonable interpretation that a lower socioeconomic class consumer whose income placed corresponding to the first concave segment wishes to shift himself up to a higher socioeconomic class whose income placed corresponding to the last concave segment, the convex segment with increasing marginal utility represented a transitional stage between the two classes.

Although the shape of the utility $v$ in our model is also $M$-shaped, there are some differences we would like to highlight. First, in terms of the meaning of utility, it refers to the utility of ``income'' in the context of \cite{FS48}. Whereas the ``utility'' investigated underline refers to the utility of the relative growth rate $\zeta=X^{\pi}(T)/x_{0}e^{\bc}$. As one may note, the form of the relative utility underline relies heavily on the hypothesis of the piece-wise utility of wealth $u$. The shape of the relative utility $v$ is entirely determined by the parameters $\alpha,\beta,\kappa$. It is possible to have other kinds of non-concave, non-$M$-shaped shaped relative utilities $v$ if one would use general $S$-shaped or other shaped utilities $u$. 

More interestingly, as explained in \cite{FS48}, $M$-shape utility is only plausible for consumer units whose level of income is in the first concave segment of the utility function. If that was so, then, in turn, the ``F-S'' hypothesis implies that the investor's ability to invest, measured by the relative return variable $\zeta$, should take values in the first concave segment, namely the region $(0,e^{\beta-1})$. In other words,, the gross return of investor should subject to $0<{X^{\pi}(T)}/{x_0}<e^{\bc+\beta-1}$. If we set $\beta=0.88$ and $\bc$ to the common value of risk-free interest rate (normally less than $0.05$), then these investors would typically refer to those who suffer losses in the market as $e^{\bc+\beta-1}<1$.

\subsection{Concavified problem}
\noindent
One common approach to overcome the non-concavity in Problem \eqref{quantile2} is to consider its \emph{concavified} problem. This approach has been used to solve non-concave utility maximization problems by Bian et al. \cite{BCX19} and Guan et al. \cite{GLXY17} for expected utility models, \blue{by Bi et al. \cite{BYCF21} for behavioral models.}

Let us define the concave envelope of $v$ on $(0,\infty)$, that is, the smallest concave function dominating $v$ on $(0,\infty)$, denoted by $\hat{v}_{0}$. Mathematically, it is given by
\[\hat{v}_{0}(x)=\sup_{\substack{0<y\leq x\leq z\\ y\neq z}}\left(\frac{z-x}{z-y}v(y)+\frac{x-y}{z-y}v(z)\right),\quad x>0.\]
We have the following explicit expression for $\hat{v}_{0}$. 
\begin{lemma}\label{twopoints}
There are two scalars $0<a<1<b$ such that the function $\hat{v}_{0}$ is concave on $(0,\infty)$, coincides with $v$ on $(0, a]\cup[b,\infty)$, and is linear on $[a,b]$. Moreover,
\begin{align}\label{abequation}
v'(a)=v'(b)=\frac{v(b)-v(a)}{b-a}.
\end{align}
\end{lemma}

The locations of $a$ and $b$ are demonstrated in \citefig{fig4}. They are determined by the utility parameters $\alpha,\beta$, and $\kappa$. 
The global concave envelope function $\hat{v}_{0}$ is different from $v$ on the interval $(a,b)$, shown in the red dot line.
\begin{figure}[H]
\begin{center}
\begin{picture}(280,160)
\linethickness{0.6pt}
\put(40,89){\vector(1,0){190}} 
\put(50,0){\vector(0,1){150}} 
\put(215,78){$x$}
\qbezier(54,14)(57,39)(96,51)
\qbezier(96,51)(136,60)(144,88)
\qbezier(144,88)(149,120)(209,138)
\put(200,122){$v(x)$}
\put(145,78){$1$}

\linethickness{0.9pt}
{{\color{red}\qbezier[80](69, 41)(113, 79)(158, 116)}}
\qbezier[25](68,41)(68, 65)(68,88)
\put(65,92){$a$}
\qbezier[14](158,116)(158,103)(158,90)
\put(156,78){$b$}
\end{picture}
\caption{The function $\hat{v}_{0}$ in red dot line on $(a,b)$.}
\label{fig4}
\end{center}
\end{figure}
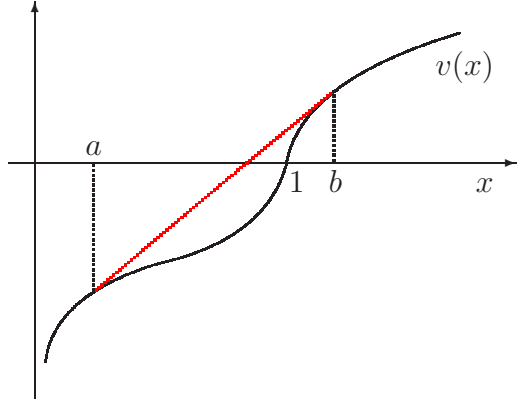

A naive way to solve Problem \eqref{quantile2} is described as follows. We first try to solve the following \emph{globally concavified problem}:
\begin{align}\label{quantile3}
\max_{G\in\setq } &\quad \int_{0}^{1}\hat{v}_{0}(G(p))\ddp \\
\mathrm{s.t.} &\quad \int_{0}^{1}G(p)\varphi'(p)\ddp=1,\quad G(0)\geq \qlow. \nn
\end{align}
We then try to show that the optimal solution to Problem \eqref{quantile3} is also optimal to Problem \eqref{quantile2}. Unfortunately, this approach fails due to the lower bound constraint $Q(0)\geq \qlow$.
In fact, one can easily see from the formulation of Problem \eqref{quantile2} that its optimal solution depends on the utility function $v$ only on $[\qlow, \infty)$. However, in Problem \eqref{quantile3}, $\hat{v}_{0}$ depends on $v$ on $(0,\infty)$, so these two problems are unlikely to share the same optimal solution. It is natural to consider the local concave envelope of $v$ on $[\qlow,\infty)$ rather than the global one $\hat{v}_{0}$ on $(0,\infty)$.
\par
The above discussion motivates us to consider the concave envelope function of $v$ on $[\qlow,\infty)$ that is denoted by $\hatv$. It is the smallest concave function dominating $v$ on $[\qlow,\infty)$, called the local concave envelope of $v$ and given by
\begin{align}\label{hatvdef}
\hatv(x)=\sup_{\substack{\qlow\leq y\leq x\leq z\\ y\neq z}}\left(\frac{z-x}{z-y}v(y)+\frac{x-y}{z-y}v(z)\right),\quad x\geq \qlow.
\end{align}
We now introduce the \emph{locally concavified problem}: 
\begin{align}\label{q1}
\max_{G\in\setq } &\quad \int_{0}^{1}\hatv(G(p))\ddp \\
\mathrm{s.t.} &\quad \int_{0}^{1}G(p)\varphi'(p)\ddp=1,\quad G(0)\geq \qlow. \nn
\end{align}
This is a concave optimization problem
which is easier to solve compared to the non-concave optimization Problem \eqref{quantile2}.
The two problems look different but we will show that they share the optimal solution and optimal value, 
\blue{
provided that they are well-posed (that is, their optimal values are finite); see \citeprop{wellposedness}.
}
\par
Before proving the above claim, let us first study the properties of the local concave envelope function $\hatv$.
Clearly, $\hatv\leq \hat{v}_{0}$ on $[\qlow,\infty)$. It is also easy to show that they are identical on $[\qlow,\infty)$ if and only if $\qlow\leq a$; see \citefig{fig4.5}.

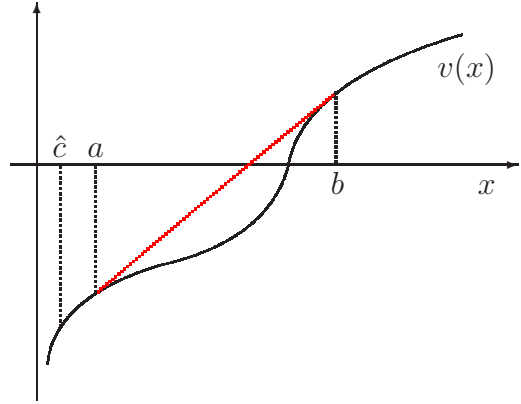
\begin{figure}[H]
\begin{center}
\begin{picture}(280,160)
\linethickness{0.6pt}
\put(40,89){\vector(1,0){190}} 
\put(50,0){\vector(0,1){150}} 
\put(215,78){$x$}
\qbezier(54,14)(57,39)(96,51)
\qbezier(96,51)(136,60)(144,88)
\qbezier(144,88)(149,120)(209,138)
\put(200,122){$v(x)$}

\linethickness{0.9pt}
{{\color{red}\qbezier[80](69, 41)(113, 79)(158, 116)}}
\qbezier[25](68,41)(68, 65)(68,88)
\put(65,92){$a$}
\qbezier[14](158,116)(158,103)(158,90)
\put(156,78){$b$}

\qbezier[30](55,28)(55,58)(55,88)
\put(52,92){$\qlow$} 
\end{picture}
\caption{The functions $\hat{v}_{0}$ and $\hatv$ coincide, when $\qlow\leq a$.}
\label{fig4.5}
\end{center}
\end{figure}

If $a<\qlow$, then the function $\hatv$ coincides with $v$ on $[d,\infty)$ and is linear on $[\qlow,d]$ for some $1<d<b$. Moreover, $\hatv< \hat{v}_{0}$ on $[\qlow, b]$ and $\hatv>v$ on $(\qlow, d)$; see \citefig{fig5}. Similar as before, we can find the value of $d$ via an algebraic equation
\[v'(d)=\frac{v(d)-v(\qlow)}{d-\qlow}, \quad\qlow>a.\]

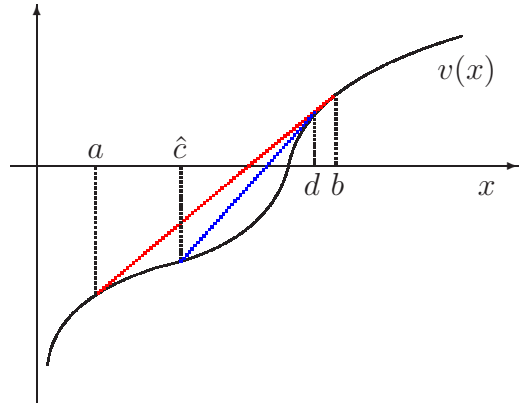
\begin{figure}[H]
\begin{center}
\begin{picture}(280,160)
\linethickness{0.6pt}
\put(40,89){\vector(1,0){190}} 
\put(50,0){\vector(0,1){150}} 
\put(215,78){$x$}
\qbezier(54,14)(57,39)(96,51)
\qbezier(96,51)(136,60)(144,88)
\qbezier(144,88)(149,120)(209,138)
\put(200,122){$v(x)$}

\linethickness{0.9pt}
{{\color{red}\qbezier[80](69, 41)(113, 79)(158, 116)}}
\qbezier[25](68,41)(68, 65)(68,88)
\put(65,92){$a$}
\qbezier[14](158,116)(158,103)(158,90)
\put(156,78){$b$}

\qbezier[20](100,54)(100,71)(100,88)
\put(97,92){$\qlow$}
{{\color{blue}\qbezier[47](100,53)(125,81)(150, 109)}}
\qbezier[12](150,90)(150,99)(150,108)
\put(146,78){$d$}
\end{picture}
\caption{The function $\hat{v}_{0}$ in red dot line and $\hatv$ in blue dot line, when $\qlow>a$.}
\label{fig5}
\end{center}
\end{figure}

From an economic perspective, the larger the risk tolerance level $c$ (i.e., the smaller the value of $\qlow$), the smaller the local concave envelope function $\hatv$ as well as the optimal value of Problem \eqref{quantile2}.
\par
Because $\hatv'$ is continuous and decreasing on $[\qlow,\infty)$, we may define its right-continuous inverse function as
\begin{align}\label{Leftinverse}
\inv(x):=\inf\{y\geq \qlow\mid \hatv'(y)\leq x\}\vee \qlow,\quad x>0.
\end{align}
Its properties are summarized in the following lemma. 
\begin{lemma}\label{Iprop}
The function $\inv$ is decreasing and right-continuous, and satisfies the following properties.
\begin{enumerate}
\item If $a\leq \qlow$, then
\begin{align}\label{I1}
\inv(x)=\begin{cases}
(v')^{-1}(x), &\quad \text{ if $0<x<v'(d)$};\\
\qlow,&\quad \text{ if $x\geq v'(d)$}.
\end{cases}
\end{align} 

\item
If $a>\qlow$, then
\begin{align}\label{I2}
\inv(x)=\begin{cases}
(v')^{-1}(x), &\quad \text{ if $0<x<v'(a)$};\\
a,&\quad \text{ if $x=v'(a)$};\\
(v')^{-1}(x), &\quad \text{ if $v'(a)<x<v'(\qlow)$};\\
\qlow,&\quad \text{ if $x\geq v'(\qlow)$}.
\end{cases}
\end{align}

\item For any $x>0$, \[\hatv(\inv(x))=v(\inv(x)). \]
\item For any $x>0$, \[\max_{y\geq \qlow}(v(y)-xy)=\max_{y\geq \qlow}(\hatv(y)-xy)=\hatv(\inv(x))-x\inv(x).\]
\end{enumerate}
\end{lemma}
In particular, $I$ has a unique jump at $x=v'(d)$ when $a\leq \qlow$ or at $x=v'(a)$ when $a> \qlow$, and it does not take values in $(\qlow,d)$ when $a\leq \qlow$ or in $(a,b)$ when $a> \qlow$.

\subsection{Optimal solution}

We now use employ the method of Xu \cite{X16} to solve all the problems formulated in the previous sections, including Problems \eqref{Optimalcontrol0}, \eqref{quantile2} and \eqref{q1}.

Let $\hatphi$ be the concave envelope of $\varphi$ on $[0,1]$, that is,
\begin{align}\label{deltadef}
\hatphi(p):=\sup_{\substack{0\leq x\leq p\leq y\leq 1\\ x\neq y}}\left(\frac{y-p}{y-x}\varphi(x)+\frac{p-x}{y-x}\varphi(y)\right),\quad p\in[0,1].
\end{align} 
Because $\varphi(1)>\varphi(p)$ for all $p\in(0,1)$, we conclude that $\hatphi$ is strictly increasing on $(0,1)$. 
Let $\hatphi'$ denote the right-continuous derivative of $\hatphi$, that is,
\[\hatphi'(p):=\lim_{\ep\to 0+}\frac{\hatphi(p+\ep)-\hatphi(p)}{\ep},\quad p \in(0,1).\]
Then $\hatphi'>0$ on $(0,1)$. 

\blue{
The following proposition characterizes the well-posedness of Problems \eqref{Optimalcontrol0}, \eqref{quantile2} and \eqref{q1}. 
\begin{prop}\label{wellposedness}
The following statements are equivalent.
\begin{enumerate}
\item Problem \eqref{Optimalcontrol0} is well-posed.
\item Problem \eqref{quantile2} is well-posed.
\item Problem \eqref{q1} is well-posed.
\item It holds that 
\begin{align}\label{well1} 
\int_0^1 \big(\max\{-\log \hatphi'(p),\;0\}\big)^{\alpha} \ddp<\infty.
\end{align}
\end{enumerate}
\end{prop}
}

The following theorem presents the optimal solution to our model.

\begin{thm}[Optimal solution]\label{vt}
Suppose \eqref{well1} holds. 
If there exists a constant $\lambda>0$ such that
\begin{align}\label{lambda1}
\int_{0}^{1}\inv(\lambda \hatphi'(p))\hatphi'(p)\ddp=1,
\end{align} 
then $$G^{*}(p)=\inv(\lambda \hatphi'(p)),\quad p \in(0,1),$$ is optimal to Problems \eqref{q1} and \eqref{quantile2}. Moreover, 
$$Q^{*}(p)=\inv(\lambda \hatphi'(1-w(1-p))),\quad p \in(0,1),$$ is optimal to Problem \eqref{quantile1}. 

Furthermore, let $U\in\setu$. Then there exists an $\{\mathcal{F}_t\}$-progressively measurable process $Z$ such that 
\[\BE{\rho(T) e^{\bc} G^{*}(1-w(U))\;\big|\;\mathcal{F}_t}=1+\int_0^t Z(s)'\dd W(s), \quad t\in[0,T].\]
Furthermore,
\[\pi^*(t)=(\sigma'(t))^{-1}\frac{\rho(t) Z(t)}{\BE{\rho(T) e^{\bc} G^{*}(1-w(U))\;\big|\;\mathcal{F}_t}},\quad t\in[0,T],\]
is optimal to Problem \eqref{Optimalcontrol0} and $$R^{\pi^{*}}(T)=\bc+\log (G^{*}(1-w(U)))$$ is the corresponding optimal terminal growth rate. 
\end{thm}

\blue{
\begin{remark}
Theorem \ref{vt} characterizes the optimal solution assuming the existence of a Lagrange multiplier $\lambda$ that solves \eqref{lambda1}. In Appendix \ref{existenceofLagrangemultiplier} we discuss when this assumption holds.
\end{remark}
}

Because $I$ does not take values in $(\qlow,d)$ when $a\leq \qlow$ or in $(a,b)$ when $a> \qlow$, so does $G^{*}$. This indicates that the optimal terminal relative growth rate and wealth will not take values close to the benchmark. From an economic point of view, the investor shall take relatively large or small returns rather than those close to the benchmark. In other words, she would seek relatively large returns at the risk of getting relatively small returns.

In Berkelaar et al. \cite{BKP04} and Jin and Zhou \cite{JZ08}, the optimal terminal relative wealth turns out to consist of a pricing kernel-dependent gain part and a constant loss part with a strict gap between them. 
In our model, when the agent's risk tolerance level $c$ is no more than the threshold $-\log a$ (i.e., $a\leq\qlow$), the same phenomenon happens by \eqref{I1}. However, when the risk tolerance level exceeds this threshold, 
we see from \eqref{I2} that the loss part is no more a constant. Instead, it is also benchmark-adjusted pricing kernel dependent and takes values from $a$ to $\qlow$. This is a new financial phenomenon observed in our model. It shows that risk tolerance level plays a critical role in determining the optimal strategy.

Zhang et al. \cite{ZJZ11} considered a CPT model with a lower bound constraint on the terminal relative wealth. Their model is similar to ours, except that they maximize the utility of terminal relative wealth and use separate probability weighting functions for gain and loss parts. Different from our case, the loss part in their model consists of a constant moderate loss and a constant maximum loss due to different probability weighting functions for gain and loss parts. These two constants become the same when the probability weighting functions for gain and loss parts are the same (see \eqref{parawealth1} below). 

In the next section, we make a comparison between their results and ours.

\section{A comparison}

We rewrite Problem \eqref{Optimalcontrol} as
\begin{align}\label{paramodel}
\max_{\pi\in\seta} & \quad \BV{u(\log (X^{\pi}(T)/x_{0})-\bc)}.\\
\mathrm{s.t.} &\quad\BE{\rho X^{\pi}(T)}\leq x_0,\quad X^{\pi}(T)\geq x_{0}e^{\bc -c}. \nn
\end{align}
In Problem \eqref{paramodel}, the terminal return $\rr^{\pi}(T)$ in \eqref{Optimalcontrol} has been replaced by the traditional terminal wealth $X^{\pi}(T)$. 

Zhang et al. \cite{ZJZ11} considered a CPT model with a lower bound constraint on the terminal relative wealth.
For comparison, we assume the agent takes the same terminal benchmark wealth at time $T$ as ours, given by $x_0e^{\bc}$; also, she uses the same lower bound. 
Let $\barc$ denote the constant maximum loss given by 
\[\barc=x_{0}e^{\bc}-x_{0}e^{\bc -c}.\]
Their problem can be formulated as 
\begin{align}\label{paramodel1}
\max_{\pi\in\seta'} & \quad \BV{u(X^{\pi}(T)-x_0e^{\bc})}.\\
\mathrm{s.t.} &\quad\BE{\rho X^{\pi}(T)}\leq x_{0},\quad X^{\pi}(T)-x_0e^{\bc}\geq -\barc, \nn
\end{align} 
when the probability weighting functions for gain and loss parts are identical as in Problem \eqref{paramodel}. 
Problems \eqref{paramodel1} and \eqref{paramodel} share the same constraint on the terminal wealth $X^{\pi}(T)$, but they have different objective functionals. The admissible portfolio sets of these two problems are also slightly different, but this will not affect our comparison.

After some tedious calculations, we can show that under our numerical setting to be specified below, the optimal terminal wealth to Problem \eqref{paramodel1} is
\begin{align}\label{parawealth1}
X^{\pi}(T)-x_0e^{\bc}=\begin{cases}
\;(u')^{-1}\Big( \frac{\lambda_1\rho}{w'(F_{\rho}(\rho))}\Big),&\quad \mbox{if } \frac{\lambda_1\rho}{w'(F_{\rho}(\rho))} \leq \alpha l^{\alpha-1};\\[10pt]
\; -\barc,&\quad \mbox{if } \frac{\lambda_1\rho}{w'(F_{\rho}(\rho))}>\alpha l^{\alpha-1},
\end{cases}
\end{align}
where $l$ solves
$$\alpha l^{\alpha-1} =\frac{l^\alpha + \kappa L^\beta }{l+L},$$
and $\lambda_1>0$ is the Lagrange multiplier that solves the budget constraint $\BE{\rho X^{\pi}(T)}= x_{0}$.

The optimal terminal wealth to Problem \eqref{paramodel} is given by
\begin{enumerate}
\item If $a\leq \qlow$, then
\begin{align}\label{parawealth2}
X^{\pi}(T)/x_0e^{\bc}=\begin{cases}
\;(v')^{-1} \Big( \frac{\lambda_2 e^{\bc} \rho}{w'(F_{\rho}(\rho))} \Big),&\quad \mbox{if} \frac{\lambda_2 e^{\bc} \rho}{w'(F_{\rho}(\rho))} \leq \frac{\alpha (\log d)^{\alpha-1}}{d};\\
\; \qlow,&\quad \mbox{if } \frac{\lambda_2 e^{\bc} \rho}{w'(F_{\rho}(\rho))} > \frac{\alpha (\log d)^{\alpha-1}}{d}.
\end{cases}
\end{align}
\item If $a>\qlow$, then
\begin{align}\label{parawealth3}
X^{\pi}(T)/x_0e^{\bc}=\begin{cases}
(v')^{-1} \Big( \frac{\lambda_2 e^{\bc} \rho}{w'(F_{\rho}(\rho))} \Big) , &\quad \text{if } \frac{\lambda_2 e^{\bc} \rho}{w'(F_{\rho}(\rho))} \leq \frac{\kappa \beta (- \log a)^{\beta-1}}{ a};\\
(v')^{-1} \Big( \frac{\lambda_2 e^{\bc} \rho}{w'(F_{\rho}(\rho))} \Big), &\quad \text{if } \frac{\kappa \beta (- \log a)^{\beta-1}}{a} < \frac{\lambda_2 e^{\bc} \rho}{w'(F_{\rho}(\rho))} \leq \frac{\kappa \beta (- \log \qlow)^{\beta-1}}{ \qlow};\\
\qlow,&\quad \text{if } \frac{\lambda_2 e^{\bc} \rho}{w'(F_{\rho}(\rho))} > \frac{\kappa \beta (- \log \qlow)^{\beta-1}}{ \qlow}.
\end{cases}
\end{align}
\end{enumerate}
where $\lambda_2>0$ is the Lagrange multiplier that solves the budget constraint.

The investor in Problem \eqref{paramodel1} classifies the future states into two scenarios: the good states when $\frac{\lambda_1\rho}{w'(F_{\rho}(\rho))} \leq \alpha l^{\alpha-1}$, in which case the payoff is decreasing in the state price density $\rho$; the bad states when $\frac{\lambda_1\rho}{w'(F_{\rho}(\rho))}>\alpha l^{\alpha-1}$, in which case the investor accepts a fixed loss. Moreover, there is a jump discontinuity in the terminal wealth at $\frac{\lambda_1\rho}{w'(F_{\rho}(\rho))}=\alpha l^{\alpha-1}$.

By contrast, the optimal terminal wealth to Problem \eqref{paramodel} exhibits two patterns depending on the relationship between $a$ and $\qlow$. If $a \leq \qlow$ (e.g., $c=0.1$ or $c=0.2$), the optimal terminal wealth resembles that to \eqref{paramodel1}: the good states when $\frac{\lambda_2 e^{\bc} \rho}{w'(F_{\rho}(\rho))} \leq \frac{\alpha (\log d)^{\alpha-1}}{d}$, in which case the payoff is decreasing in the state price density $\rho$; the bad states when $\frac{\lambda_2 e^{\bc} \rho}{w'(F_{\rho}(\rho))} > \frac{\alpha (\log d)^{\alpha-1}}{d}$, in which case the investor accepts a fixed loss; and there is a jump discontinuity in the terminal wealth at $\frac{\lambda_2 e^{\bc} \rho}{w'(F_{\rho}(\rho))} = \frac{\alpha (\log d)^{\alpha-1}}{d}$. 

If $a > \qlow$ (e.g., $c=0.3$), the investor in our model classifies the future states into three scenarios: the good states when $\frac{\lambda_2 e^{\bc} \rho}{w'(F_{\rho}(\rho))} \leq \frac{\kappa \beta (- \log a)^{\beta-1}}{ a}$ and intermediate states when $\frac{\kappa \beta (- \log a)^{\beta-1}}{a} < \frac{\lambda_2 e^{\bc} \rho}{w'(F_{\rho}(\rho))} \leq \frac{\kappa \beta (- \log \qlow)^{\beta-1}}{ \qlow}$, in which cases the payoff is decreasing in the state price density $\rho$ but there is a jump discontinuity at $\frac{\lambda_2 e^{\bc} \rho}{w'(F_{\rho}(\rho))} = \frac{\kappa \beta (- \log a)^{\beta-1}}{ a}$; the bad states when $\frac{\lambda_2 e^{\bc} \rho}{w'(F_{\rho}(\rho))} > \frac{\kappa \beta (- \log \qlow)^{\beta-1}}{ \qlow}$, in which case the investor accepts a fixed loss.

For ease of illustration and comparison, we make the following assumptions in our numerical examples. The investment horizon is 1 year, i.e., $T=1$, and the initial wealth is $x_0=1$. The investment opportunity set is deterministic with $r=0.02$ and $|\theta|=0.2$, so $\rho$ is lognormally distributed, i.e., $\ln \rho \sim N(\mu_{\rho},\sigma_{\rho}^2)$ with $\mu_\rho = -(r+\frac{|\theta|^2}{2})T=-0.04$ and $\sigma_\rho = \sqrt{|\theta|^2T}=0.2$. We follow \cite{TK92} to set $\alpha=\beta=0.88$ and $\kappa=2.5$. In addition, we assume the benchmark growth rate $\bc$ is given by $\bc = (r + g)T$, where $g$ represents the growth rate in excess of the risk-free rate. In our numerical examples, $c$ will take values in $\{0.1,0.2,0.3\}$ and $g$ will take values in $\{0, 0.05, -0.05\}$. 

We consider three types of probability weighting functions:
\begin{enumerate}
\item The identity function $w(p)=p, ~ p \in [0,1]$, i.e., there is no probability distortion;

\item The power function $w(p)=\sqrt{p}, ~ p \in [0,1]$;

\item The inverse $S$-shaped probability weighting function proposed in Jin and Zhou \cite{JZ08}:
\begin{equation*}
w(p) = \left\{
\begin{aligned}
k e^{(\bar{a}+\bar{b})\Phi ^{-1}(\bar{p})+\frac{\bar{a}^2}{2}} \Phi \left(\Phi ^{-1}(p)+\bar{a}\right), \quad & p\le \bar{p},\\
A + ke^{\frac{\bar{b}^2}{2}} \Phi \left(\Phi ^{-1}(p)-\bar{b} \right) , \quad & p> \bar{p},\\
\end{aligned}
\right.
\end{equation*}
where $\Phi$ is the standard normal probability distribution, $\bar{a}$ and $\bar{b}$ are two nonnegative parameters, and 
$$k = \frac{1}{e^{\frac{\bar{b}^2}{2}}\Phi \left(-\Phi ^{-1}(\bar{p})+\bar{b} \right) + e^{(\bar{a}+\bar{b})\Phi ^{-1}(\bar{p})+\frac{\bar{a}^2}{2}} \Phi \left(\Phi ^{-1}(\bar{p})+\bar{a}\right)}, \quad A = 1 - ke^{\frac{\bar{b}^2}{2}}.$$
We choose the following parameterization: $\bar{p}=0.3$, $\bar{a}=1.6\sigma_{\rho}$, $\bar{b}=0.8 \sigma_{\rho}$.

\end{enumerate}

\begin{figure}[htbp]
\centering

\includegraphics[width=5in]{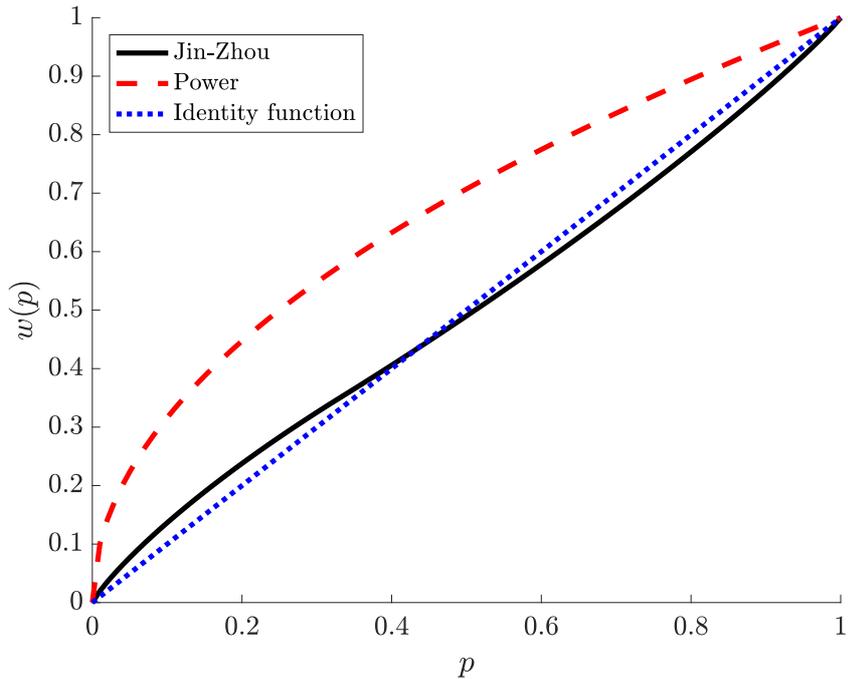}

\caption{Three types of probability weighting functions}\label{figure:distortion}
\end{figure}

Figure \ref{figure:distortion} displays the three types of probability weighting functions under our parameterization.

\begin{figure}[htbp]
\centering
\begin{minipage}[t]{2in}
\centering
\centerline{$c=0.1$ and $g=0$}
\includegraphics[width=2.2in]{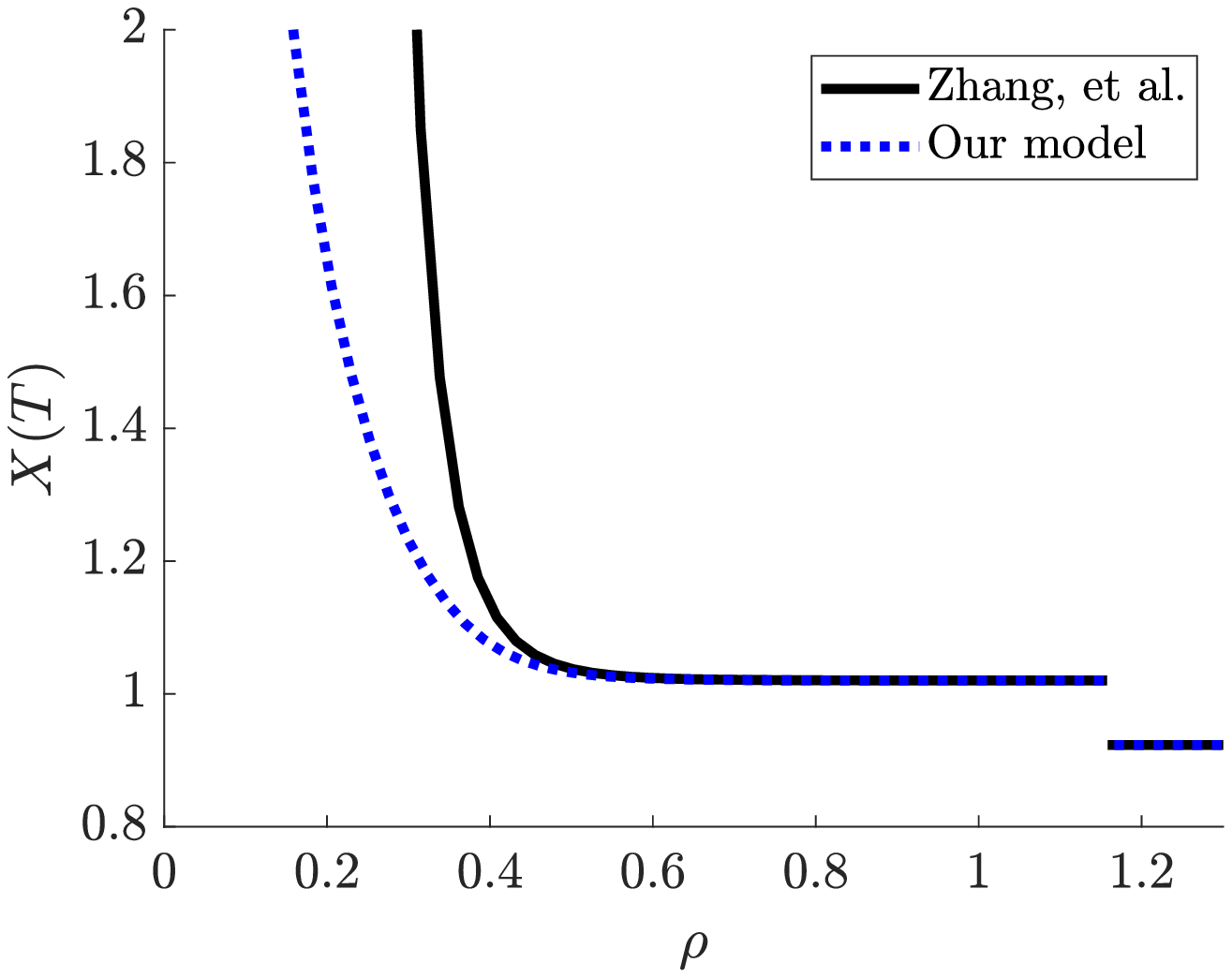}
\end{minipage}
\begin{minipage}[t]{2in}
\centering
\centerline{$c=0.1$ and $g=0.05$}
\includegraphics[width=2.2in]{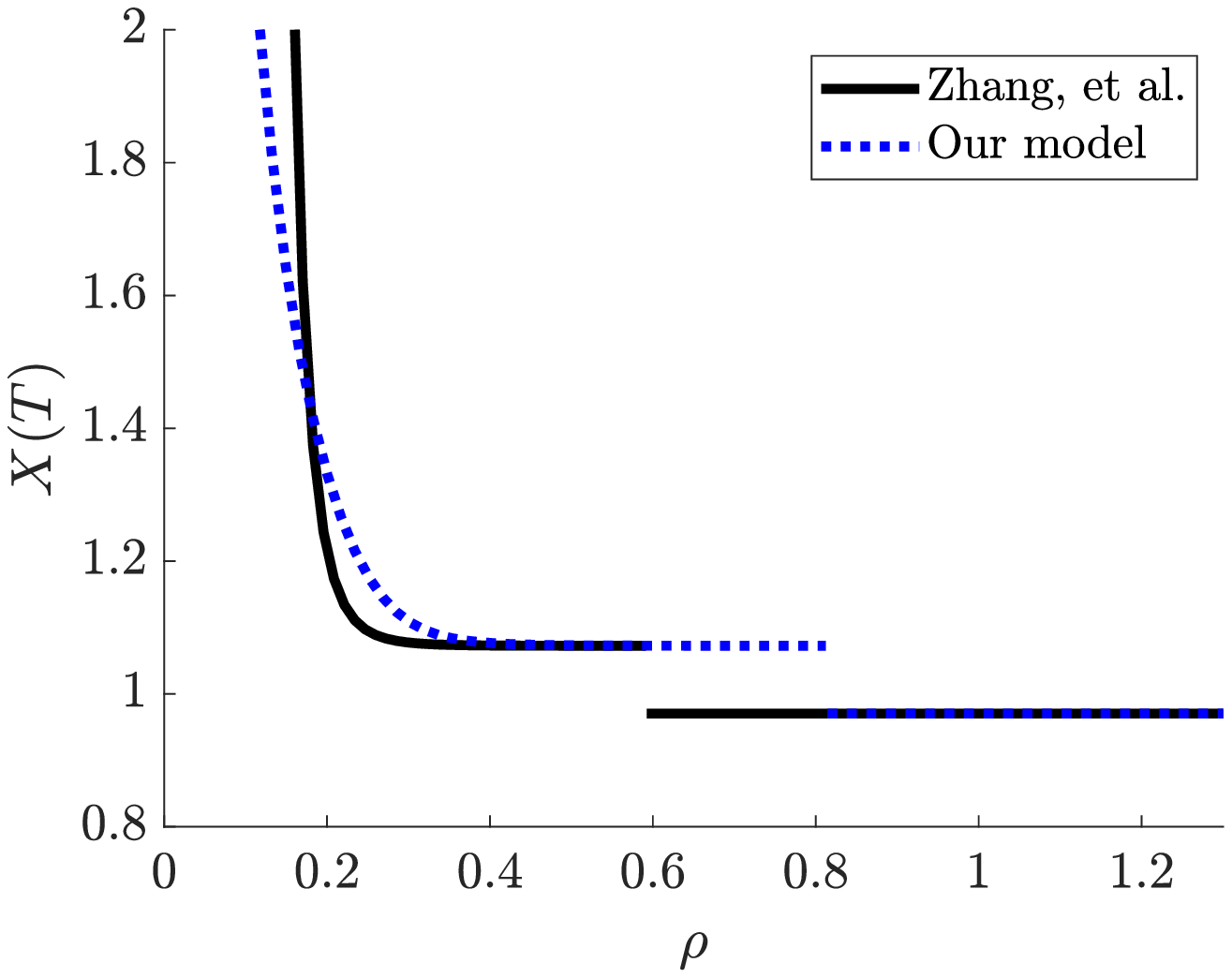}
\end{minipage}
\begin{minipage}[t]{2in}
\centering
\centerline{$c=0.1$ and $g=-0.05$}
\includegraphics[width=2.2in]{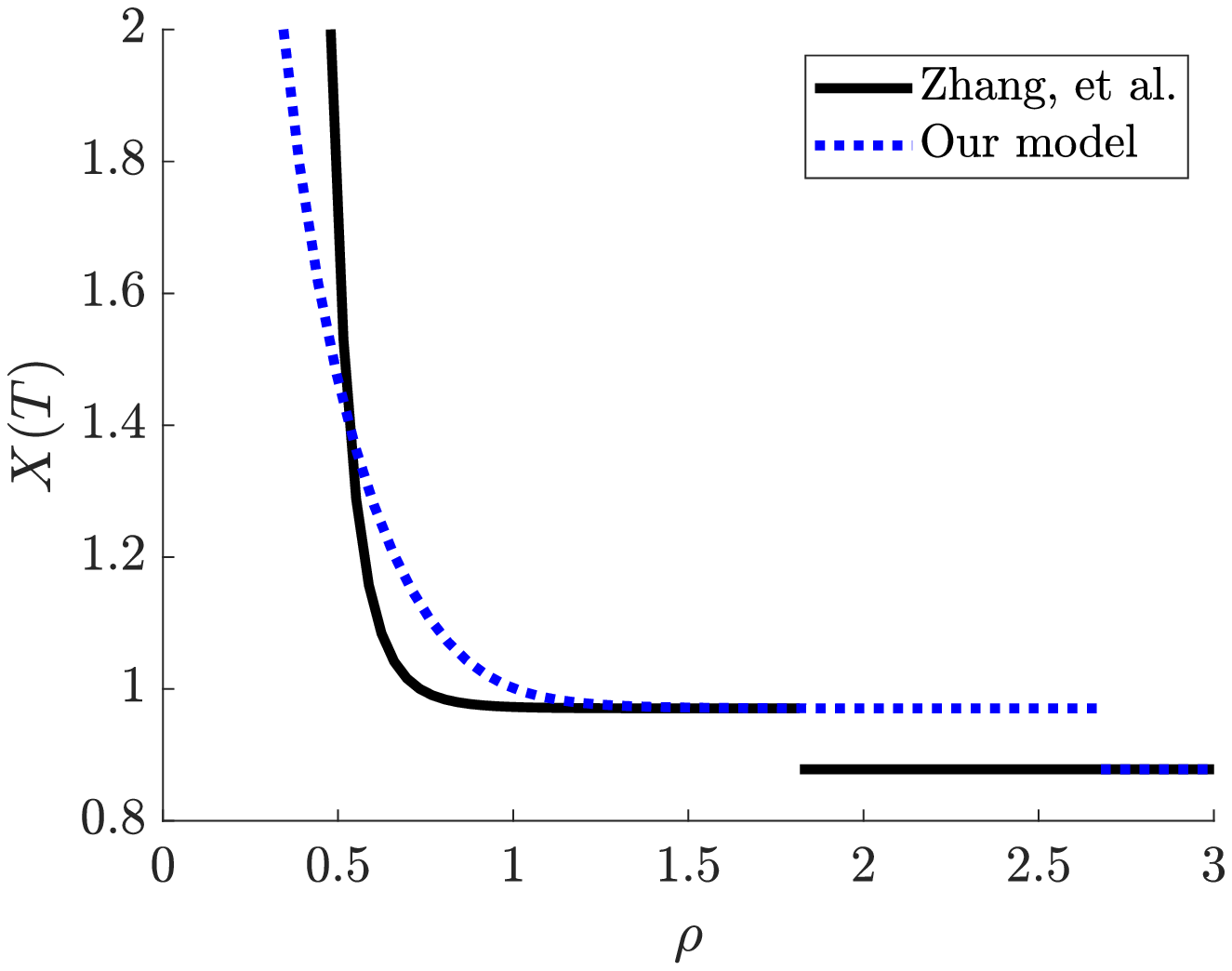}
\end{minipage}

\quad

\quad

\begin{minipage}[t]{2in}
\centering
\centerline{$c=0.2$ and $g=0$}
\includegraphics[width=2.2in]{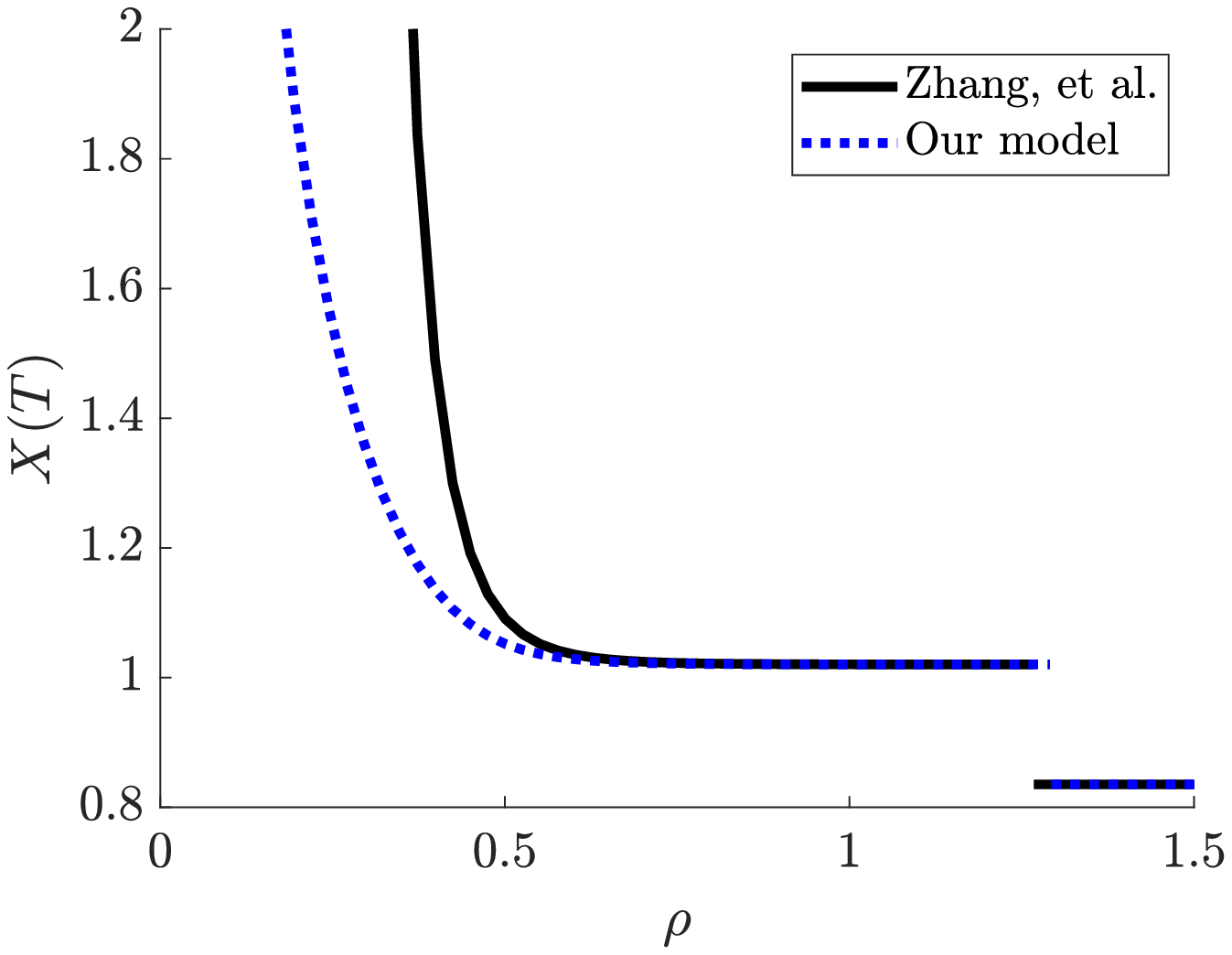}
\end{minipage}
\begin{minipage}[t]{2in}
\centering
\centerline{$c=0.2$ and $g=0.05$}
\includegraphics[width=2.2in]{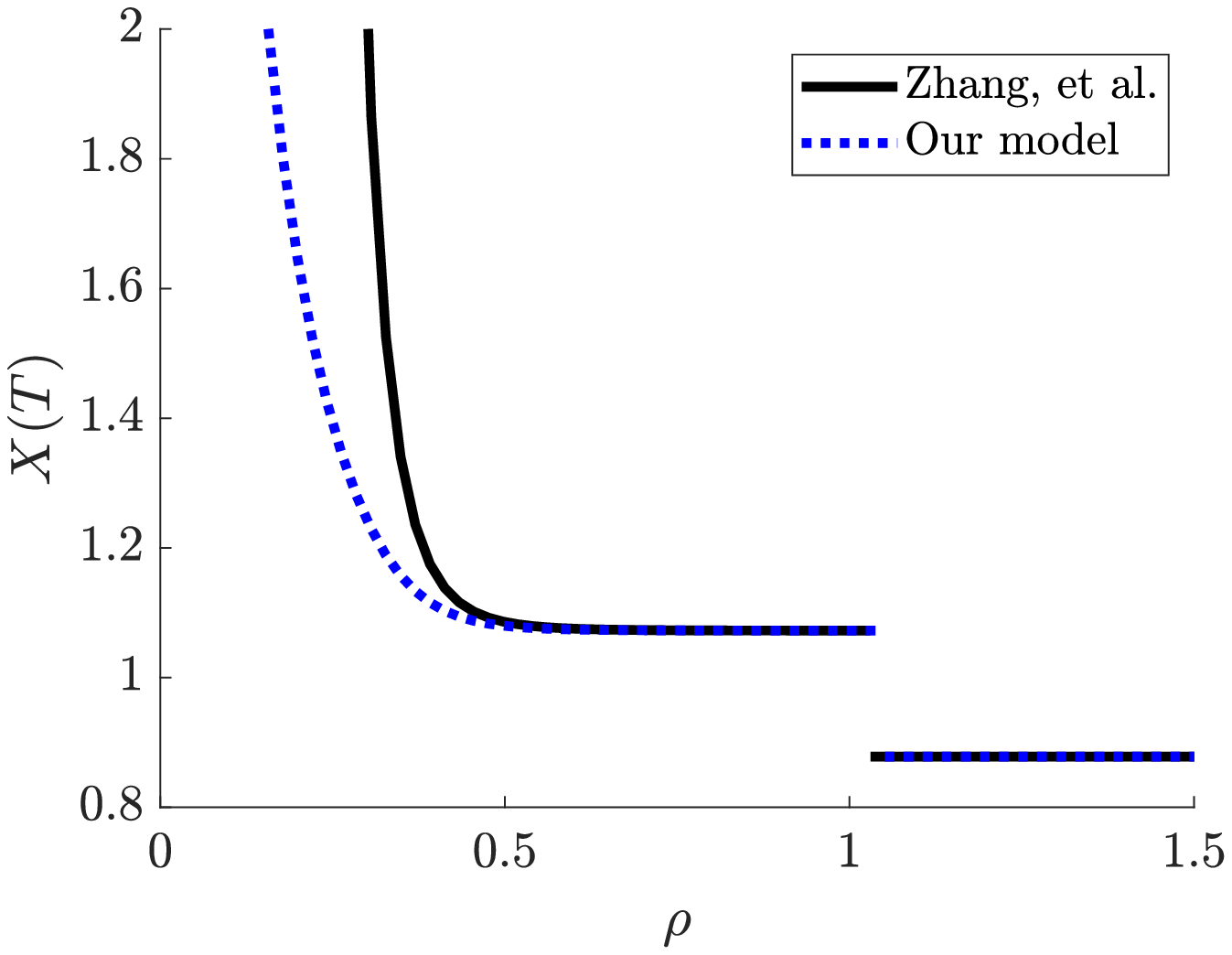}
\end{minipage}
\begin{minipage}[t]{2in}
\centering
\centerline{$c=0.2$ and $g=-0.05$}
\includegraphics[width=2.2in]{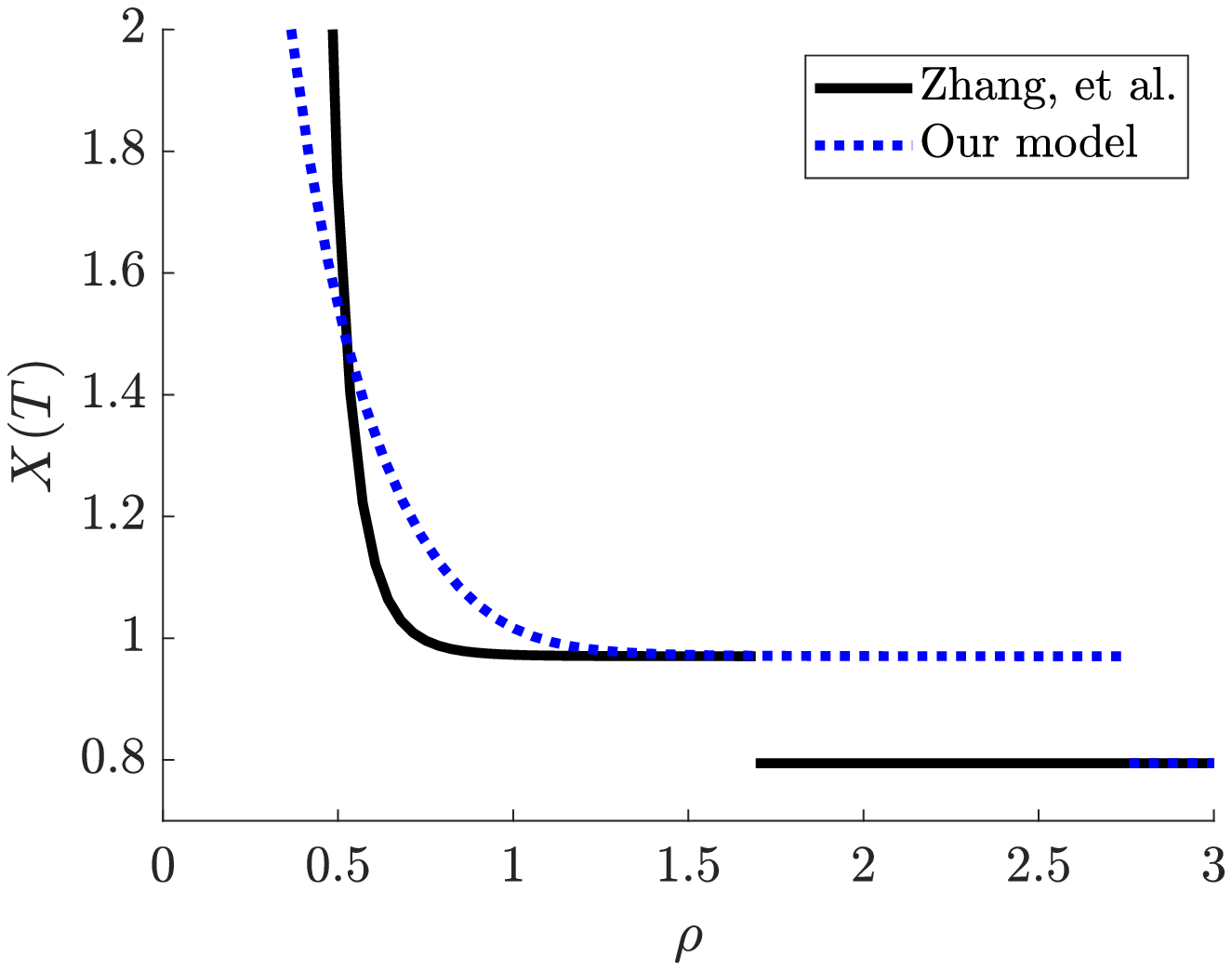}
\end{minipage}

\quad

\quad

\begin{minipage}[t]{2in}
\centering
\centerline{$c=0.3$ and $g=0$}
\includegraphics[width=2.2in]{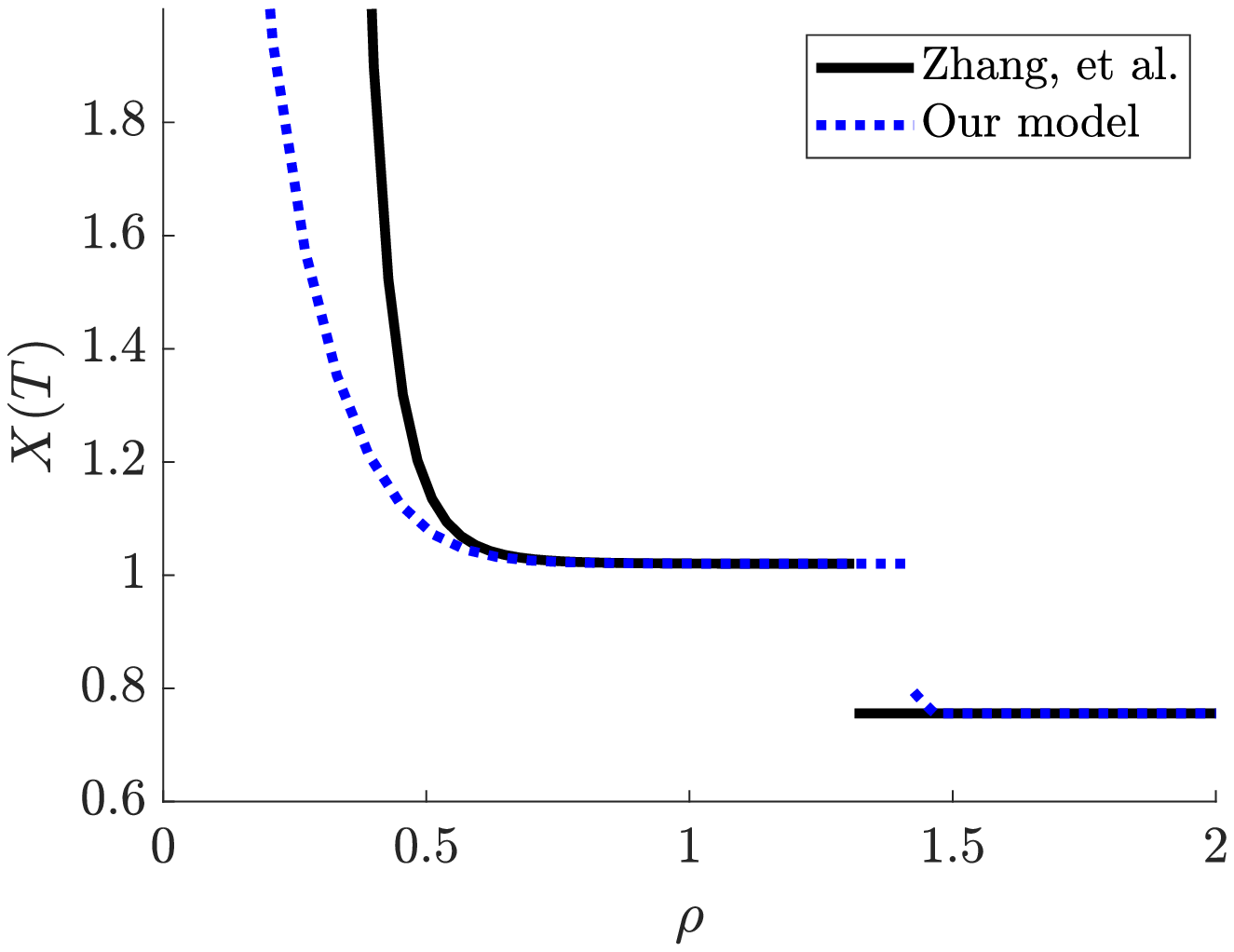}
\end{minipage}
\begin{minipage}[t]{2in}
\centering
\centerline{$c=0.3$ and $g=0.05$}
\includegraphics[width=2.2in]{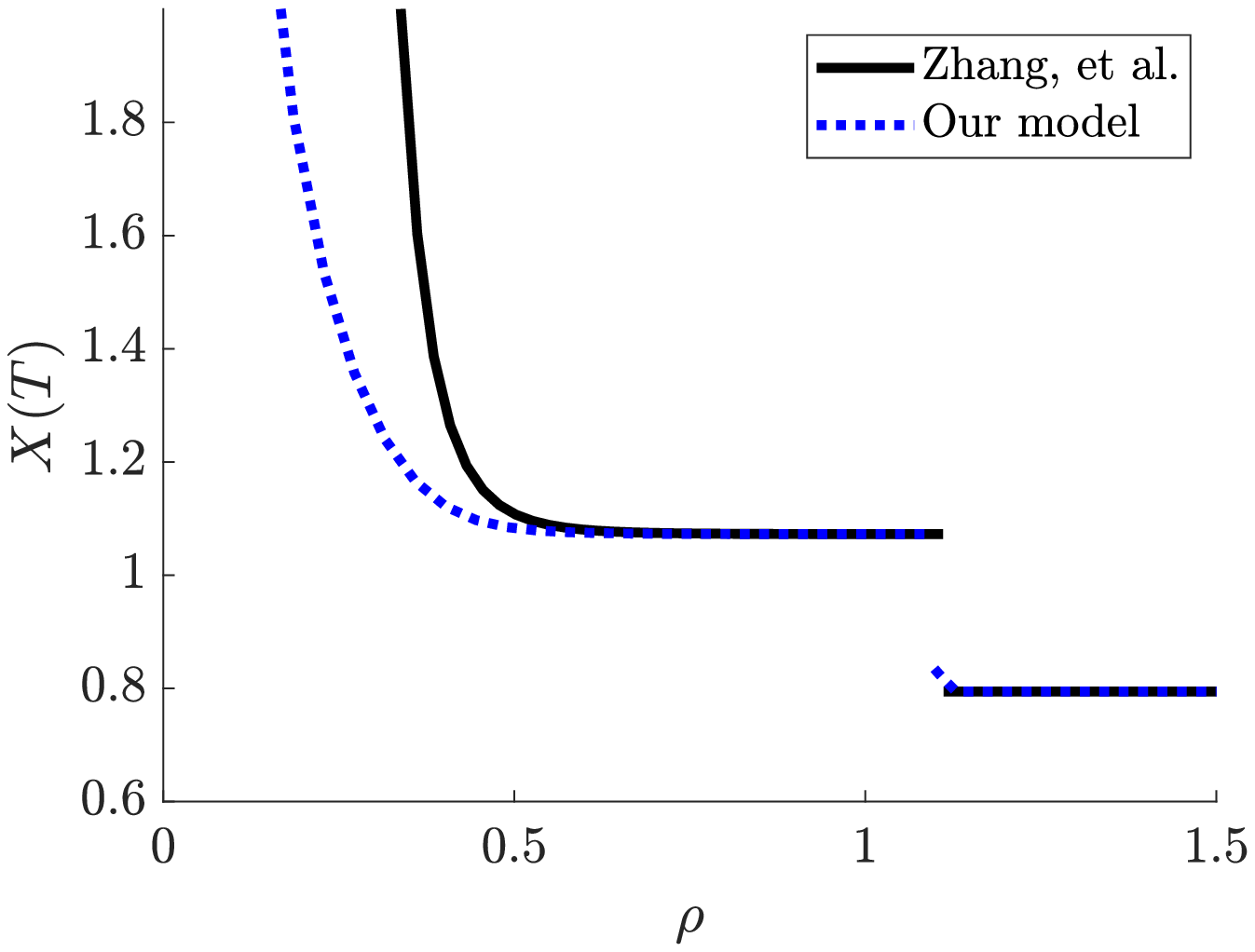}
\end{minipage}
\begin{minipage}[t]{2in}
\centering
\centerline{$c=0.3$ and $g=-0.05$}
\includegraphics[width=2.2in]{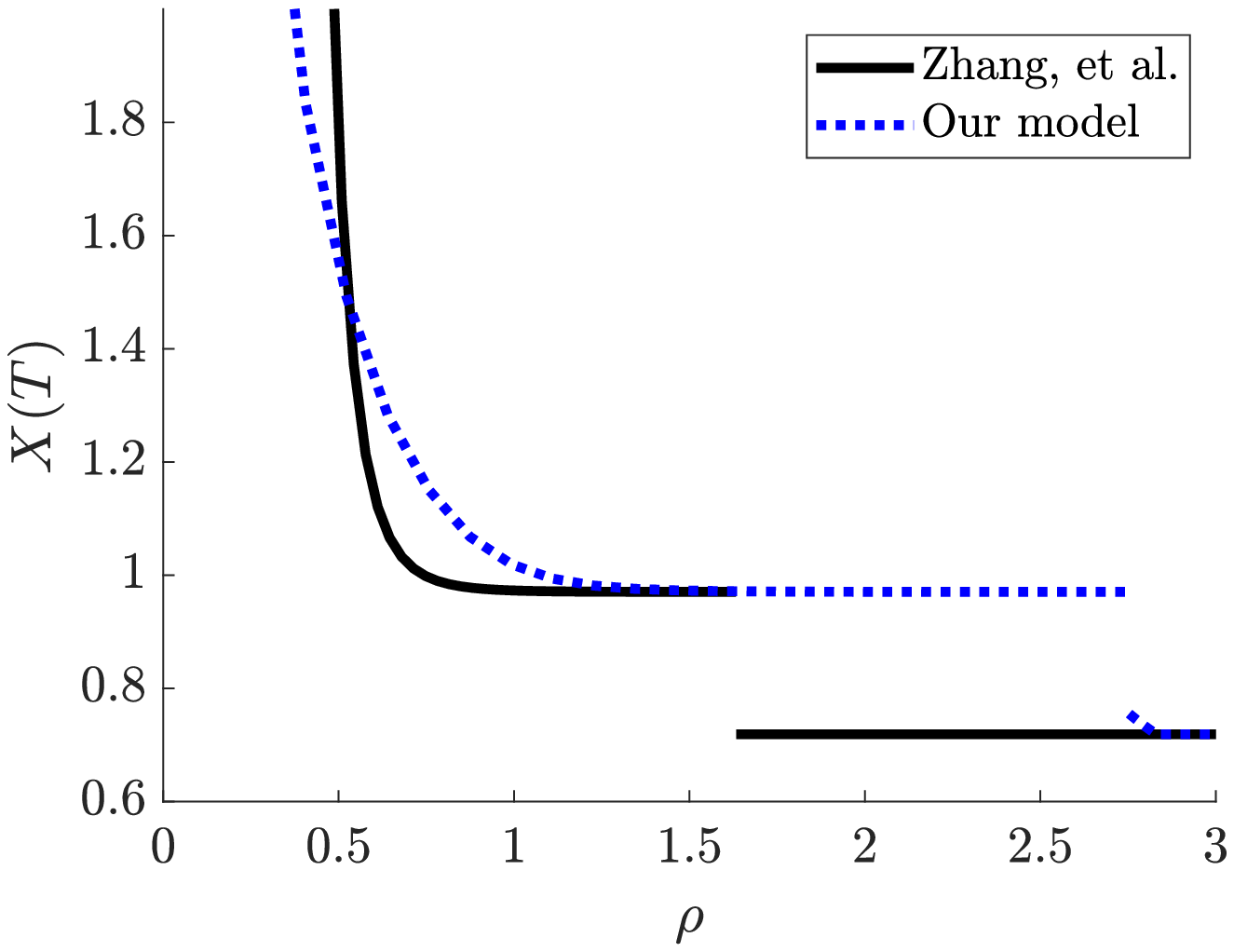}
\end{minipage}
\caption{Optimal terminal wealth as functions of $\rho$ under the identity weighting function. }\label{figure:no_weighting}
\end{figure}

\begin{figure}[htbp]
\centering
\begin{minipage}[t]{2in}
\centering
\centerline{$c=0.1$ and $g=0$}
\includegraphics[width=2.2in]{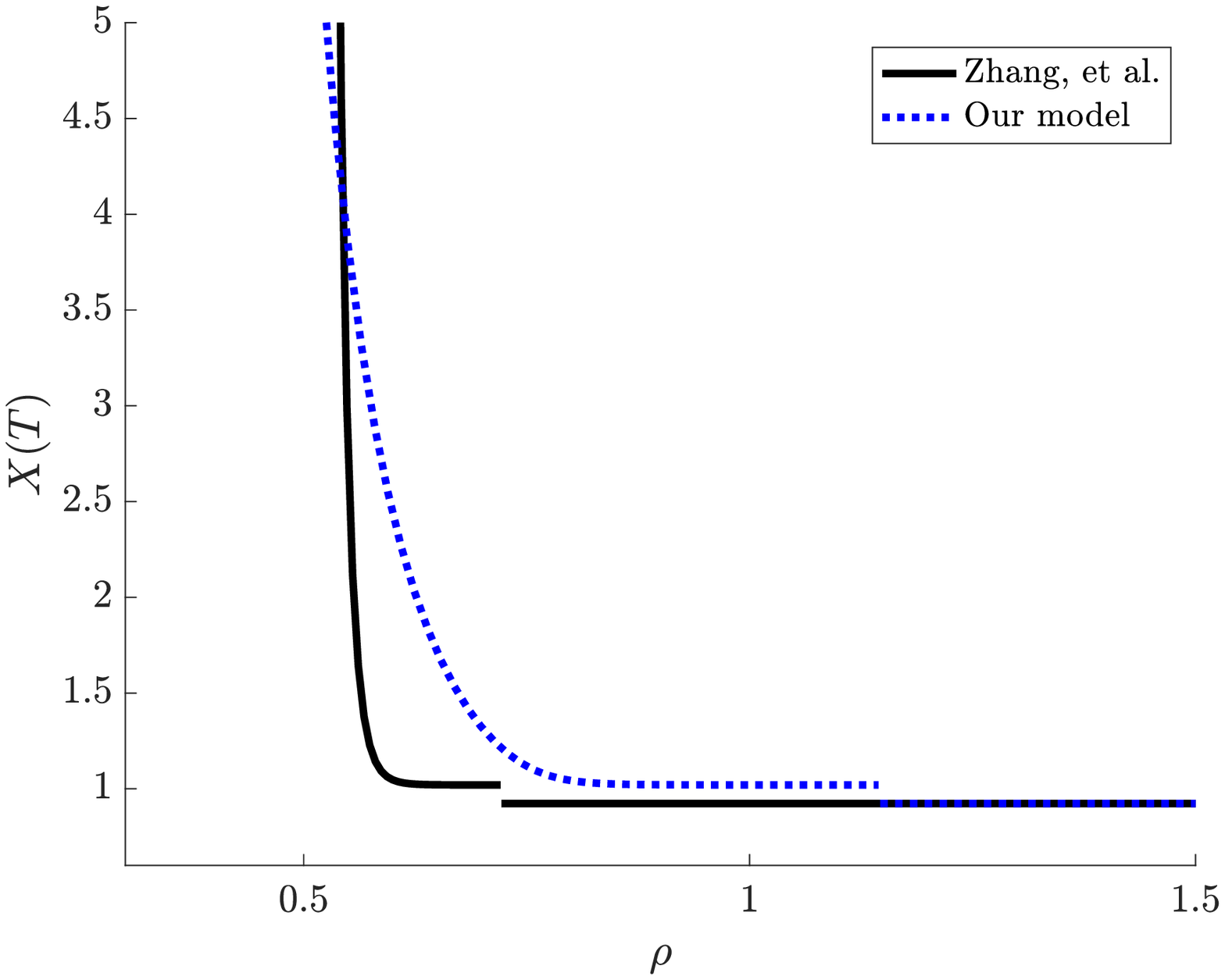}
\end{minipage}
\begin{minipage}[t]{2in}
\centering
\centerline{$c=0.1$ and $g=0.05$}
\includegraphics[width=2.2in]{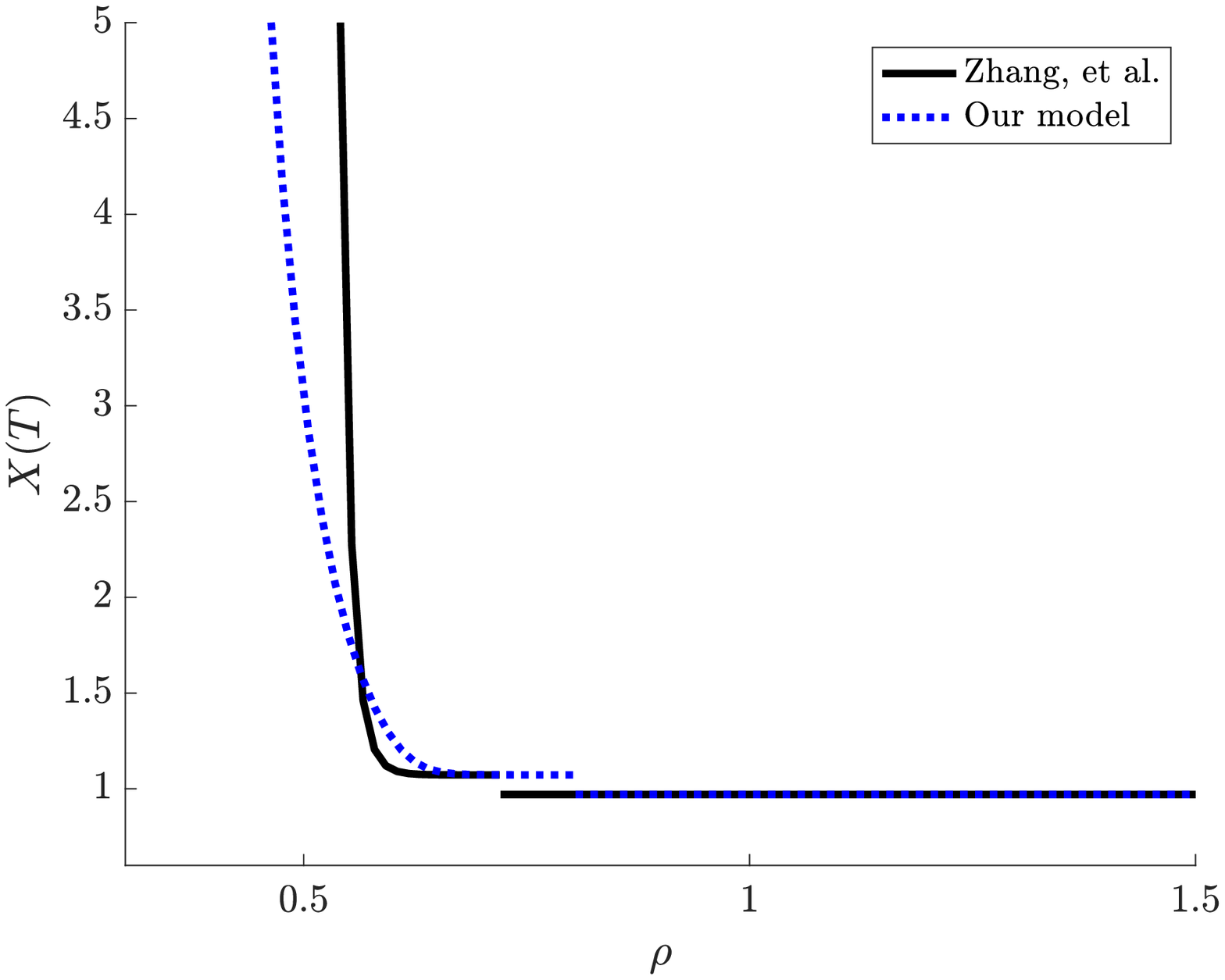}
\end{minipage}
\begin{minipage}[t]{2in}
\centering
\centerline{$c=0.1$ and $g=-0.05$}
\includegraphics[width=2.2in]{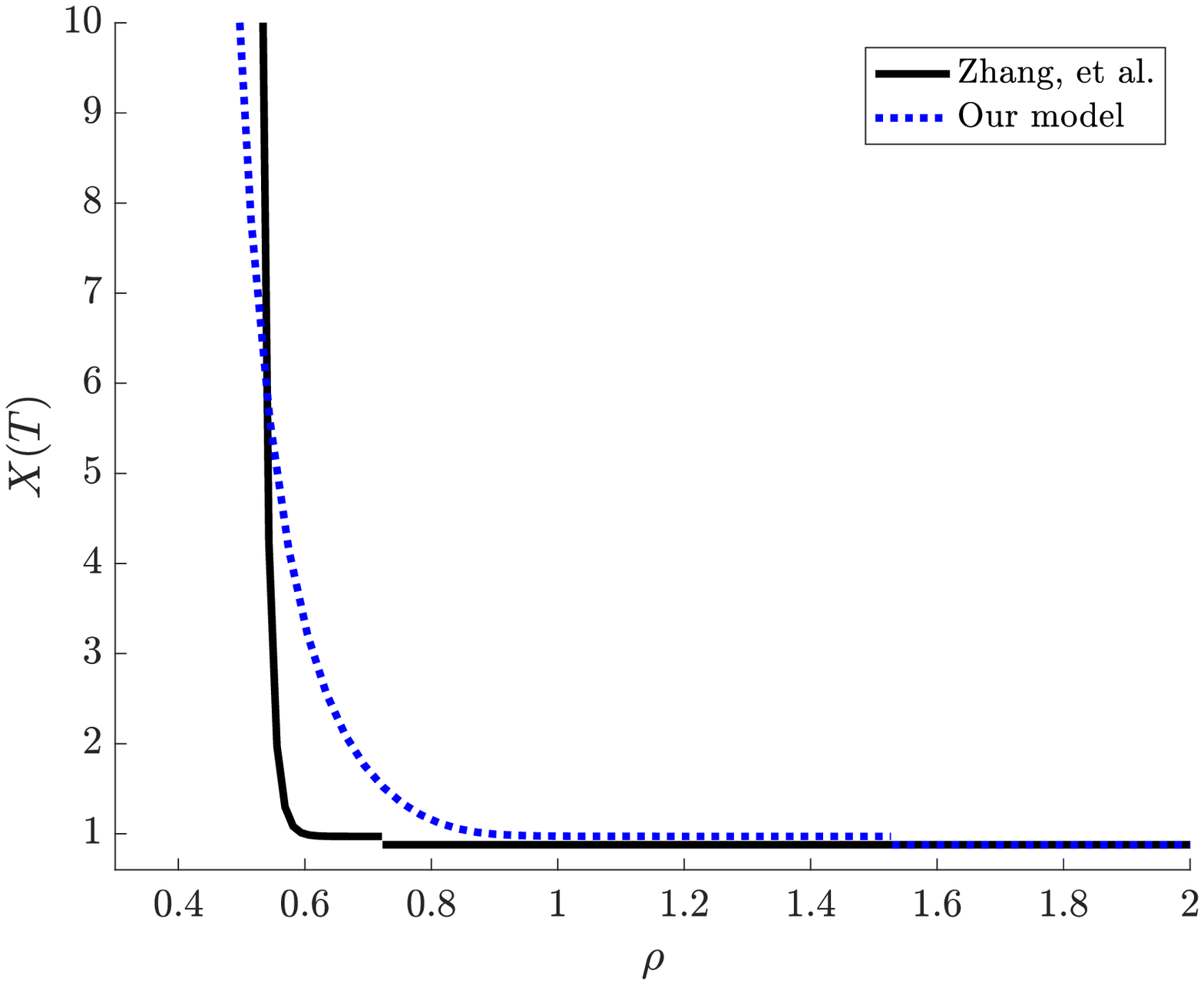}
\end{minipage}

\quad

\quad

\begin{minipage}[t]{2in}
\centering
\centerline{$c=0.2$ and $g=0$}
\includegraphics[width=2.2in]{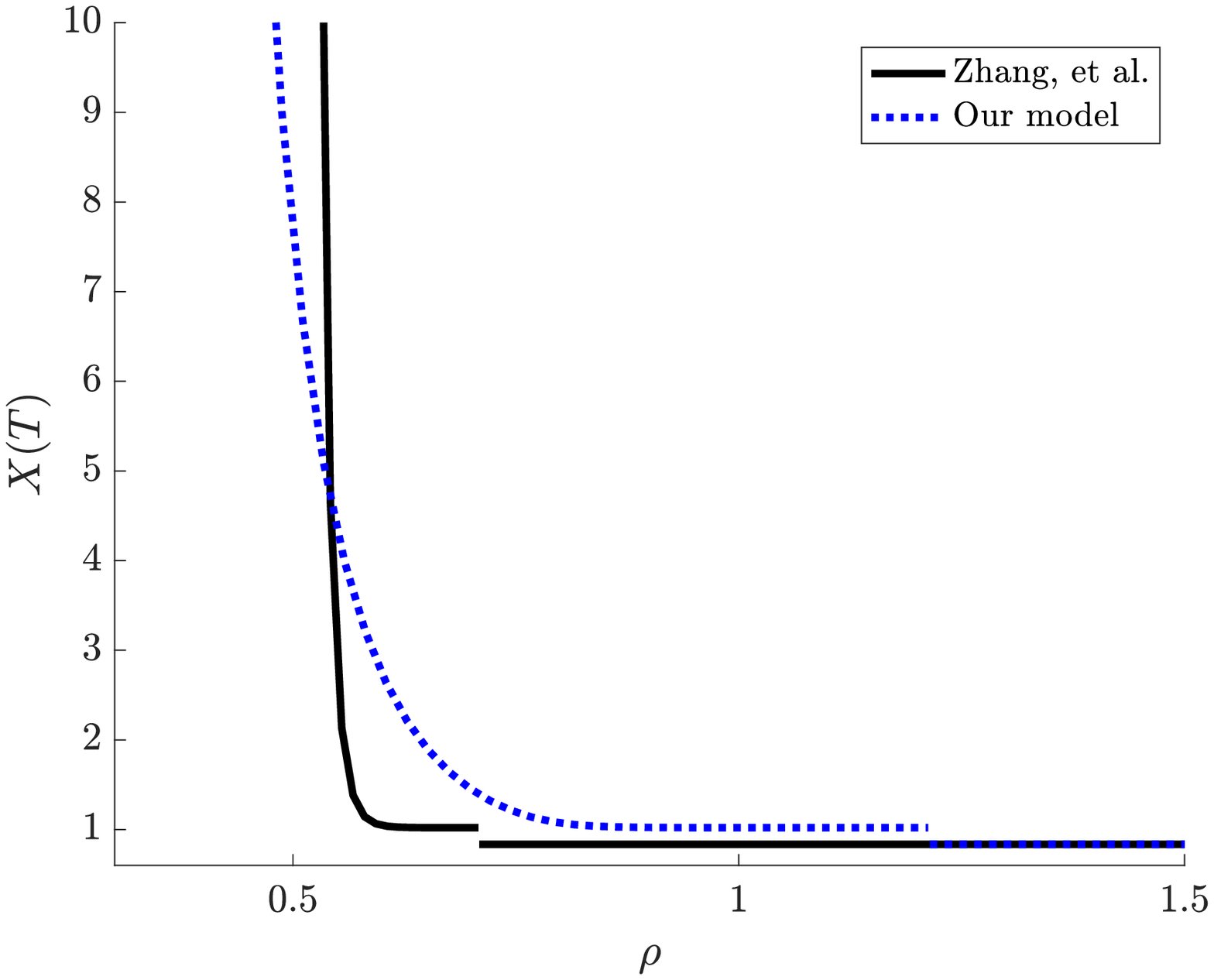}
\end{minipage}
\begin{minipage}[t]{2in}
\centering
\centerline{$c=0.2$ and $g=0.05$}
\includegraphics[width=2.2in]{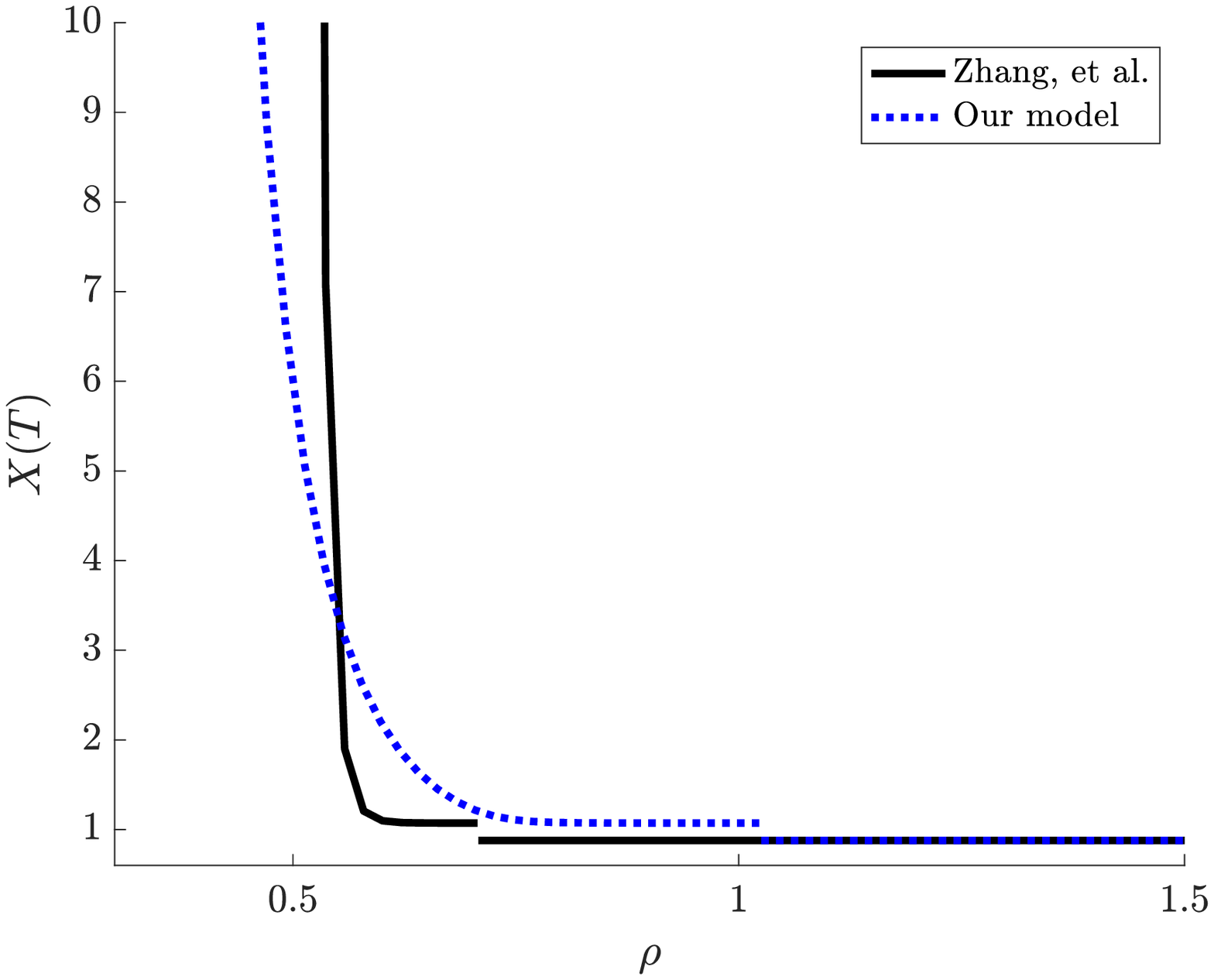}
\end{minipage}
\begin{minipage}[t]{2in}
\centering
\centerline{$c=0.2$ and $g=-0.05$}
\includegraphics[width=2.2in]{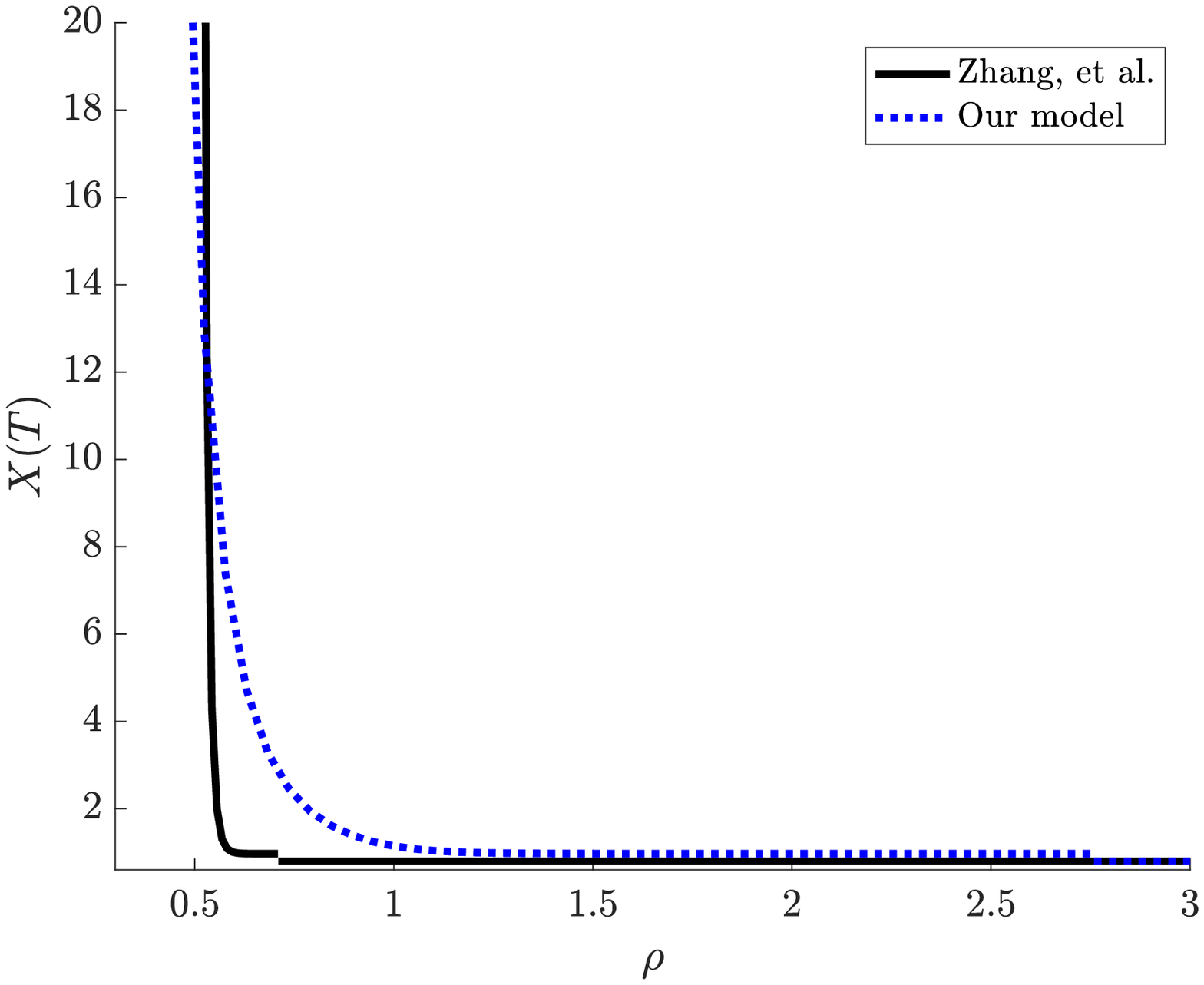}
\end{minipage}

\quad

\quad

\begin{minipage}[t]{2in}
\centering
\centerline{$c=0.3$ and $g=0$}
\includegraphics[width=2.2in]{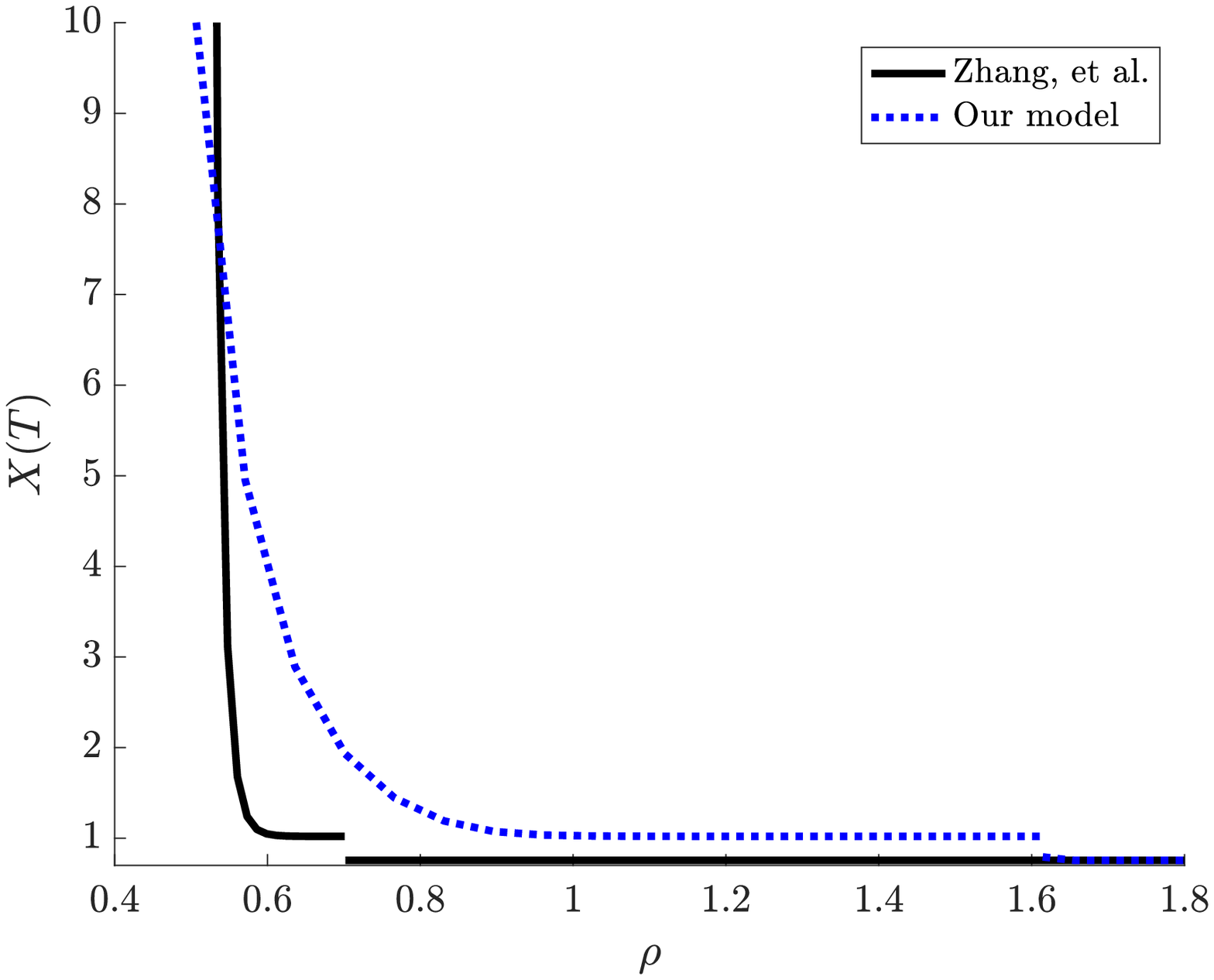}
\end{minipage}
\begin{minipage}[t]{2in}
\centering
\centerline{$c=0.3$ and $g=0.05$}
\includegraphics[width=2.2in]{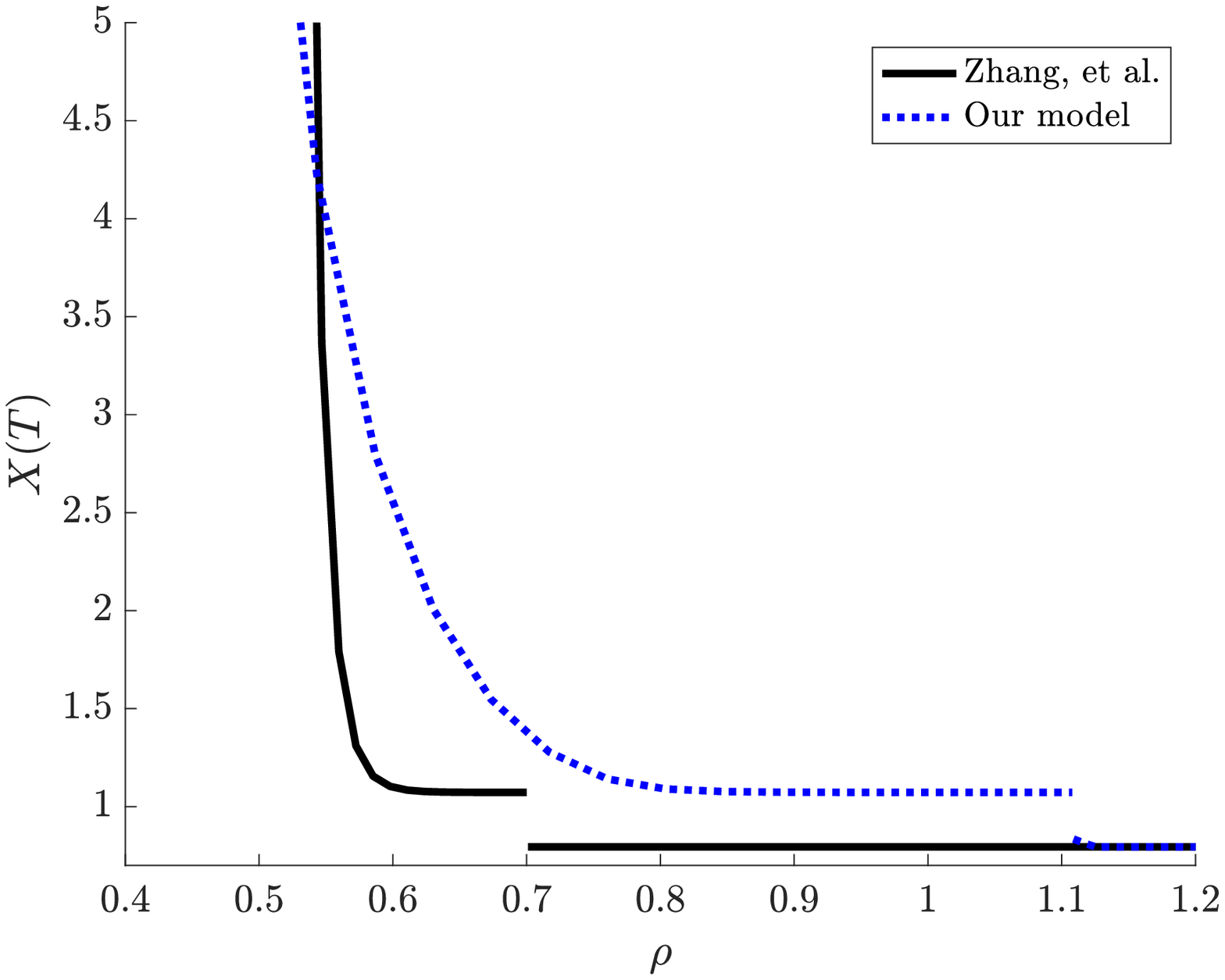}
\end{minipage}
\begin{minipage}[t]{2in}
\centering
\centerline{$c=0.3$ and $g=-0.05$}
\includegraphics[width=2.2in]{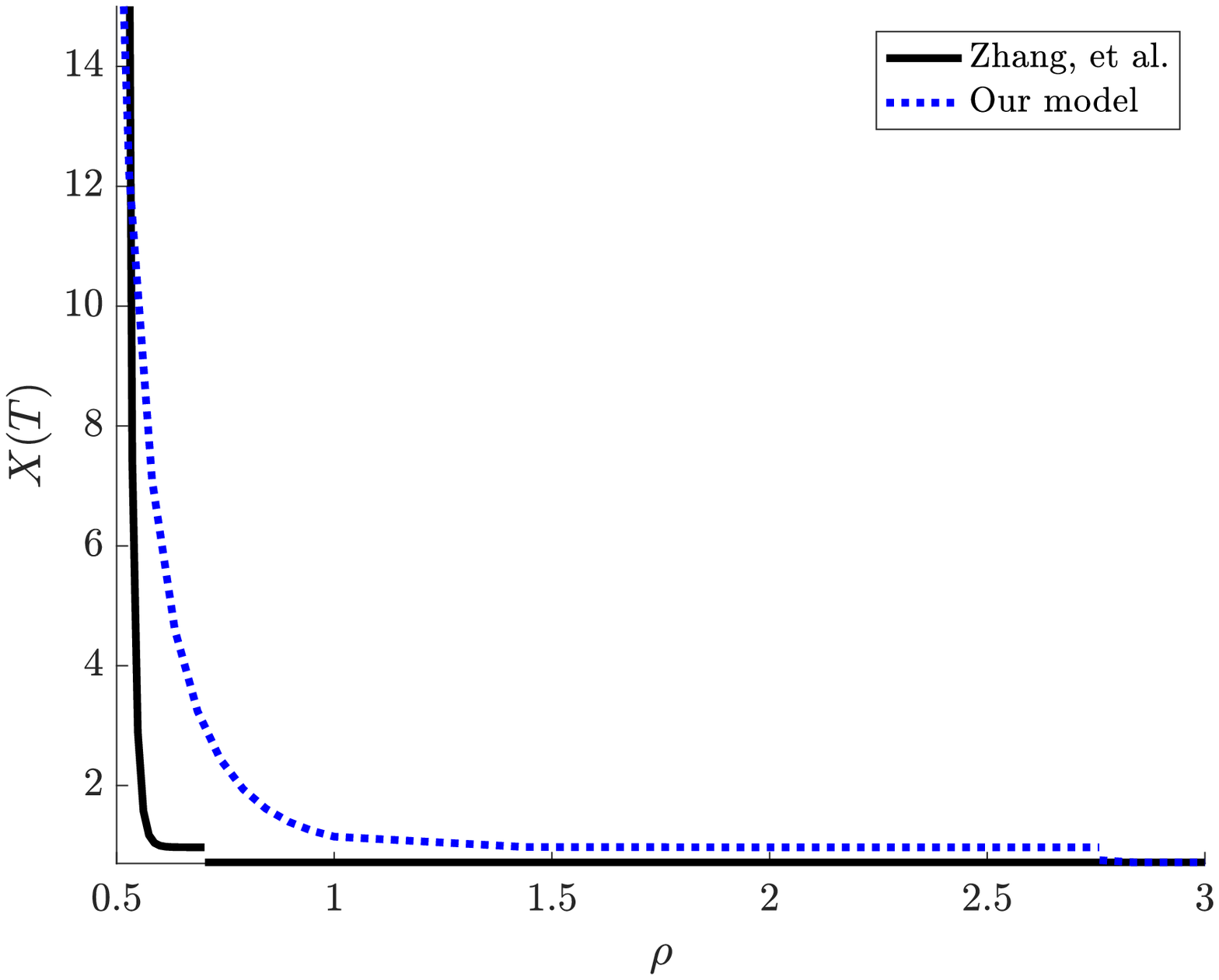}
\end{minipage}
\caption{Optimal terminal wealth as functions of $\rho$ under the power weighting function. }\label{figure:power}
\end{figure}

\begin{figure}[htbp]
\centering
\begin{minipage}[t]{2in}
\centering
\centerline{$c=0.1$ and $g=0$}
\includegraphics[width=2.2in]{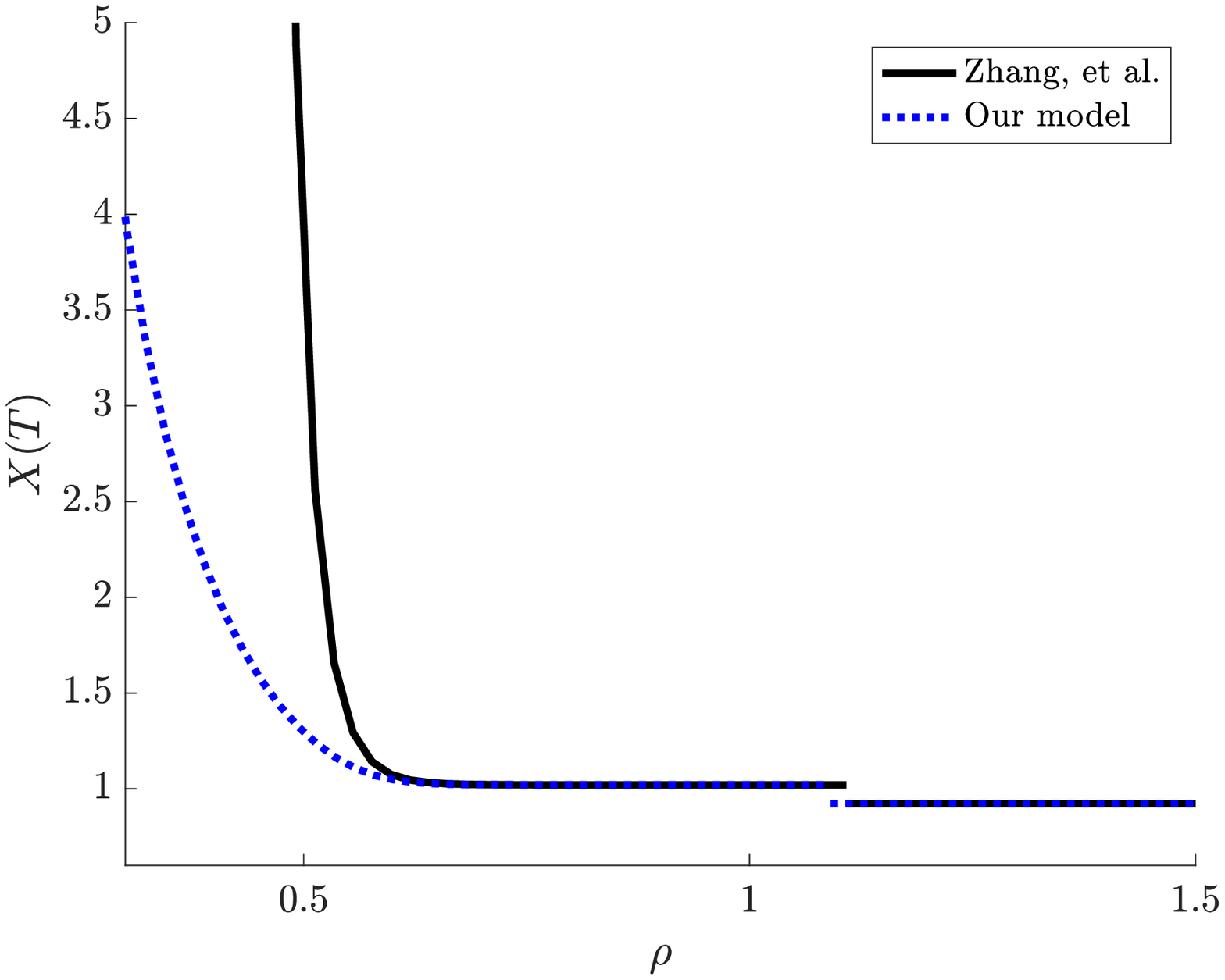}
\end{minipage}
\begin{minipage}[t]{2in}
\centering
\centerline{$c=0.1$ and $g=0.05$}
\includegraphics[width=2.2in]{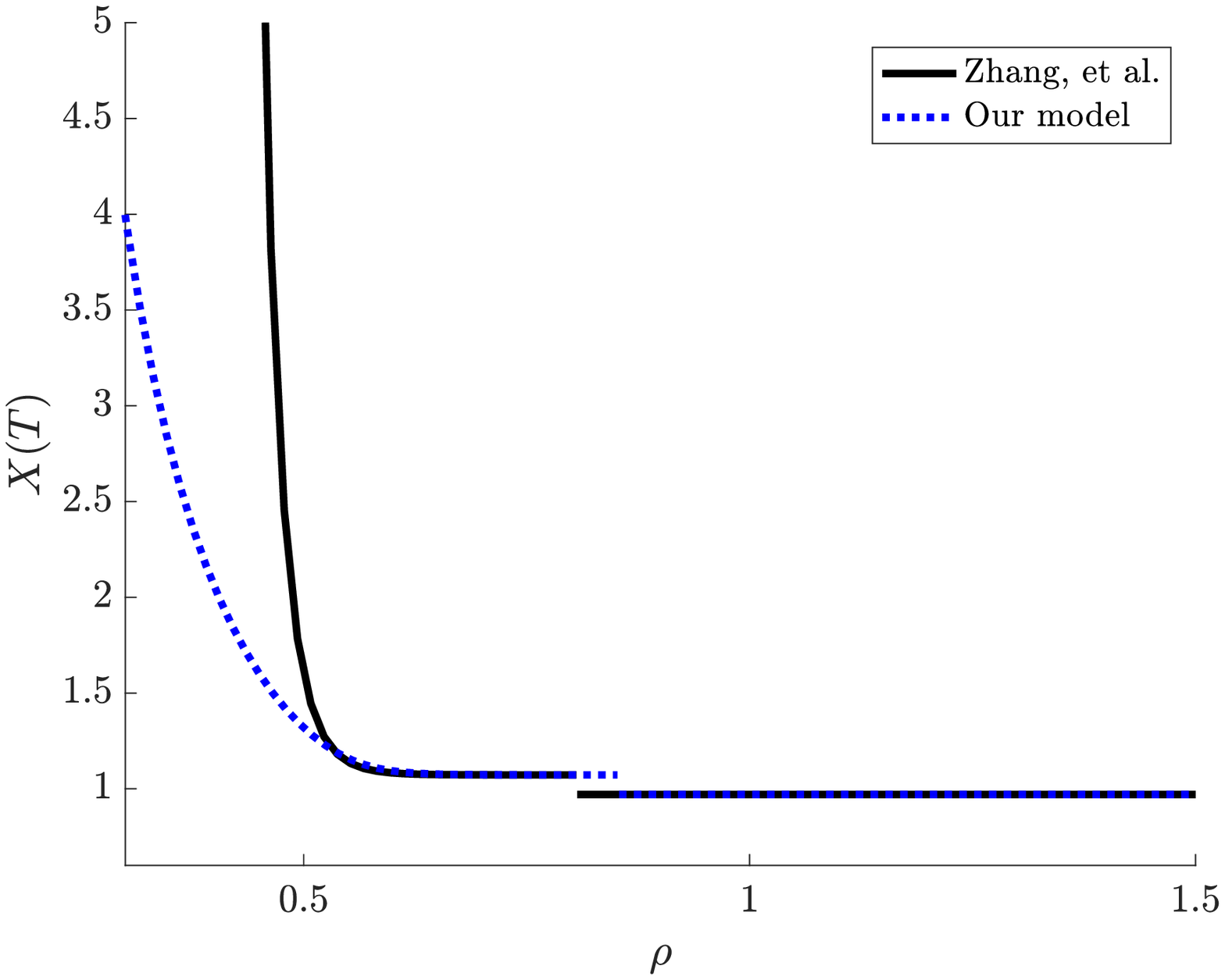}
\end{minipage}
\begin{minipage}[t]{2in}
\centering
\centerline{$c=0.1$ and $g=-0.05$}
\includegraphics[width=2.2in]{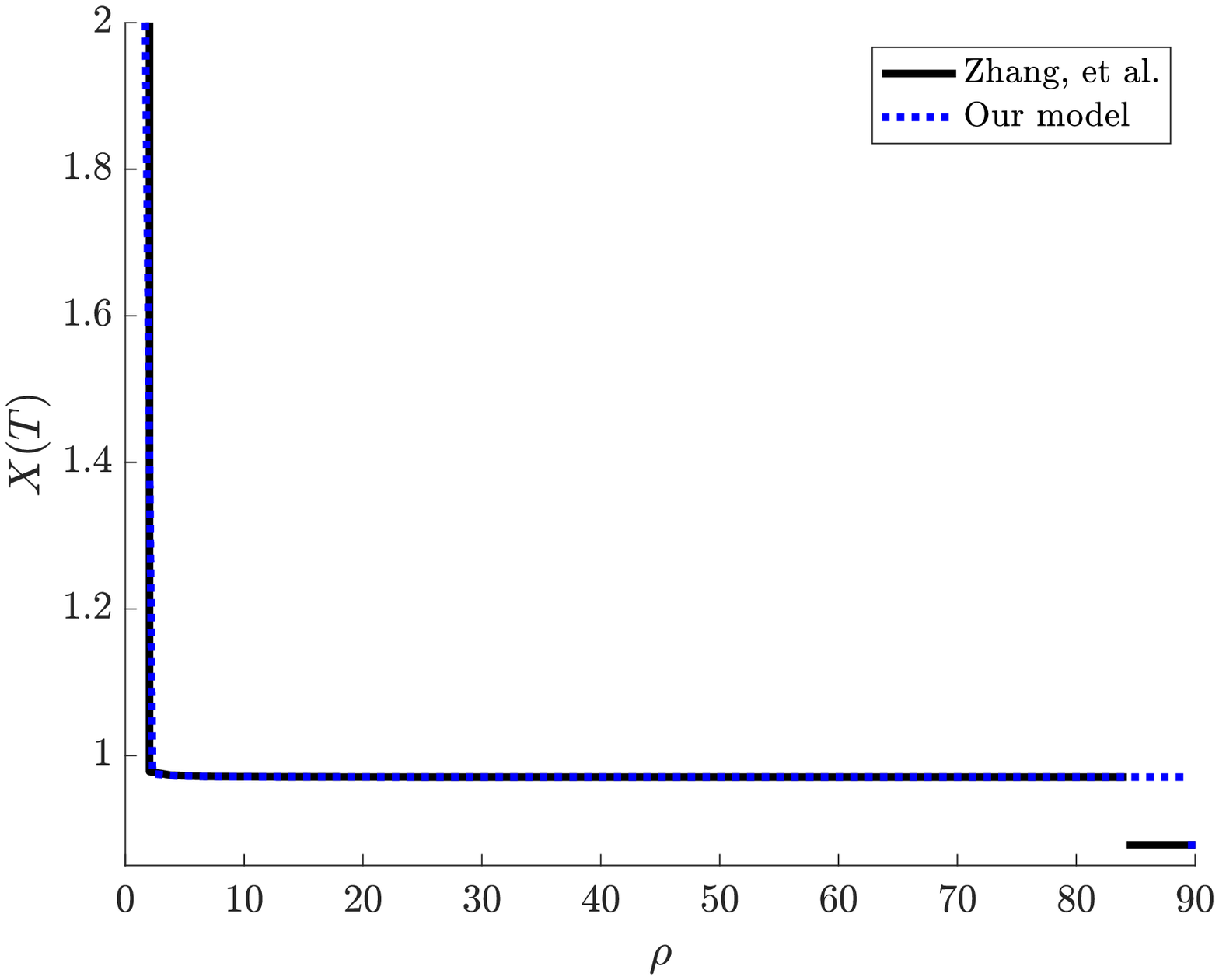}
\end{minipage}

\quad

\quad

\begin{minipage}[t]{2in}
\centering
\centerline{$c=0.2$ and $g=0$}
\includegraphics[width=2.2in]{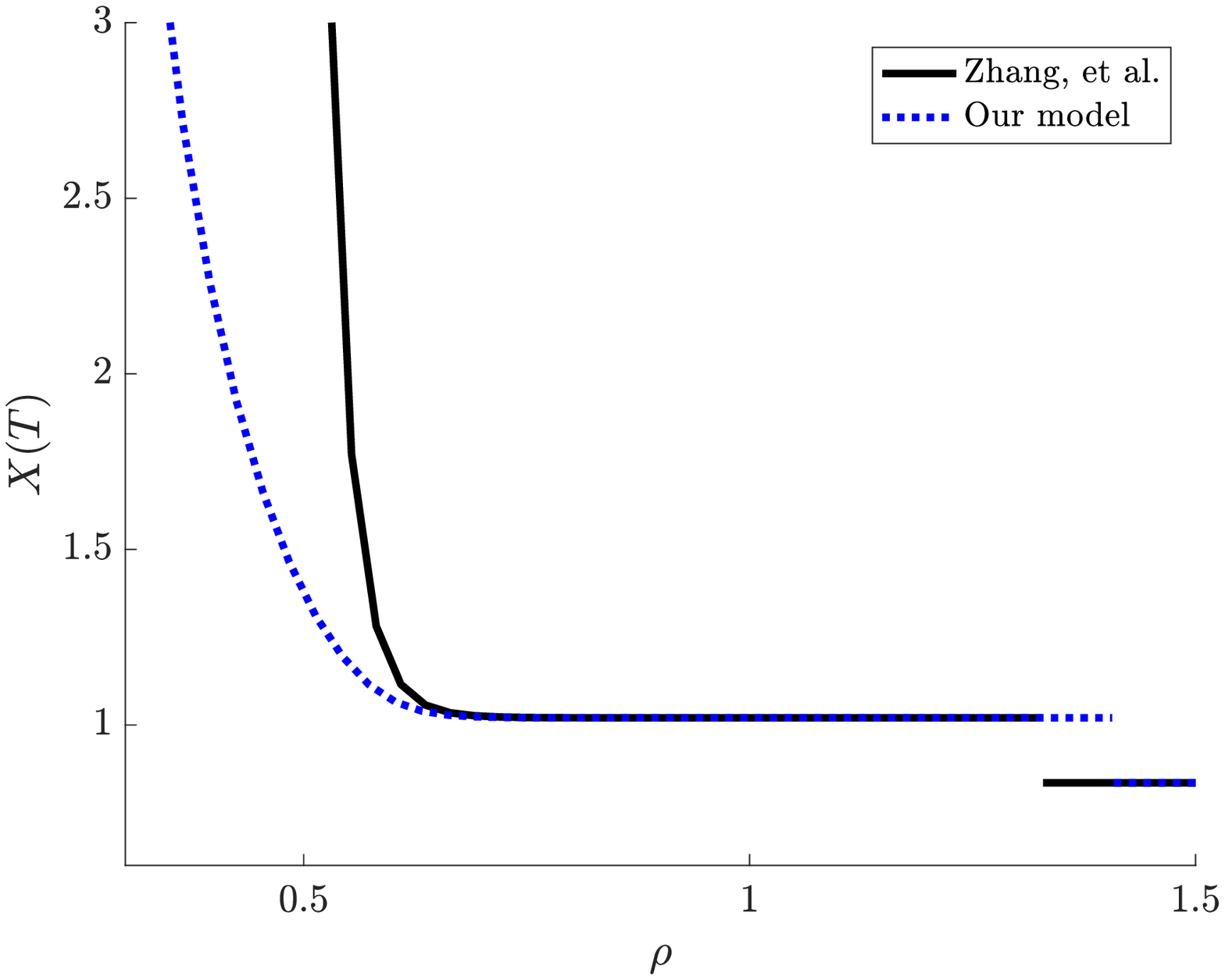}
\end{minipage}
\begin{minipage}[t]{2in}
\centering
\centerline{$c=0.2$ and $g=0.05$}
\includegraphics[width=2.2in]{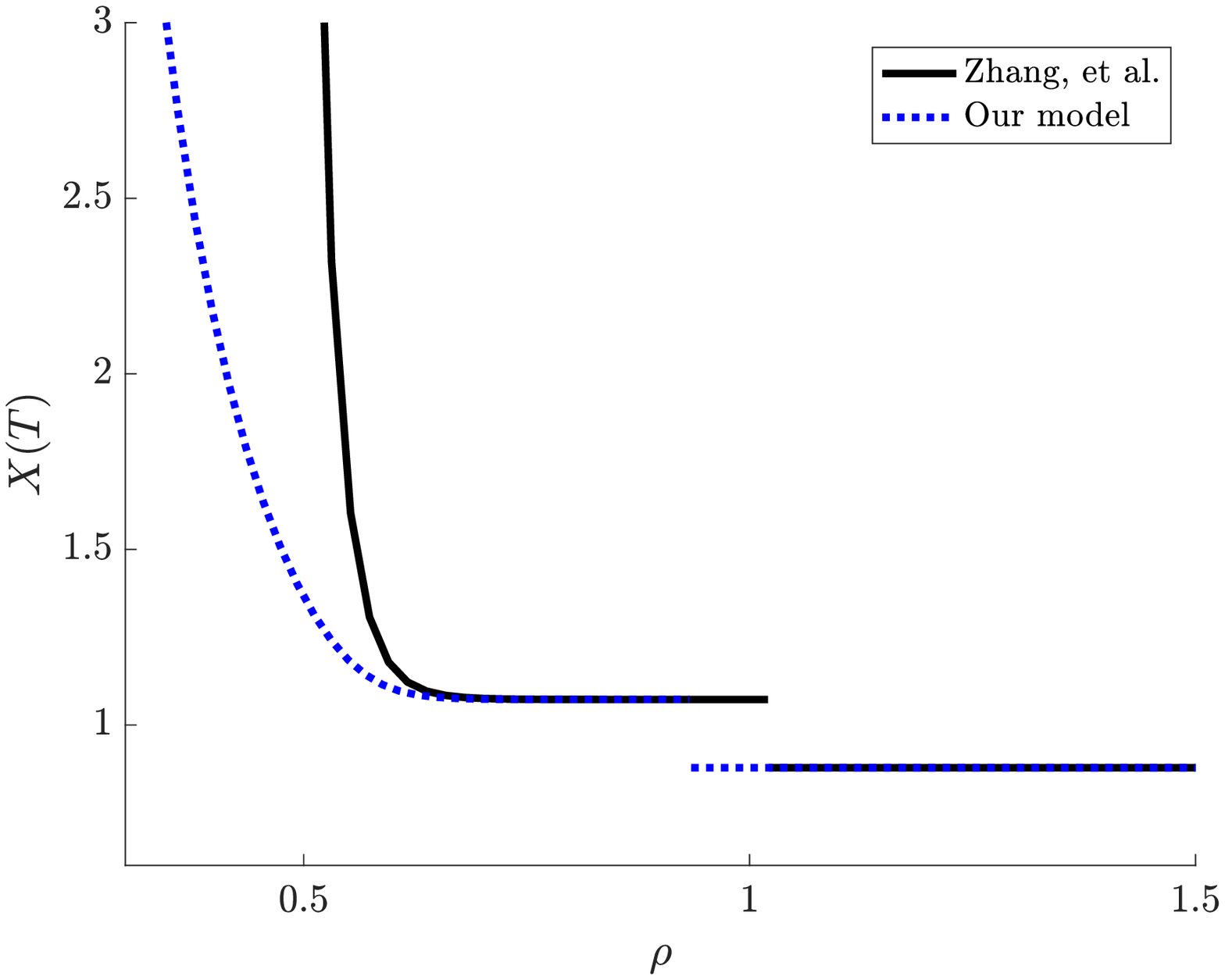}
\end{minipage}
\begin{minipage}[t]{2in}
\centering
\centerline{$c=0.2$ and $g=-0.05$}
\includegraphics[width=2.2in]{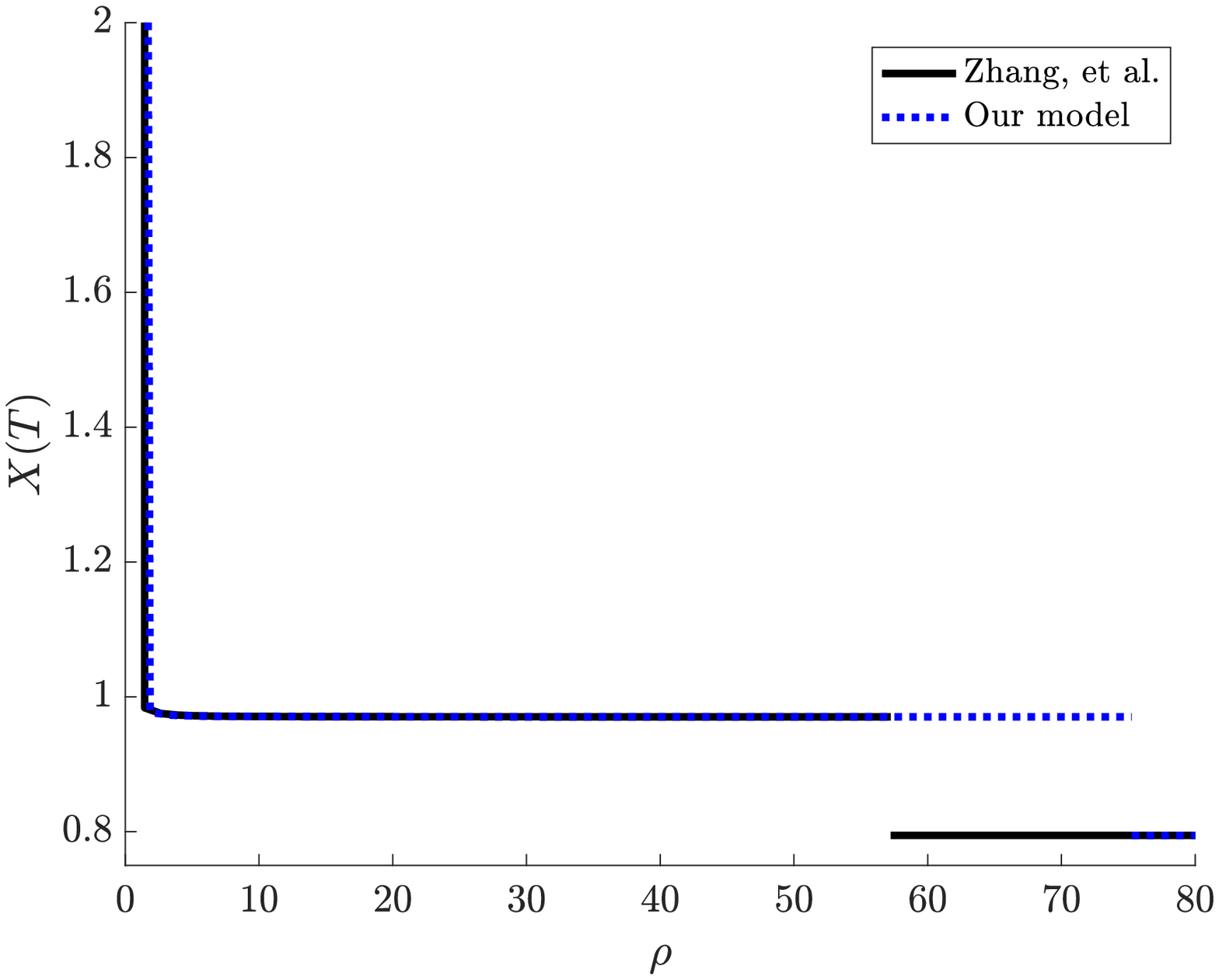}
\end{minipage}

\quad

\quad

\begin{minipage}[t]{2in}
\centering
\centerline{$c=0.3$ and $g=0$}
\includegraphics[width=2.2in]{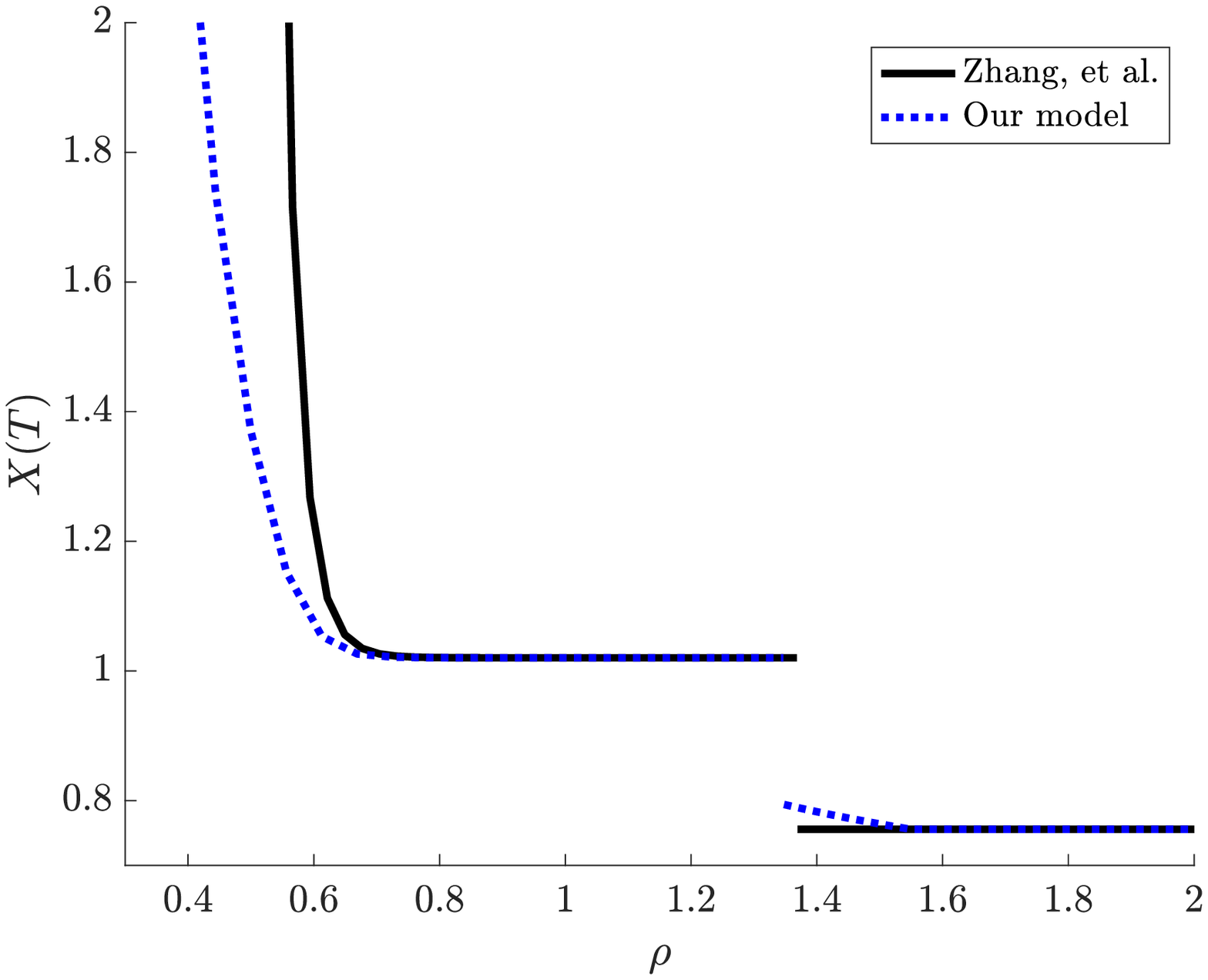}
\end{minipage}
\begin{minipage}[t]{2in}
\centering
\centerline{$c=0.3$ and $g=0.05$}
\includegraphics[width=2.2in]{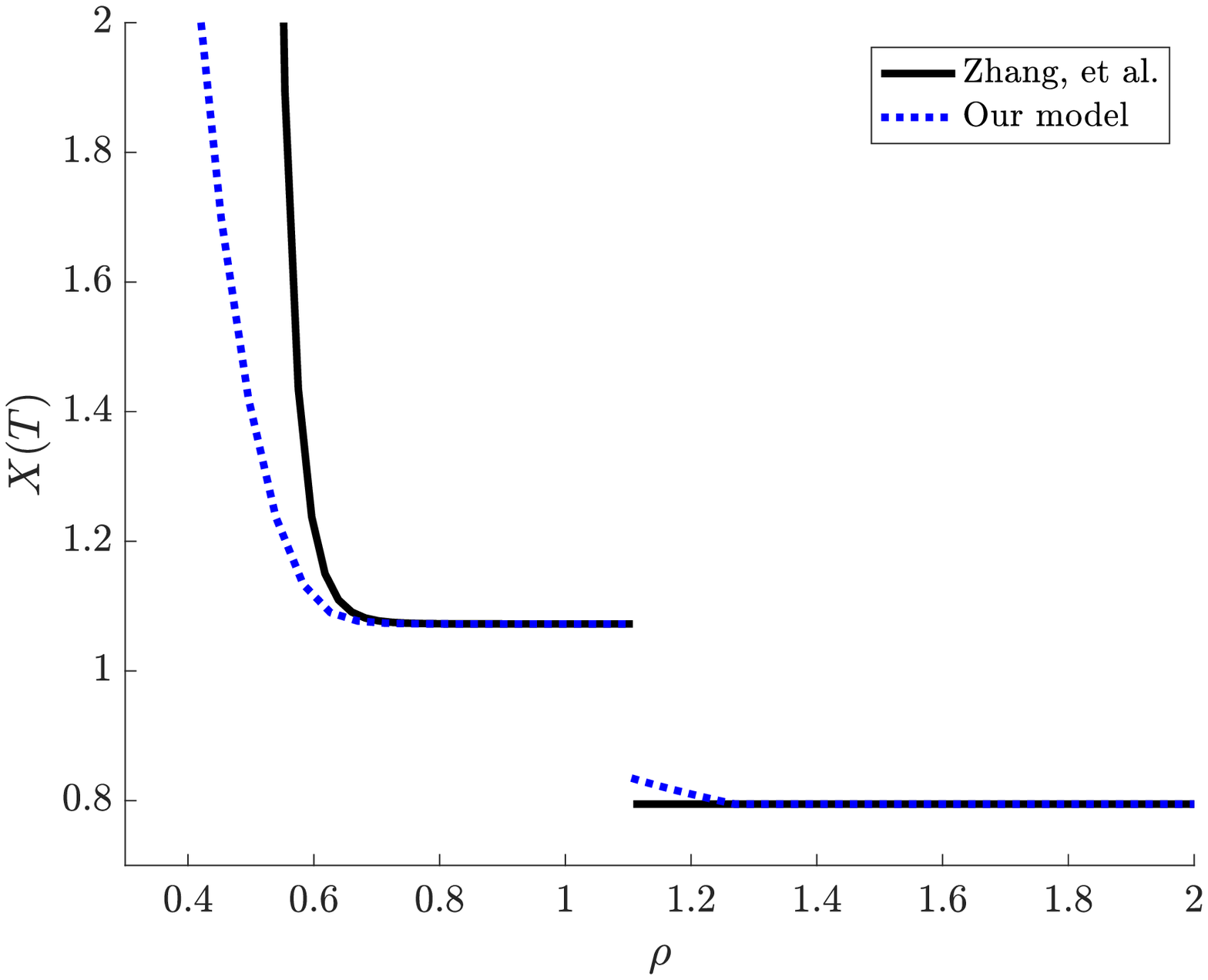}
\end{minipage}
\begin{minipage}[t]{2in}
\centering
\centerline{$c=0.3$ and $g=-0.05$}
\includegraphics[width=2.2in]{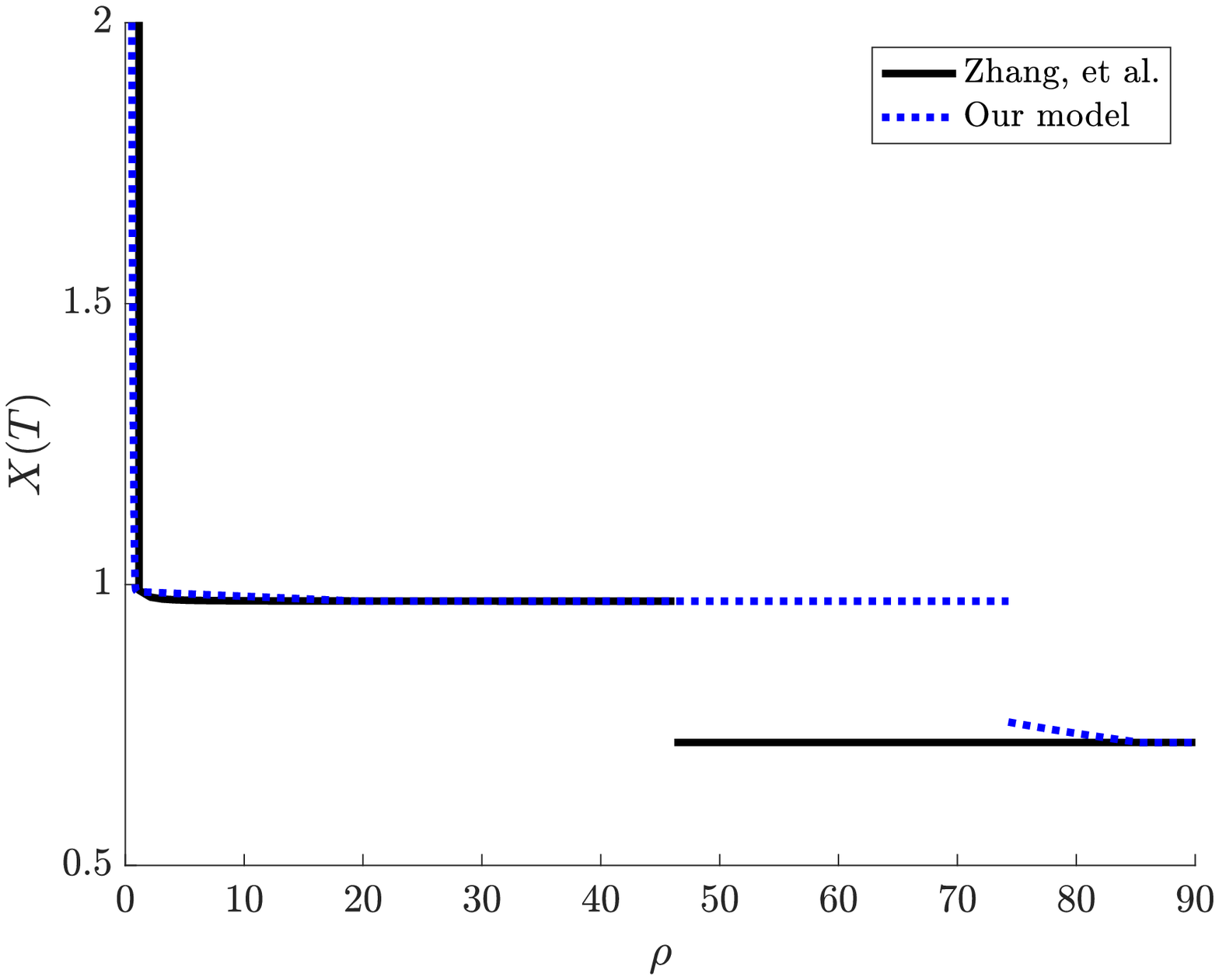}
\end{minipage}
\caption{Optimal terminal wealth as functions of $\rho$ under the Jin-Zhou weighting function. }\label{figure:JZ}
\end{figure}

Figures \ref{figure:no_weighting}, \ref{figure:power}, and \ref{figure:JZ} display the optimal terminal wealth to \eqref{paramodel} and \eqref{paramodel1} as functions of $\rho$ when the weighting function is given by the identity weighting function, the power weighting function, and the Jin-Zhou weighting function, respectively, in different scenarios. 

\bigskip

In all scenarios under the three types of probability weighting functions, Figures \ref{figure:no_weighting}, \ref{figure:power} and \ref{figure:JZ} show that our terminal wealth is higher than or at least the same as Zhang et al.'s in the intermediate and bad states (i.e. $\rho$ is not very small). By contrast, our one is lower than Zhang et al.'s in the good states (i.e. $\rho$ is very small). From an economic viewpoint, the performance of our terminal wealth is more stable than Zhang et al.'s, and it is less sensitive to the pricing kernel (i.e. the curve is flatter). It suggests that our strategy is more suitable for loss-sensitive investors than Zhang et al.'s, whereas their strategy is more suitable for those investors betting for extremely good states. 

\blue{
Once we have the expression for the optimal terminal wealth, we can solve for the time-$t$ wealth via $$X(t) = \frac{1}{\rho(t)}\BE{\rho(T) X(T)\;\big|\;\mathcal{F}_t},$$
which can be expressed as a deterministic function of $t$ and $\rho (t)$, say $\Psi(t,\rho(t))$. It is then straightforward to verify that the optimal portfolio weight is given by
$$\pi (t) = - \Psi_{x}(t,\rho(t)) \cdot \frac{\rho (t)}{X(t)} \left( \sigma (t)' \right)^{-1} \theta. $$
It is well-known that $\left( \sigma (t)'
 \right)^{-1} \theta$ is the (instantaneous) mean-variance portfolio, or the GOP, i.e., it is optimal for an investor maximizing the logarithm utility of terminal wealth. Therefore, $-\Psi_{x}(t,\rho(t)) \cdot {\rho (t)}/{X(t)}$ represents the risk exposure relative to the GOP.

Figure \ref{figure:no_weighting_strategy} presents the optimal risk exposures as functions of $\rho (t)$ when $t=0.5$ and the weighting function is given by the identity weighting function.\footnote{Optimal risk exposures under other distortion functions can be obtained in a similar fashion.} As $\rho (t)$ increases, the optimal risk exposures in both our model and Zhang et al.'s first decrease, then increase and finally decrease. In all scenarios, when $\rho (t)$ is small, the risk exposure in our model is significantly lower than that in Zhang et al.'s. When $g$ is small, the risk exposure in our model is slightly higher than Zhang et al.'s when $\rho (t)$ takes intermediate values.

\begin{figure}[htbp]
\centering
\begin{minipage}[t]{2in}
	\centering
	\centerline{$c=0.1$ and $g=0$}
	\includegraphics[width=2.2in]{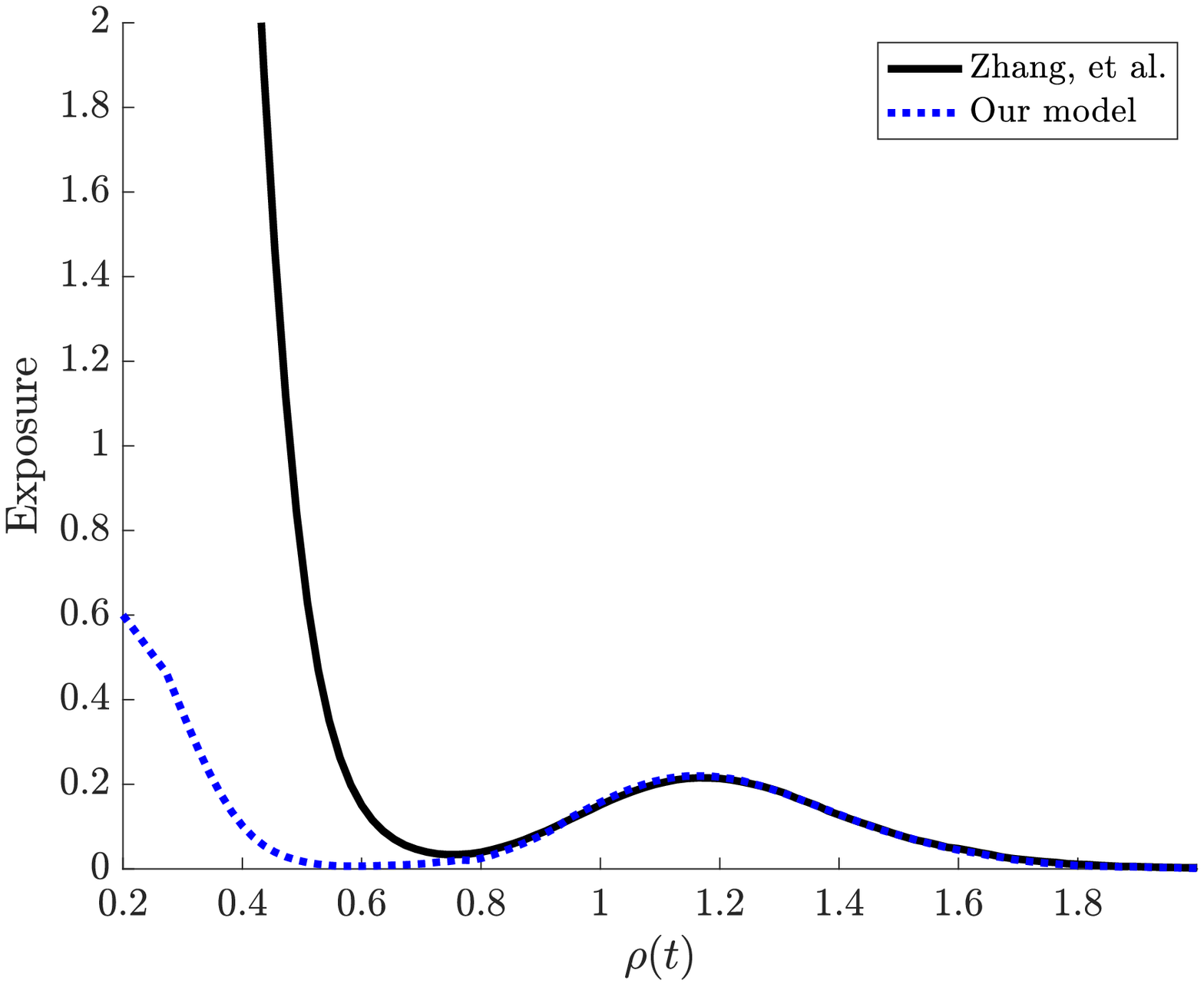}
\end{minipage}
\begin{minipage}[t]{2in}
	\centering
	\centerline{$c=0.1$ and $g=0.05$}
	\includegraphics[width=2.2in]{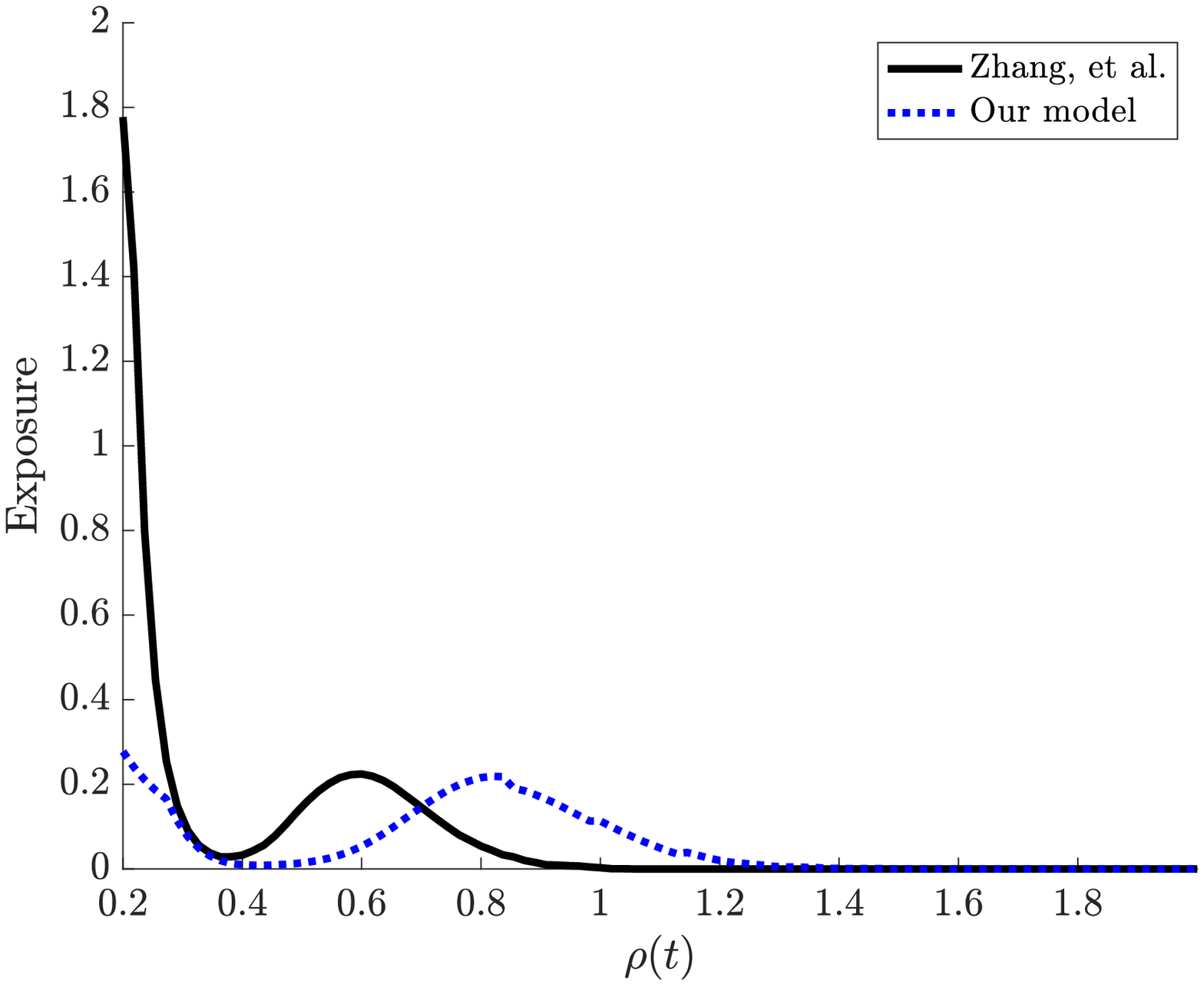}
\end{minipage}
\begin{minipage}[t]{2in}
	\centering
	\centerline{$c=0.1$ and $g=-0.05$}
	\includegraphics[width=2.2in]{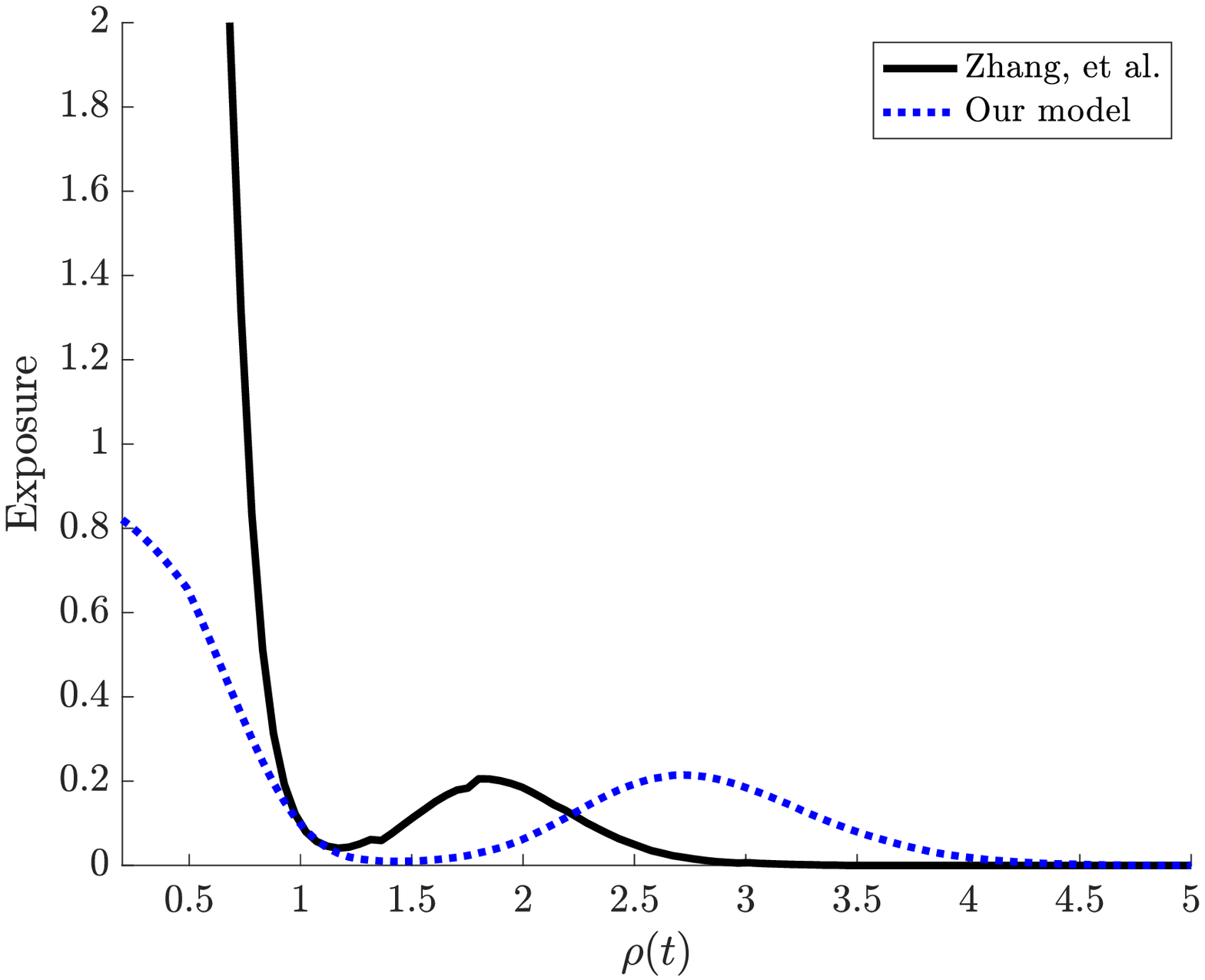}
\end{minipage}

\quad

\quad

\begin{minipage}[t]{2in}
	\centering
	\centerline{$c=0.2$ and $g=0$}
	\includegraphics[width=2.2in]{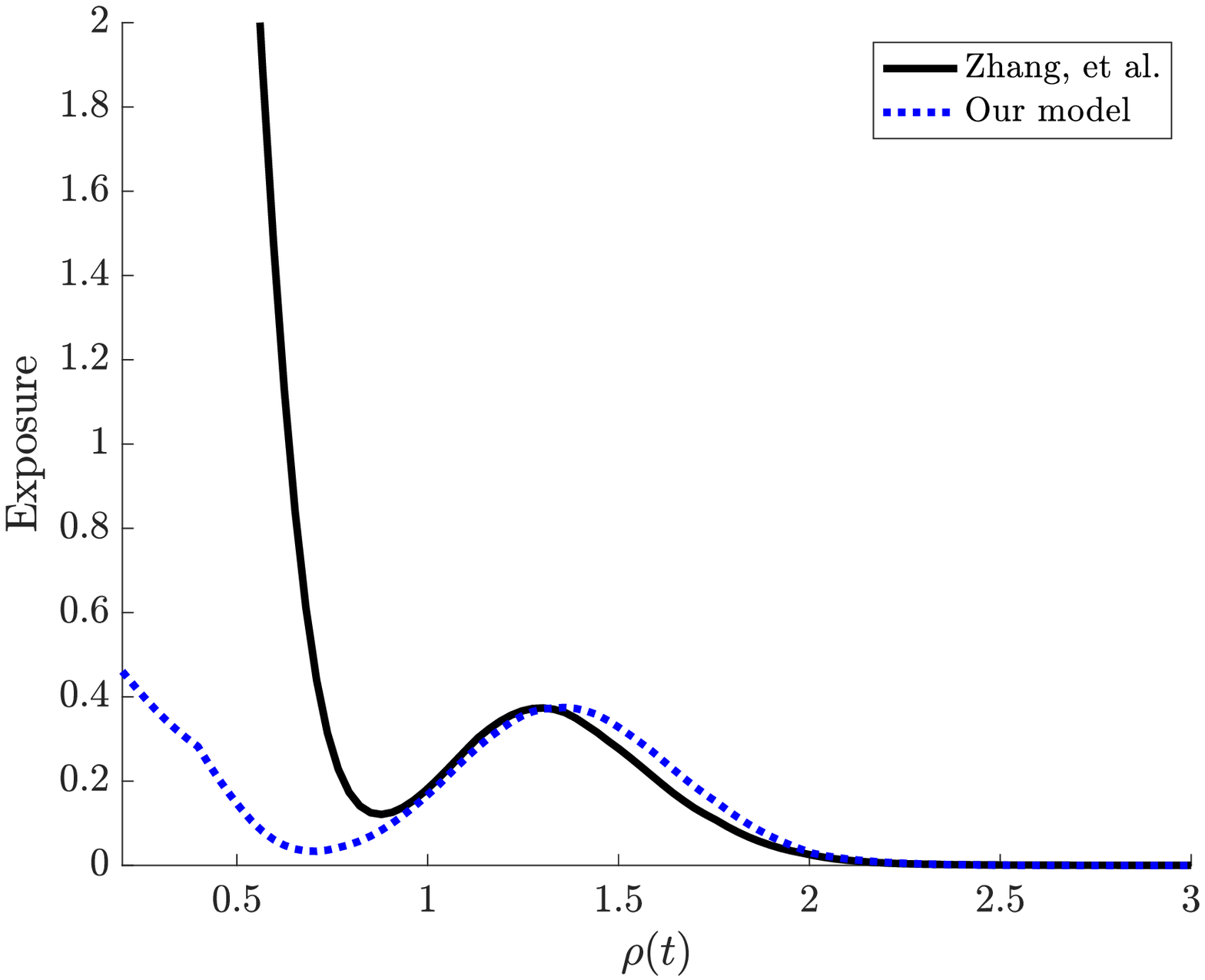}
\end{minipage}
\begin{minipage}[t]{2in}
	\centering
	\centerline{$c=0.2$ and $g=0.05$}
	\includegraphics[width=2.2in]{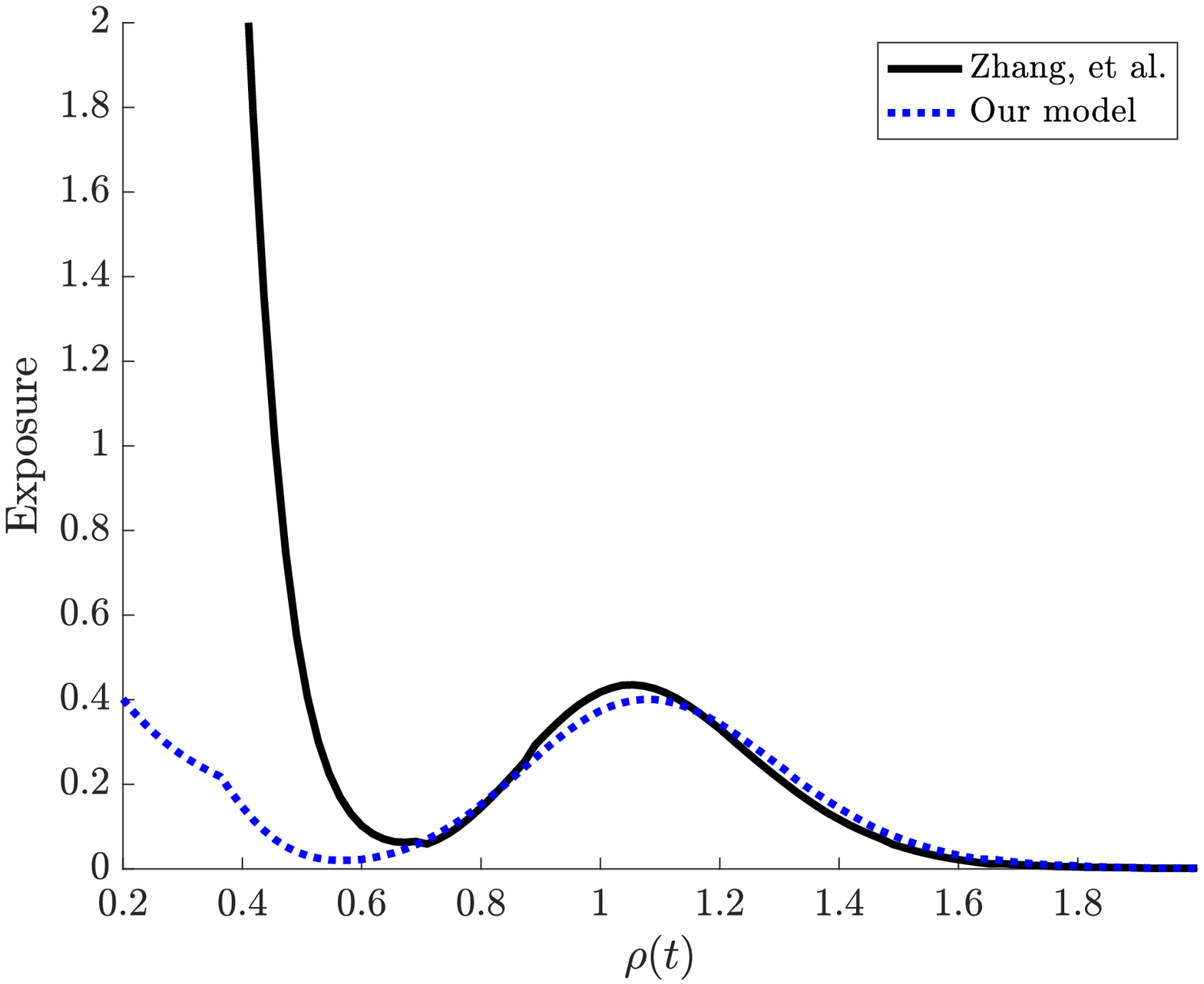}
\end{minipage}
\begin{minipage}[t]{2in}
	\centering
	\centerline{$c=0.2$ and $g=-0.05$}
	\includegraphics[width=2.2in]{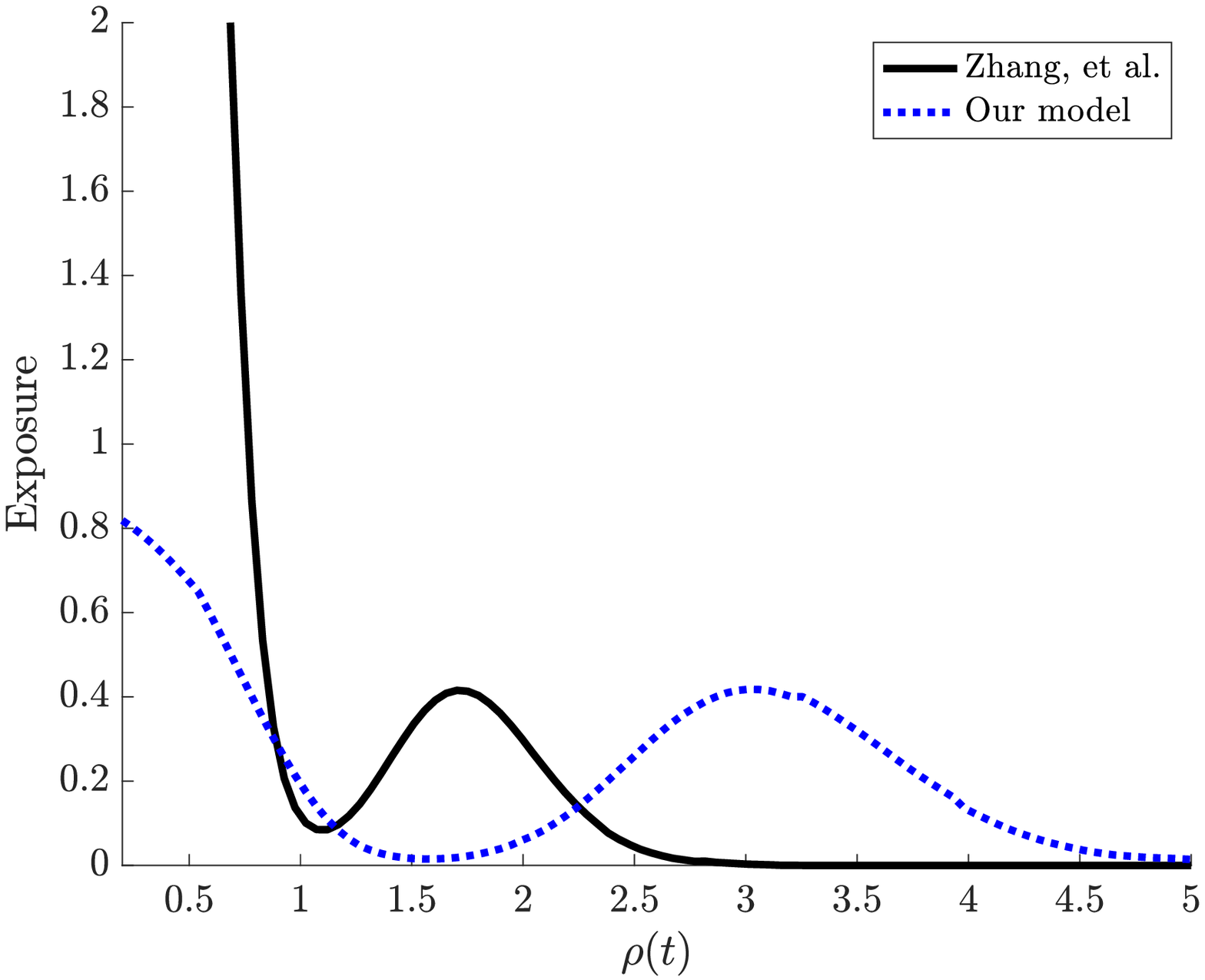}
\end{minipage}

\quad

\quad

\begin{minipage}[t]{2in}
	\centering
	\centerline{$c=0.3$ and $g=0$}
	\includegraphics[width=2.2in]{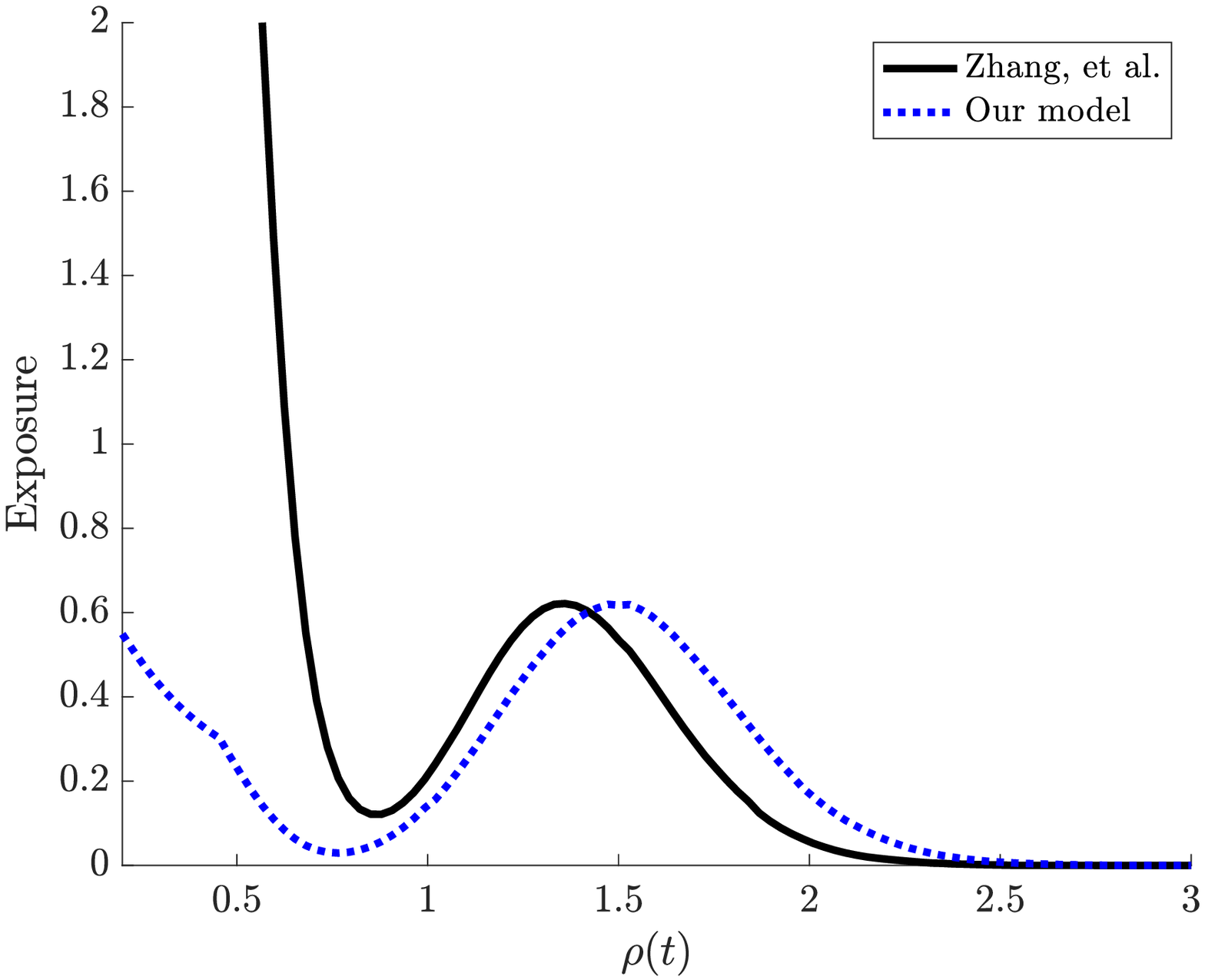}
\end{minipage}
\begin{minipage}[t]{2in}
	\centering
	\centerline{$c=0.3$ and $g=0.05$}
	\includegraphics[width=2.2in]{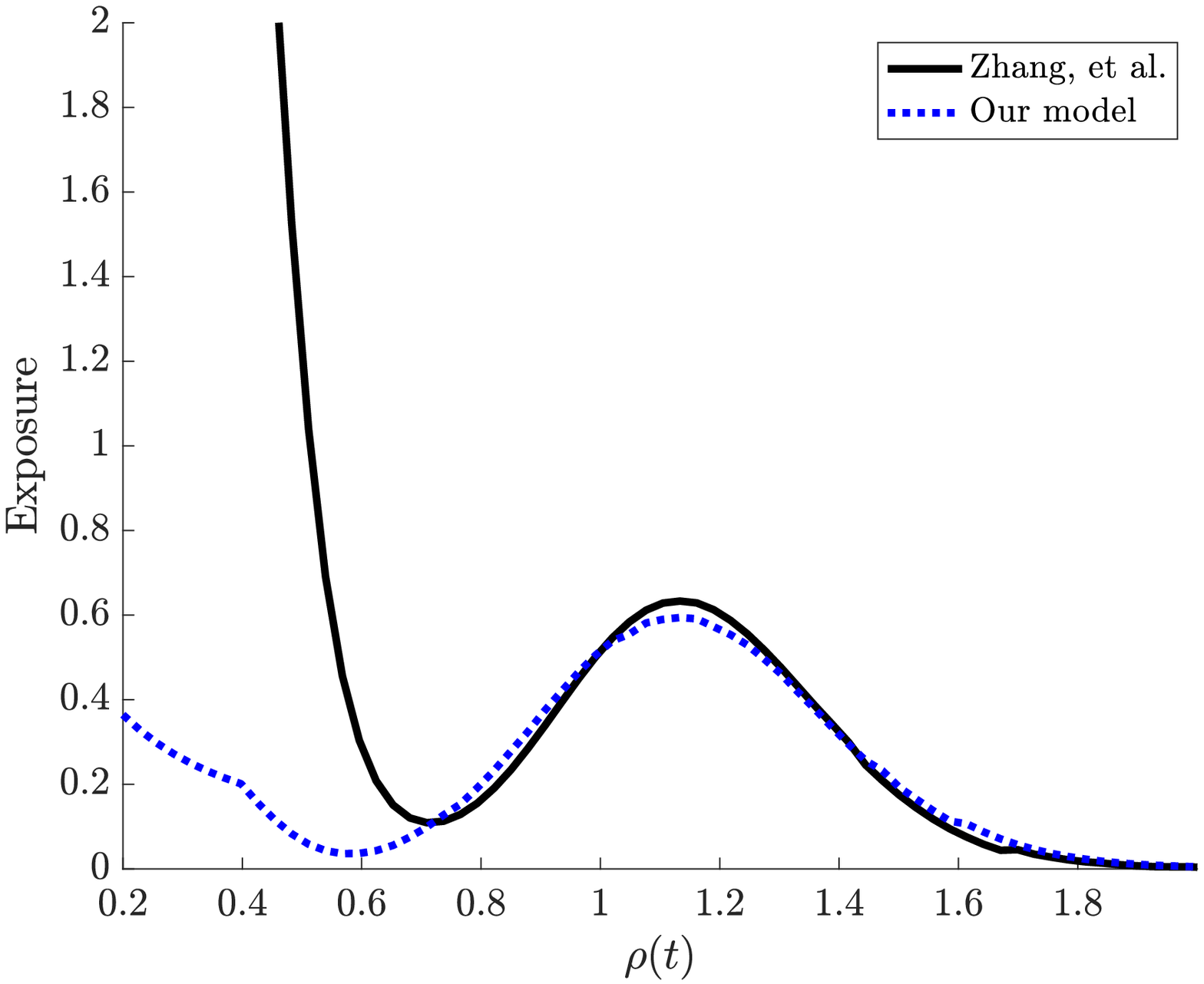}
\end{minipage}
\begin{minipage}[t]{2in}
	\centering
	\centerline{$c=0.3$ and $g=-0.05$}
	\includegraphics[width=2.2in]{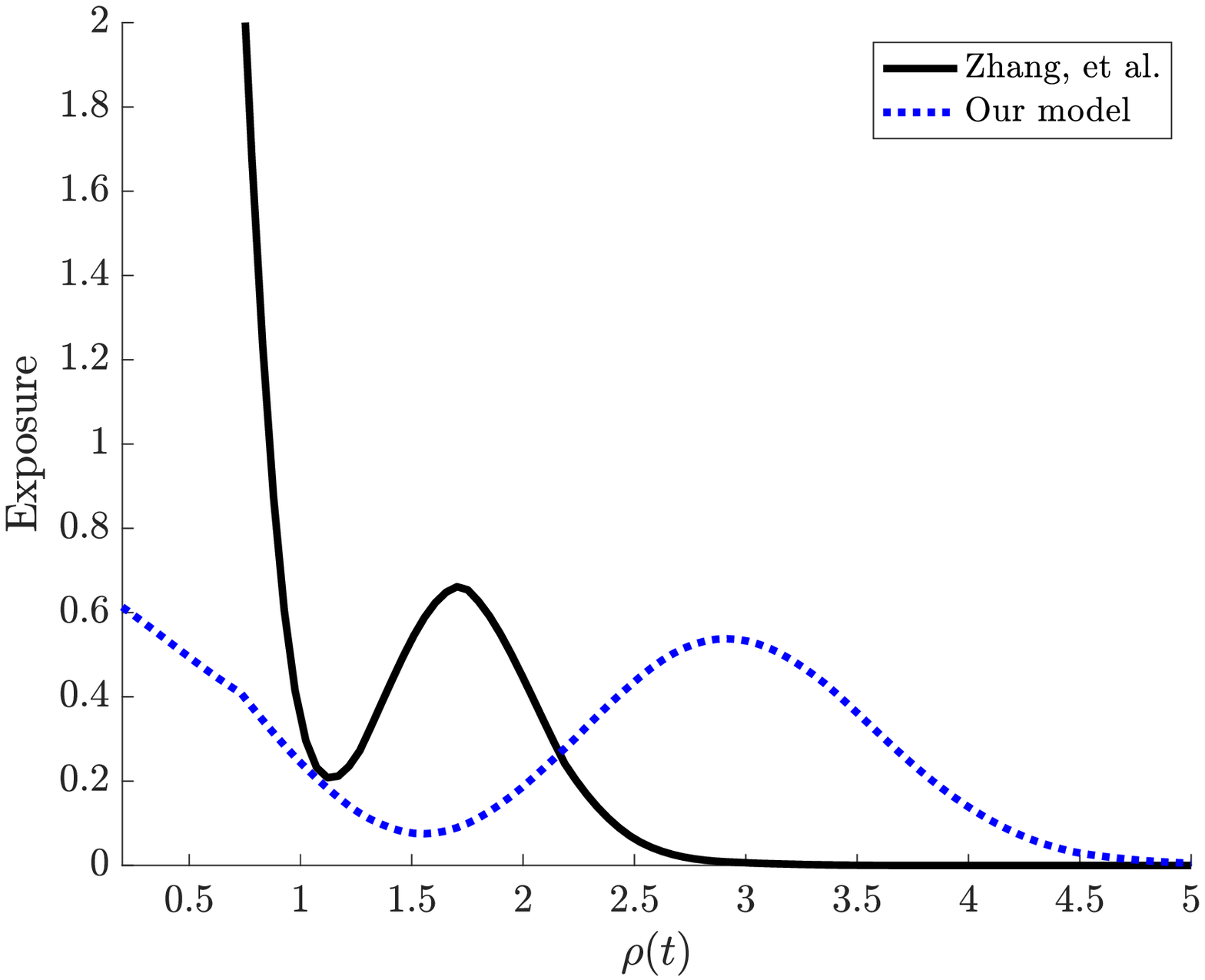}
\end{minipage}
\caption{Optimal exposures as functions of $\rho (t)$ under the identity weighting function; $t=0.5$.}\label{figure:no_weighting_strategy}
\end{figure}

}

\newpage 
\section*{Appendix}
\appendix

\section{Proofs}\label{proofs}

\begin{proofof}{\citelem{martingalemethod}} 
By Karatzas and Shreve (\cite{KS98}, p.24, Theorem 6.6), there exits an $\{\mathcal{F}_t\}$-progressively measurable process $k$ such that $X(\cdot)$ is tame (that is, the process $\rho(t)X(t)$ is lower bounded by a r.v. which has a finite expectation) and 
\begin{align*} 
\begin{cases}
\dd X(t) =\big[ r(t) X(t) + k(t)^{\prime}\sigma(t)\theta(t)\big]\dt + k(t)^{\prime}\sigma(t) \dd W(t),\quad t\geqslant 0, \\
X(0)=x_{0},\quad 
X(T) =e^{\xi}>0.
\end{cases}
\end{align*}
By It\^{o}'s lemma,
\begin{align*}
\dd\; (\rho(t)X(t))=\rho(t)\big(k(t)^{\prime}\sigma(t)-X(t)\theta(t)^{\prime}\big)\dd W(t).
\end{align*}
Hence, $\rho(t)X(t)$ is a local martingale. 
Thus, there exists a sequence of stopping times $\{\tau_n\}$ such that $\lim_n\tau_n=\infty$ and 
\[ \rho(t) X(t)= \BE{\rho(T\wedge \tau_n) X(T\wedge \tau_n)\;|\;\mathcal{F}_t}.\]
Because $X(\cdot)$ is tame, by Fatou's lemma, we can take limit to get
\[ \rho(t) X(t)= \lim_n\BE{\rho(T\wedge \tau_n) X(T\wedge \tau_n)\;|\;\mathcal{F}_t}
\geq\BE{\rho(T) X(T)\;|\;\mathcal{F}_t}>0.\]
Hence, $X(t)>0$ and we can define $\pi(t)=\frac{k(t)}{X(t)}$. Then
\begin{align*} 
\begin{cases}
\dd X(t) =X(t) \big[(r(t)+ \pi(t)^{\prime}\sigma(t)\theta(t))\dt + \pi(t)^{\prime}\sigma(t) \dd W(t)\big],\quad t\geq 0, \\
X(0)=x_{0},\quad 
X(T) =e^{\xi}>0.
\end{cases}
\end{align*}
It is not hard to check that $\pi$ fulfills the requirement. 
\end{proofof}

\begin{proofof}{\citelem{martingalemethod2}} 
Clearly $R^{\pi}(T)$ is a feasible solution to Problem \eqref{Optimalcontrol1} for any admissible portfolio $\pi$. If $\xi^{*}$ is optimal to Problem \eqref{Optimalcontrol1}, 
then because the objective functional is strictly increasing in $\xi$ in Problem \eqref{Optimalcontrol1}, we must have $ \BE{\rho e^{\xi^{*}}}=1$.
In view of \citelem{martingalemethod}, there exits an admissible portfolio $\pi^{*}$ such that $R^{\pi^{*}}(T)=\xi^{*}$. This means $\pi^{*}$ is optimal to Problem \eqref{Optimalcontrol}. 

On the other hand, if $\pi^{*}$ is an optimal portfolio to Problem \eqref{Optimalcontrol}, then $\rr^{\pi^{*}}(T)$ is a feasible solution to Problem \eqref{Optimalcontrol1}.
If it is not optimal, then there exists $\xi^{*}$ such that 
\[\BE{\rho e^{\xi^{*}}}\leq 1,\quad \xi^{*}-\bc \geq -c,\]
and
\[ \BV{u\left( \xi^{*}-\bc \right)}> \BV{u\left( \rr^{\pi^{*}}(T)-\bc \right)}.\]
Let \[\widehat{\xi}=\log \Big(e^{\xi^{*}}+\tfrac{1-\BE{\rho e^{\xi^{*}}}}{\BE{\rho}}\Big).\] Then $\BE{\rho e^{\widehat{\xi}}}=1$ and thanks to \citelem{martingalemethod}, there exists an admissible portfolio $\hat\pi$ such that $\rr^{\hat\pi}(T)=\widehat{\xi}$. Note that $\rr^{\hat\pi}(T)\geq \xi^{*}$, so 
\[ \rr^{\hat\pi}(T)-\bc \geq \xi^{*}-\bc \geq -c,\]
and
\[ \BV{u\left( \rr^{\hat\pi}(T)-\bc \right)}\geq \BV{u\left( \xi^{*}-\bc \right)}> \BV{u\left( \rr^{\pi^{*}}(T)-\bc \right)},\]
contradicting the optimality of $\pi^{*}$, thereby completing the proof.
\end{proofof}

\begin{proofof}{\citeprop{quantileformulation}} 
We need the following well-known Hardy-Littlewood inequality (also known as the Hoeffding-Frechet bounds). 
\begin{lemma}\label{HLi}
For any two r.v.s $X$ and $Y$, we have 
\[\int_{0}^{1}Q_{X}(p)Q_{Y}(1-p)\ddp\leq \BE{XY}\leq \int_{0}^{1}Q_{X}(p)Q_{Y}(p)\ddp,\] 
where the upper (resp. lower) bound is achieved if and only if $X=Q_{X}(U)$ and $Y=Q_{Y}(U)$ (resp. $X=Q_{X}(U)$ and $Y=Q_{Y}(1-U)$) a.s. for some r.v. $U$ uniformly distributed on $(0,1)$. 
\end{lemma}

We now prove \citeprop{quantileformulation}. 
Suppose $\zeta^{*}$ is an optimal solution to Problem \eqref{Obj3}.
If there is no r.v. $U\in\setu$ such that $\zeta^{*}=Q_{\zeta^{*}}(1-U)$ a.s., then by \citelem{HLi}, we have \[\int_{0}^{1}Q_{\zeta^{*}}(p)Q_{\eta}(1-p)\ddp<\BE{\zeta^{*}\eta}\leq 1.\] 
Note that
\[0<\int_{0}^{1}Q_{\eta}(1-p)\ddp=\BE{\eta}<\infty,\]
so there exists a constant $\ep>0$ such that 
\[\int_{0}^{1}(Q_{\zeta^{*}}(p)+\ep)Q_{\eta}(1-p)\ddp=1.\] 
Set $\bar\zeta=Q_{\zeta^{*}}(1-U)+\ep$ for any $U\in\setu$. Then
\[\BE{\bar\zeta\eta}=\BE{(Q_{\zeta^{*}}(1-U)+\ep)Q_{\eta}(U)}=\int_{0}^{1}(Q_{\zeta^{*}}(1-p)+\ep)Q_{\eta}(p)\ddp=1.\] 
Clearly $\bar\zeta>Q_{\zeta^{*}}(1-U)\geq \qlow$ and 
\[\BV{v(\bar\zeta)}> \BV{v(Q_{\zeta^{*}}(1-U))}=\BV{v(\zeta^{*})},\]
contradicting the optimality of $\zeta^{*}$ to Problem \eqref{Obj3}. Therefore, there is a r.v. $U\in\setu$ such that $\zeta^{*}=Q_{\zeta^{*}}(1-U)$ a.s.

Now suppose $Q_{\zeta^{*}}$ is not optimal to Problem \eqref{quantile1}. 
Let $\barQ$ be an optimal solution to Problem \eqref{quantile1}. Define $\bar\zeta=\barQ(1-U)$ for $U\in\setu$. 
Then
\begin{align*} 
\BE{\bar\zeta\eta}=\BE{\barQ(1-U)Q_{\eta}(U)}=\int_{0}^{1}\barQ(1-p)Q_{\eta}(p)\ddp=1,\quad \bar\zeta\geq \barQ(0)\geq \qlow.\nn
\end{align*}
Because $\barQ$ is optimal to Problem \eqref{quantile1} but $Q_{\zeta^{*}}$ is not, we have
\[ \BV{v(\bar\zeta)}=\int_{0}^{1}v(\barQ(p))w'(1-p)\ddp>\int_{0}^{1}v(Q_{\zeta^{*}}(p))w'(1-p)\ddp=\BV{v(\zeta^{*})}, \]
contradicting the optimality of $\zeta^{*}$ to Pproblem \eqref{Obj3}. 

The reverse direction can be proved in a similar way. We leave it to the interested readers.
\end{proofof}

\begin{proofof}{\citelem{twopoints}}
Let $$
f_{1}(y)=\inf \{y x-v(x): x \geq 1\}
$$
and $$
f_{2}(y)=\inf \{y x-v(x): 0<x \leq 1\}
$$
Then both $f_1$ and $f_2$ are continuous functions in $(0,\infty)$.
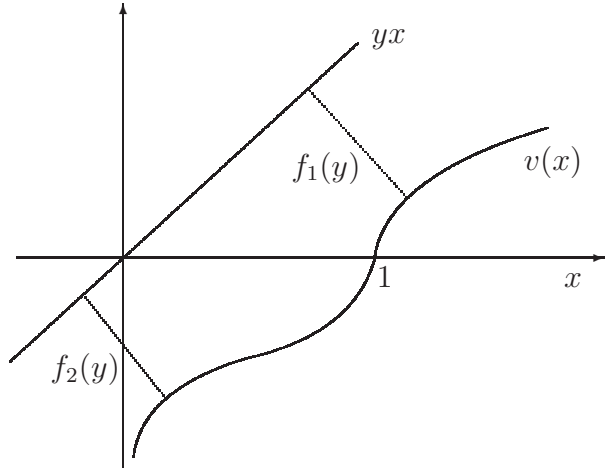
\begin{figure}[H]
\begin{center}
\begin{picture}(250,180)
\linethickness{0.6pt}
\put(10,89){\vector(1,0){220}} 
\put(50,10){\vector(0,1){175}} 
\put(215,78){$x$}
\qbezier(54,14)(57,39)(96,51)
\qbezier(96,51)(136,60)(144,88)
\qbezier(144,88)(148,120)(209,138)
\put(200,122){$v(x)$}
\put(145,78){$1$}

\qbezier(8, 50)(73, 110)(138, 170)
\put(143, 170){$y x$}

\qbezier[40](156, 112) (138, 132) (120, 152)
\put(113, 120){$f_1(y )$}

\qbezier[40](66, 36) (51, 55) (36, 74)
\put(23, 44){$f_2(y )$}

\linethickness{0.9pt} 
\end{picture}
\caption{The values of $f_1$ and $f_2$ in dot line.}
\end{center}
\end{figure}

One can show $f_1(y)>f_2(y)$ when $y$ is sufficiently large, and $f_1(y)<f_2(y)$ when $y$ is small, so that $f_1(y)=f_2(y)$ for some $y$. Also there exist $b>1$ and $0<a<1$ such that $f_1(y)=y b- v(b)$ and $f_2(y)=y a- v(a)$. It thus follows $y a- v(a)=y b- v(b)$. Since $a$ and $ b$ minimize $x\mapsto y x-v(x)$, respectively, on $(0,1)$ and $(1,\infty)$, the first order condition gives $v'(b)=y$ and $v'(a)=y$. This validates the equation \eqref{abequation}.
Let $\hat{v}_{0}$ coincide with $v$ on $(0, a]\cup[b,\infty)$ and be linear on $[a,b]$. It is not hard to verify that $\hat{v}_{0}$ is the concave envelope of $v$ on $(0,\infty)$. 
This completes the proof. 
\end{proofof}

\begin{proofof}{\citelem{Iprop}}
The first three properties are easy to verify, let us prove the last property only. In fact, $\hatv$ is concave and for any $x>0$, we have
\[\sup_{y\geq \qlow}(\hatv(y)-xy)=\hatv(\inv(x))-x\inv(x)=v(\inv(x))-x\inv(x)\leq \sup_{y\geq \qlow}(v(y)-xy),\]
where the second equation is due to the third property. 
The reverse inequality
\[\sup_{y\geq \qlow}(\hatv(y)-xy)\geq \sup_{y\geq \qlow}(v(y)-xy)\]
is trivial as $\hatv\geq v$ on $[\qlow,\infty)$, so the above two inequalities are indeed identities.
\end{proofof}

\begin{proofof}{\citeprop{wellposedness}} 
$1\Longleftrightarrow2.$
The change of variables argument shows that Problems \eqref{Optimalcontrol0} and \eqref{quantile2} are equivalent, so are their well-posedness. 

$2\Longleftrightarrow3.$
Because $v$ and $\hatv$ only differ on $[a,b]$, there exists a constant $\ep>0$ such that $v\leq\hatv\leq v+\ep$. Consequently, the well-posedness of Problems \eqref{quantile2} and \eqref{q1} are equivalent.

$2\Longleftrightarrow4.$ It is easy to show that there exists a constant $\ep>0$ such that $(\log(x+1))^{\alpha}-\ep\leq v(x)\leq (\log(x+1))^{\alpha}+\ep$ for all $x\in[\qlow,\infty)$. Hence, Problem \eqref{q1} is well-posed if and only if so is the following problem
\begin{align}\label{q1a}
\max_{G\in\setq } &\quad \int_{0}^{1}(\log (G(p)+1))^{\alpha}\ddp \\
\mathrm{s.t.} &\quad \int_{0}^{1}G(p)\varphi'(p)\ddp=1,\quad G(0)\geq \qlow. \nn
\end{align} 
Note $G\geq \qlow$ and the mapping $x\mapsto (\log(x+1))^{\alpha}$ is strictly concave on $[\qlow,\infty)$, so the above problem is a concave optimization problem. 
According to Xu \cite{X16}, Problem \eqref{q1a} 
is well-posed if and only if so is the following problem
\begin{align}\label{q1b}
\max_{G\in\setq } &\quad \int_{0}^{1}(\log (G(p)+1))^{\alpha}\ddp \\
\mathrm{s.t.} &\quad \int_{0}^{1}G(p)\hatphi'(p)\ddp=1,\quad G(0)\geq \qlow. \nn
\end{align} 
Moreover, Problem \eqref{q1b} is the quantile formulation of the following problem 
\begin{align*}
\max_{\xi} &\quad \BE{(\log (\xi+1))^{\alpha}} \\
\mathrm{s.t.} &\quad \BE{\xi\barho}=1,\quad \xi\geq \qlow, \nn
\end{align*} 
where $\barho=\hatphi'(U)$ and $U$ is a r.v. uniformly distributed on $(0,1)$.
Clearly, it does not affect the well-posedness of the above problem if we replace the last constraint $\xi\geq \qlow$ with $\xi\geq 0$.
By \cite[Theorem 3.1]{JXZ08}, this problem is well-posed if and only if there exits a constant $\lambda>0$ such that 
\begin{align}\label{cond1}
\BE{(\log (g(\lambda\barho)+1))^{\alpha}+g(\lambda\barho)\barho}<\infty,
\end{align}
where $g:(0,\infty)\to (0,\infty)$ is the inverse function of the continuous strictly decreasing function 
$$x\mapsto \frac{\alpha}{x+1} (\log (x+1))^{\alpha-1},\quad x>0.$$ 
Because $g$ is decreasing, 
\begin{align*} 
&\quad\;\BE{\big\{(\log (g(\lambda\barho)+1))^{\alpha}+g(\lambda\barho)\barho\big\}\idd{\lambda\barho\geq 1/2}}\\
&\leq \BE{\big\{(\log (g(1/2)+1))^{\alpha}+g(1/2)\barho\big\}\idd{\lambda\barho\geq 1/2}}\\
&\leq (\log (g(1/2)+1))^{\alpha}+g(1/2)\BE{\barho}<\infty.
\end{align*}
Hence, the condition \eqref{cond1} is equivalent to 
\begin{align}\label{cond2}
\BE{\big\{(\log (g(\lambda\barho)+1))^{\alpha}+g(\lambda\barho)\barho\big\}\idd{\lambda\barho<1/2}}<\infty.
\end{align}

From 
\[\frac{\alpha}{g(x)+1} (\log (g(x)+1))^{\alpha-1}=x,\] 
we get $$\lim_{x\to 0} g(x)= +\infty.$$ It thus follows
\begin{align*}
\lim_{x\to 0} x g(x)=\lim_{x\to 0}\frac{\alpha g(x)}{g(x)+1} (\log (g(x)+1))^{\alpha-1}=0,
\end{align*}
and hence,
\begin{align}\label{g1} 
\BE{g(\lambda\barho)\barho\idd{\lambda\barho<1/2}}<\infty.
\end{align}
We also have 
\begin{align*}
\lim_{x\to 0}\frac{\log (g(x)+1)}{-\log x}
=\lim_{x\to 0}\frac {\log (g(x)+1)}{\log (g(x)+1)-\log \alpha-(\alpha-1)\log \log (g(x)+1)}
=1.
\end{align*}
Hence, the condition \eqref{cond2} is equivalent to 
\begin{align*} 
\BE{(-\log (\lambda\barho))^{\alpha} \idd{\lambda\barho<1/2}}<\infty.
\end{align*}
Because 
\begin{align*} 
\lim_{x\to 0}\frac{-\log (\lambda x)}{-\log x}=1,
\end{align*}
the above estimate is equivalent to 
\begin{align*} 
\BE{ ( \max\{-\log\barho,\; 0\})^{\alpha} }<\infty,
\end{align*}
namely
\begin{align*} 
\int_0^1 \big(\max\{-\log \hatphi'(p),\;0\}\big)^{\alpha} \ddp<\infty.
\end{align*}
This completes the proof. 
\end{proofof}

\begin{proofof}{\citethm{vt}}
Because $\inv$ is right-continuous and $\hatphi'$ is right-continuous, we see $G^{*}$ is right-continuous. 
From the definition of $\inv$, we have $G^{*}\geq \qlow$. Together with \eqref{lambda1}, we infer that $G^{*}$ is a feasible solution to Problems \eqref{quantile2} and \eqref{q1}. Moreover, by Xia and Zhou \cite{XZ16} or Xu \cite{X16}, it is optimal to Problem \eqref{q1}. By virtue of \citelem{Iprop}, 
\[\hatv(G^{*}(p))=\hatv(\inv(\lambda \hatphi'(p)))=v(\inv(\lambda \hatphi'(p)))=v(G^{*}(p)).\]
Therefore, for any feasible solution $G$ to Problem \eqref{quantile2}, thanks to $v\leq \hatv$ on $[\qlow,\infty)$, we have
\[\int_{0}^{1}v(G(p))\ddp \leq \int_{0}^{1}\hatv(G(p))\ddp \leq \int_{0}^{1}\hatv(G^{*}(p))\ddp
=\int_{0}^{1}v(G^{*}(p))\ddp. \]
This means $G^{*}$ is also optimal to Problem \eqref{quantile2}. 

For any $U\in\setu$, we have $Q_{\eta}(U)=\eta$, so
\begin{align*} 
\BE{G^{*}(1-w(U))\eta} &=\BE{G^{*}(1-w(U))Q_{\eta}(U)}\\
&=\int_{0}^{1}G^{*}(1-w(p))Q_{\eta}(p)\ddp =\int_{0}^{1}Q^{*}(1-p)Q_{\eta}(p)\ddp= 1.
\end{align*} 
where the last equation is due to the fact that $Q^*$ is a feasible solution to Problem \eqref{quantile1}.
Hence, 
$$\BE{\rho(T) e^{\bc} G^{*}(1-w(U))}=\BE{\eta G^{*}(1-w(U))}=1. $$
By the martingale representation theorem, there exists an $\{\mathcal{F}_t\}$-progressively measurable process $Z$ such that 
\[\int_{0}^{T} \Vert Z(s)\Vert^2 \ds<\infty, \text { a.s.,}\]
and
\begin{align*} 
\BE{\rho(T) e^{\bc} G^{*}(1-w(U))\;\big|\;\mathcal{F}_t}&=\BE{\rho(T) e^{\bc} G^{*}(1-w(U))}+\int_0^t Z(s)'\dd W(s)\\
&=1+\int_0^t Z(s)'\dd W(s).
\end{align*} 
Let \[Y(t)=\frac{1}{\rho(t)}\BE{\rho(T) e^{\bc} G^{*}(1-w(U))\;\big|\;\mathcal{F}_t}.\]
It follows that $$\rho(t) Y(t)=1+\int_0^t Z(s)'\dd W(s),$$ 
or
\[\dd \big(\rho(t) Y(t)\big)=Z(t)'\dd W(t)= Y(t)\pi^*(t)^{\prime}\sigma(t)\dd W(t).\]
Then It\^{o}'s lemma yields 
\[\dd Y(t) =Y(t) \big[(r(t)+ \pi^*(t)^{\prime}\sigma(t)\theta(t))\dt + \pi^*(t)^{\prime}\sigma(t) \dd W(t)\big]. \]
Notice 
$$Y(0)=\BE{\rho(T) e^{\bc} G^{*}(1-w(U))}=1. $$
It is not hard to verify that $(\pi^*, X^{\pi^*})\equiv (\pi^*, x_0Y)$ satisfies the wealth process \eqref{wealthprocess}. 
Hence $\pi^{*}$ is an optimal portfolio to Problem \eqref{Optimalcontrol0} and $$R^{\pi^{*}}(T)=\log (X^{\pi^{*}}(T)/x_0)=\log Y(T)=\bc+\log (G^{*}(1-w(U)))$$ is the corresponding optimal terminal growth rate.
\end{proofof}

\section{Change of variable}\label{changeofvariable}
The mapping between $Q$ and $G$ is bijective since $\nu$ is continuous and strictly increasing. Therefore, $G$ is a quantile function if and only if so is $Q$. Also 
\begin{align*}
\int_{0}^{1}v(Q(p))w'(1-p)\ddp&=\int_{0}^{1}v(Q(p))\dd\; (1-w(1-p))\\
&=\int_{0}^{1}v(Q(\nu(p)))\dd\; (1-w(1-\nu(p)))\\
&=\int_{0}^{1}v(G(p))\ddp,
\end{align*}
and
\begin{align}
\int_{0}^{1}Q(p)Q_{\eta}(1-p)\ddp=\int_{0}^{1}Q(\nu(p))Q_{\eta}(1-\nu(p))\dd \nu(p)=\int_{0}^{1}G(p)\varphi'(p)\ddp. 
\end{align} 
Notice $Q(0)\geq \qlow$ if and only $G(0)\geq \qlow$. Therefore, Problem \eqref{quantile1} is, in terms of $G$, equivalent to Problem \eqref{quantile2}. 

\blue{
\section{Existence of the Lagrange multiplier}\label{existenceofLagrangemultiplier}

Theorem \ref{vt} characterizes the optimal solution assuming the existence of a Lagrange multiplier $\lambda$ that solves \eqref{lambda1}. We now discuss the existence of the Lagrange multiplier. This is not a trivial task because, even in the absence of probability distortion, the Lagrange multiplier may not always exist; see Jin, Xu and Zhou \cite{JXZ08} for a comprehensive study. 

Let \[f(\lambda)=\int_{0}^{1}I(\lambda\hatphi'(p))\hatphi'(p)\ddp,\]
and $\ep=\lim_{p\to1}\hatphi'(p)$. If $\ep>0$, then since $I$ and $\hatphi'$ are decreasing, 
\[f(\lambda)\leq I(\lambda\ep)\int_{0}^{1}\hatphi'(p)\ddp=I(\lambda\ep)<\infty.\]
If $\ep=0$, then when $p$ is sufficiently close to 1, 
\begin{align*}
I(\lambda\hatphi'(p))\hatphi'(p)=(v')^{-1}(\lambda\hatphi'(p))\hatphi'(p)=(g(\lambda\hatphi'(p))+1)\hatphi'(p),
\end{align*}
where $g$ is defined in the proof of \citeprop{wellposedness}. Because the problem \eqref{q1} is well-posed, the function $p\mapsto g(\lambda\hatphi'(p))\hatphi'(p)$ is integrable by \eqref{g1}. Therefore, $f$ is well-defined and finite on $(0,\infty)$.

Since $I$ is decreasing and right-continuous, by the monotone convergence theorem, so is $f$. Moreover, thanks to \citeassmp{feasibility} and \citelem{Iprop}, 
\[\lim_{\lambda\to 0}f(\lambda)=+\infty,\qquad \lim_{\lambda\to \infty}f(\lambda)=\hat{c}\int_{0}^{1}\hatphi'(p)\ddp=\hat{c} \BE{\eta}<1. \]
If $f$ is continuous on $(0,\infty)$, then there exists $\lambda>0$ such that \eqref{lambda1} holds. 
Unfortunately, in our model, $f$ can be discontinuous due to the jump continuity in $I$, so the existence of $\lambda$ is not guaranteed. We note that a similar issue, i.e., the existence of the Lagrange multiplier, also arises in the non-concave quantile optimization problem studied in Bi et al. \cite{BYCF21}. Therefore, we provide heuristic derivation in our subsequence analysis and refer interested readers to Bi et al. \cite{BYCF21} for a more detailed discussion.

Recall (from Lemma \ref{Iprop}) that $I$ has a unique jump at $x=v'(d)$ when $a\leq \qlow$ or at $x=v'(a)$ when $a> \qlow$. Let 
\begin{equation*}
\tilde{v} = \left\{
\begin{aligned}
	v'(d), ~ \text{if } a\leq \qlow,\\
	v'(a), ~ \text{if } a> \qlow.
\end{aligned}
\right.
\end{equation*}
Then $I$ has a jump continuity at $\tilde{v}$ and is continuous otherwise.

For any $\lambda>0$, define the set
$$\mathcal{P}(\lambda) = \big\{ p \in [0,1] \;\big|\; \lambda \hatphi'(p) = \tilde{v}\big\}. $$
It is straightforward to verify that $f$ is continuous for $\lambda$ such that $\mathcal{P}(\lambda)$ has Lebesgue measure $0$.

Let us consider three scenarios.

In the first scenario, we assume $\varphi$ is concave (e.g., when there is no probability distortion). Thus, $\hatphi\equiv\varphi$ and $\hatphi'(p)$ is strictly decreasing. Consequently, $\mathcal{P}(\lambda)$ has measure $0$ for all $\lambda>0$ and $f$ is continuous. We can then find $\lambda>0$ such that $f(\lambda)=1$.

In the second scenario, we assume $\varphi$ is $S$-shaped so that 
$$\big\{ p \in [0,1] \;\big|\; \hatphi(p) > \varphi (p) \big\} = (0,p_0)$$ 
for some $p_0 \in (0,1)$. This is not a restrictive assumption. As discussed in Bi et al. \cite{BYCF21}, this assumption is valid for commonly used inverse $S$-shaped distortion functions when the pricing kernel $\rho$ is log-normally distributed. Recall that $\hatphi(p)$ is affine and $ \hatphi'(p) $ is constant when $\hatphi(p) > \varphi (p)$, or equivalently, $p \in (0,p_0)$. For $p \in (p_0,1)$, $ \hatphi'(p) $ is strictly decreasing. Let $$\lambda_0 = \frac{\tilde{v}}{\hatphi'(p_0)}.$$
For $\lambda \neq \lambda_0$, $\mathcal{P}(\lambda)$ has measure $0$. Therefore, $f(\lambda)$ is continuous on $(0,\lambda_0)$ and $(\lambda_0,+\infty)$. Recall that $\lim_{\lambda\to 0}f(\lambda)=+\infty$ and $\lim_{\lambda\to \infty}f(\lambda)<1$. If $\lim_{\lambda \uparrow \lambda_0} f(\lambda) < 1$ or $\lim_{\lambda \downarrow \lambda_0} f(\lambda)\ge 1$, then we can find $\lambda>0$ such that $f(\lambda)=1$. Otherwise, the Lagrange multiplier may not exist.

In the third scenario, we consider a general $\varphi$. Because
$$\big\{ p \in [0,1] \;\big|\; \hatphi(p) > \varphi (p) \big\}$$ 
 is an open set, it can be expressed as a union of countable disjoint open intervals. Therefore, $ \big\{\hatphi'(p)\;\big|\; \hatphi(p) > \varphi (p) \big\}$ is a countable set. Let
$$\Lambda = \left\{ \frac{\tilde{v}}{\hatphi'(p)} \;\bigg|\; \hatphi(p) > \varphi (p) \right\}, $$
which is also a countable set. For $\lambda \notin \Lambda$, $\mathcal{P}(\lambda)$ has measure $0$. Therefore, $f$ is continuous for $\lambda \notin \Lambda$. If
$$1 \notin \bigcup_{\lambda_0 \in \Lambda} \left[\lim_{\lambda \downarrow \lambda_0} f(\lambda) , \lim_{\lambda \uparrow \lambda_0} f(\lambda) \right),$$ 
then we can find $\lambda>0$ such that $f(\lambda)=1$. Otherwise, the Lagrange multiplier may not exist.\footnote{If there exists $\lambda_0 \in \Lambda$ such that $\lim_{\lambda \downarrow \lambda_0} f(\lambda)=1$, then the Largange multiplier may not exists but $G^{*}(p)=\lim_{\lambda \downarrow \lambda_0} \inv(\lambda \hatphi'(p)), ~ p \in(0,1),$ is optimal to \eqref{q1}.}

As remarked in Bi et al. \cite{BYCF21}, when the Lagrange multiplier does not exist, we can no longer use the Lagrange method to solve the optimization problem, and finding the optimal solution, in this case, remains an interesting topic for future research.

}

\newpage


\end{document}